\documentclass[acmsmall]{acmart}

\makeatletter
\def\@ACM@checkaffil{
    \if@ACM@instpresent\else
    \ClassWarningNoLine{\@classname}{No institution present for an affiliation}%
    \fi
    \if@ACM@citypresent\else
    \ClassWarningNoLine{\@classname}{No city present for an affiliation}%
    \fi
    \if@ACM@countrypresent\else
        \ClassWarningNoLine{\@classname}{No country present for an affiliation}%
    \fi
}
\makeatother

\usepackage{float}
\usepackage{rotating}
\usepackage{hyperxmp}
\usepackage{graphicx}
\usepackage{tikz}
\usepackage{lscape}
\usepackage{forest}
\usepackage{subcaption}
\usetikzlibrary{trees,positioning,shapes,shadows,arrows.meta}

\usepackage{pgf-pie}

\AtBeginDocument{%
  \providecommand\BibTeX{{%
    \normalfont B\kern-0.5em{\scshape i\kern-0.25em b}\kern-0.8em\TeX}}}

\acmJournal{JACM}
\acmVolume{37}
\acmNumber{4}
\acmArticle{111}
\acmMonth{8}

\begin{document}

\title{Cold Start Latency in Serverless Computing: A Systematic Review, Taxonomy, and Future Directions}

\author{Muhammed Golec}
	\email{m.golec@qmul.ac.uk}
	\affiliation{%
		\institution{Queen Mary University of London}
  \city{London}
		\country{UK}}

\author{Guneet Kaur Walia and Mohit Kumar}
	\affiliation{%
		\institution{Dr. B. R. Ambedkar National Institute of Technology, Jalandhar}
  \city{Jalandhar}
		\country{India}}

  \author{Felix Cuadrado}
	\email{felix.cuadrado@upm.es}
	\affiliation{%
		\institution{Technical University of Madrid (UPM)}
  \city{Madrid}
		\country{Spain}}

  \author{Sukhpal Singh Gill and Steve Uhlig}
	\email{s.s.gill@qmul.ac.uk}
	\affiliation{%
		\institution{Queen Mary University of London}
  \city{London}
		\country{UK}}

\authorsaddresses{%
		Authors’ addresses: Muhammed Golec, Sukhpal Singh Gill and Steve Uhlig are with the School of Electronic Engineering and Computer Science, Queen Mary University of London, United Kingdom; emails: m.golec@qmul.ac.uk, s.s.gill@qmul.ac.uk, steve.uhlig@qmul.ac.uk; Muhammed Golec is also with Abdullah Gul University, Kayseri, Turkey; Guneet Kaur Walia and Mohit Kumar are with Department of Information Technology, Dr. B. R. Ambedkar National Institute of Technology, Jalandhar, Punjab, India; emails: kumarmohit@nitj.ac.in, guneetkw.it.22@nitj.ac.in; Felix Cuadrado is with the Technical University of Madrid (UPM), Spain. email: felix.cuadrado@upm.es; Muhammed Golec would express his thanks to the Ministry of Education of the Turkish Republic, for their support and funding. F. Cuadrado has been supported by the HE ACES project, Spain (Grant No. 101093126)}

\renewcommand{\shortauthors}{Golec, et al.}

\begin{abstract}
\textcolor{black}{Recently, academics and the corporate sector have paid attention to serverless computing, which enables dynamic scalability and an economic model. In serverless computing, users only pay for the time they actually use resources, enabling zero scaling to optimise cost and resource utilisation. However, this approach also introduces the serverless cold start problem. Researchers have developed various solutions to address the cold start problem, yet it remains an unresolved research area. In this article, we propose a systematic literature review on clod start latency in serverless computing. Furthermore, we create a detailed taxonomy of approaches to cold start latency, which we use to investigate existing techniques for reducing the cold start time and frequency. We have classified the current studies on cold start latency into several categories such as caching and application-level optimisation-based solutions, as well as Artificial Intelligence (AI)/Machine Learning (ML)-based solutions. Moreover, we have analyzed the impact of cold start latency on quality of service, explored current cold start latency mitigation methods, datasets, and implementation platforms, and classified them into categories based on their common characteristics and features. Finally, we outline the open challenges and highlight the possible future directions.}
\end{abstract}



\maketitle

\section{Introduction}

\textcolor{black}{ With the developments in virtualization technology, the concept of cloud computing has witnessed developments that abstract responsibilities such as server maintenance and infrastructure management from customers and put them under the responsibility of service providers \cite{gill2022ai}. In light of these developments, one of the latest developments that integrates Backend as a Service (BaaS) and Function as a Service (FaaS) services is serverless computing \cite{jonas2019cloud}. The first example of serverless computing is the Lambda platform introduced by Amazon in 2014 \cite{villamizar2016infrastructure}. In this service model, code developers write their code as functions. These functions are processed by triggers such as  Hypertext Transfer Protocol (HTTP) in serverless computing operating in event-driven architecture. In this way, customers can focus only on code development by abstracting from server management processes such as server and container management. In addition to this benefit, it has other advantages such as serverless computing, dynamic scalability, and a pay-as-you-go economic pricing model \cite{golec2023healthfaas}. This means that resources can automatically scale up and down based on demand, and are only charged for the time the resources are used.} 

Literature reported that interest in serverless computing is growing rapidly and its market size will reach \$22 billion by 2025 \cite{wen2023rise}. The same research results reported that by 2025, around 50\% of global businesses could use serverless computing. Serverless computing, along with developing processor and sensor technologies, shapes technology by being used in a wide variety of sectors and fields. One of these areas is edge computing, where latency and bandwidth performance are increased by bringing the processing power closer to the data source \cite{chen2018edge}. In order to benefit from the advantages of serverless computing and edge computing, researchers have focused on the serverless edge computing paradigm, where these two concepts are integrated. This means lower latency and bandwidth consumption, as well as isolating infrastructure management from the customer and automatically scaling resources \cite{xie2021serverless}. Additionally, since data is processed on edge devices, this means lower security and privacy issues \cite{baresi2019towards}. Research conducted by \textit{MarketsandMarkets} shows that the total Serverless edge computing market will be approximately 100 billion dollars by the end of 2029 with the development of 5G and Internet of Things (IoT) technologies \cite{MarketsandMarkets}.

\textcolor{black}{In addition to the advantages it offers, serverless computing also brings with it challenges such as security and privacy concerns, resource management, and the cold start latency that are still waiting to be solved \cite{golec2021ifaasbus}. One of the most important of these challenges is the cold start. In the serverless paradigm, idle resources are released due to energy efficiency and favorable pricing policy. This process is known as scaling to zero \cite{castro2019rise}. Resources scaled to zero take some time to be processed for reuse when a new function arrives. This latency is called a cold start in serverless computing \cite{baldini2017serverless}. Cold start latency has negative effects such as delayed response to time-sensitive applications, inconsistent performance, increased resource consumption, and poor user experience \cite{wen2023rise}. Handling this challenge is essential for a smooth and responsive user experience. In addition, it will contribute to the rapid adaptation of companies to this service model by increasing confidence in serverless computing. As per our best knowledge, no systematic review and taxonomy study has been conducted for the cold start latency problem in serverless computing before. A comprehensive systematic review is the basis for understanding a research area and research gap. In addition, such a study can provide an overview by presenting up-to-date solutions for interested researchers and academicians.}

\subsection{Motivation and Contributions}

\textcolor{black}{Serverless computing is one of the latest service models of cloud computing  \cite{kim2019practical}. Compared to other service models, it has attracted attention with its advantages such as rapid auto-scaling, abstraction of infrastructure management from the customer, and economic model \cite{golec2023qos}. In addition to these advantages, it also brings challenges such as cold start that are still waiting to be solved. When the literature is examined, it is tried to solve the cold start problem with the studies carried out in both the academy and the private sector. As far as we know, there is no Systematic Review and Taxonomy study examining cold start studies in serverless computing in the literature. This survey aims to present research gaps and contribute to the development of serverless computing by providing researchers with an overview of the cold start problem.} \textcolor{black}{This article makes three main contributions to a better understanding of cold start and cold start current solutions in serverless computing as follows:}

\begin{itemize}

\item \textcolor{black}{We are bringing the most comprehensive and largest systematic review and taxonomy study to date on cold start and cold start solution studies in serverless computing to the literature. To do this, the authors in this paper systematically analyzed over 100 papers from academia and industrial literature, technical reports, book chapters, and reference lists. With these analysis results and the snowball technique applied, a total of 32 paper sets containing cold start solutions were obtained (Section \ref{sec:Methodology}).}

\item \textcolor{black}{One of the most important aspects of a systematic review is to identify research questions \cite{keele}. We define five research questions (Section \ref{sec:RQ}) to review and analyze the current literature on the cold start problem in serverless computing. In line with these questions;}
\begin{itemize}

   \item \textcolor{black}{Cold start and QoS relationship (Section \ref{sec:RQ1}),}
   \item \textcolor{black}{Factors affecting cold start (Section \ref{sec:RQ2}),}
   \item \textcolor{black}{Existing literature studies and classification about cold start solutions (Section \ref{sec:RQ3}),}
   \item \textcolor{black}{Platforms used by studies about cold start solutions (Section \ref{sec:RQ4}),}
   \item \textcolor{black}{The places where cold start studies are published and the dataset/code availability are analyzed (Section \ref{sec:RQ5}).}
\end{itemize}
  
\item \textcolor{black}{The clear challenges of cold start in serverless computing are discussed and future directions are identified, providing valuable information for future researchers (Section \ref{sec:challengesandfuture})}.

\end{itemize}

\subsection{Article Organization}

\begin{figure}[t]
	\centering
	\includegraphics[scale=0.7]{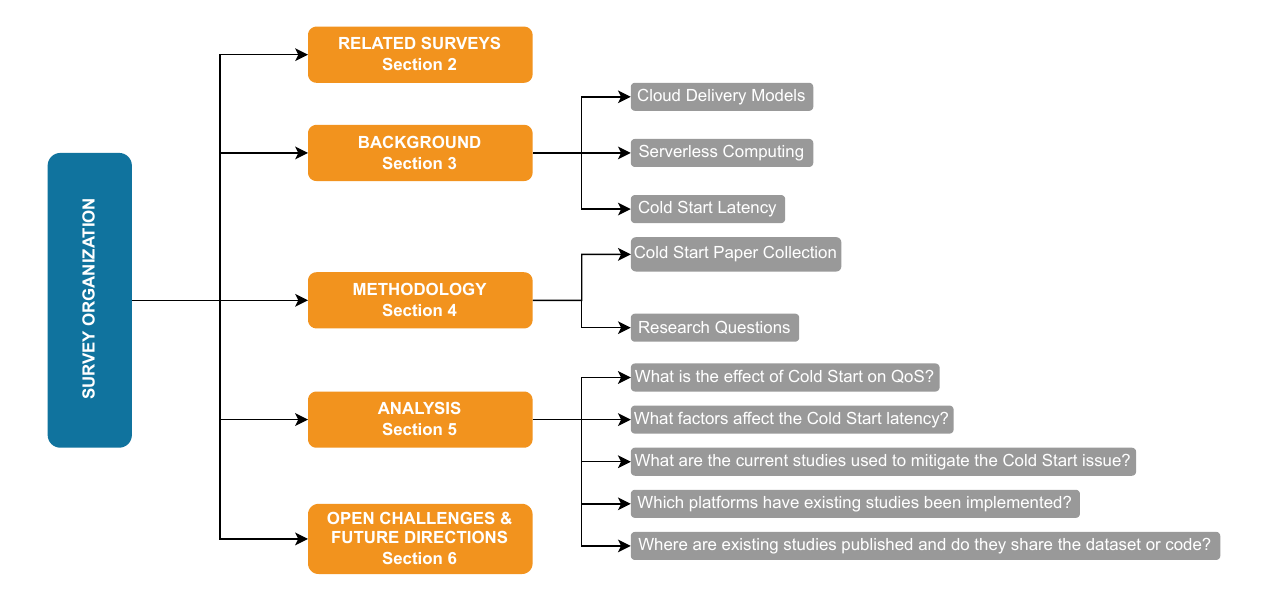}
	\Description{This figure shows the tree structure organization of the survey.}
	\caption{The Organization of the Survey}
	\label{fig:fig1}
\end{figure}

\textcolor{black}{Figure \ref{fig:fig1} shows the organization of this paper. In Section \ref{sec:related}, the relevant survey studies in the literature are explained and compared. Section \ref{sec:background} explains the basic concepts of serverless computing and cold start, including the cloud delivery model, serverless architecture, serverless platforms, serverless applications, and Quality of Service (QoS) parameters. Section \ref{sec:Methodology} explains how the research papers were collected and the research questions for this paper. In Section \ref{sec:analysis}, the data obtained from the collected papers for each research question is analyzed and solution-based classification is made. In the  section \ref{sec:challengesandfuture}, open challenges and future directions regarding cold start are highlighted for prospective researchers.} Finally, section \ref{sec:conclusions} concludes the article.

\section{RELATED SURVEYS}\label{sec:related}

\textcolor{black}{With the popularity of serverless computing, taxonomy, and survey studies on serverless computing have been conducted in recent years, examining various aspects of it. In this section, we provide an overview of the contributions and limitations of current literature studies.}

\textcolor{black}{
Vahidina et al. \cite{vahidinia2020cold} investigated cold start mitigation strategies in serverless computing in their study. Further, they also conducted experiments on cold start targeting a commercial platform. In these experiments, they determined the number of cold starts for concurrent requests, sequential requests, and concurrent with different time intervals requests, respectively. Mampage et al. \cite{mampage2022holistic} have done a review that examines the concept of resource management in serverless computing from multiple perspectives. They examined in detail the factors affecting resource management, workloads, and system design in serverless computing. They also analyzed the literature on resource management in serverless. In an another research study \cite{li2022serverless2}, serverless architecture was discussed in detail as four stack layers. These layers are virtualization, encapsule, system orchestration, and system coordination, respectively. The authors analyzed the contribution of related studies to resource management for each layer and highlighted the challenges. The authors in \cite{shafiei2022serverless} and \cite{eismann2021state} examine serverless applications in their literature study. The authors in \cite{shafiei2022serverless} classify serverless implementations in eight main areas, discussing the viability of serverless and the challenges encountered. They search for all solutions by classifying the challenges they detect under nine headings among themselves. In the research in \cite{eismann2021state}, 16 features are examined in serverless applications. Answers are sought for questions such as in which situations serverless applications are advantageous and how they should be applied. For this, by applying the mixed method 89 papers are examined. The authors compare 10 existing studies in the literature to confirm the validity of the research results. Jawaddi and Ismail \cite{jawaddi2023autoscaling} have proposed a taxonomy for autoscaling in serverless. They compared the studies on serverless autoscaling in the literature and found that the literature studies are still insufficient in terms of scaling target scope and Service Level Agreement (SLA) requirements. And they also identified that there is still a need for models such as anomaly and energy-aware autoscaling in this area as a research gap. The authors in \cite{hassan2021survey}, \cite{li2022serverless}, \cite{eismann2008review}, and \cite{wen2023rise} have proposed survey studies to identify research gaps and opportunities in serverless computing. The authors in \cite{hassan2021survey} and \cite{eismann2008review} compared serverless platforms and tools, emphasizing the advantages and disadvantages. The authors in \cite{li2022serverless} reviewed challenges and solutions related to infrastructure characteristics in serverless. They also provide a literature review to identify the challenges and opportunities found in serverless. The study in \cite{wen2023rise} covers 17 different research directions, including performance optimization and multi-cloud development on serverless. All solutions for 17 research directions are compared.} 

Cassel et al. \cite{cassel2022serverless} extensively examine the use of the IoT in serverless computing in their review. For this, 60 different articles were reviewed and the main challenges for incubating and executing functions were examined. And solutions to these challenges, such as protocols and programming languages, were explored. In \cite{marin2022serverless}, an in-depth analysis of security in serverless computing was done. The authors examine serverless architectures, highlighting security shortcomings and possible solutions. \textcolor{black}{Dittakavi's paper is far from a review and taxonomy and only provides a limited summary of cold start-related research \cite{dittakavi2023cold}. Additionally, it does not include future research directions and a comprehensive analysis of cold start solutions.} Mallick et al. \cite{mallick2024securing} conducted a systematic review on network security challenges and solutions in serverless computing. He reported that attackers are leveraging cold start to cause service interruptions and reduce system performance, which can lead to denial-of-service (DoS). 
Likewise, Ahmadi \cite{ahmadi2024challenges} examines the relationship between cold start and DoS attacks in his short review study examining network security challenges and solutions in serverless computing. Donta et al. \cite{donta2023exploring} They conducted a survey to provide a deep understanding of distributed computing continuity systems (DCCSs). This survey discusses the evolution of computing paradigms up to DCCS. Additionally, the opportunities and limitations (such as cold start) offered by each computing paradigm (cloud, edge, etc.) are examined. The authors in \cite{donta2023towards} examined the limitations of protocols used for edge networks and ways to facilitate learning in edge networks. \textcolor{black}{Huang et al. \cite{huang2018machine} address mobile communication technologies and quality wireless services in the survey. They emphasize that ML algorithms can be used to improve the efficiency and energy optimization of QoS parameters in communication systems. The same ML algorithms can be used to optimize the performance degradation in QoS parameters that occur due to cold start in serverless computing. Nazari et al \cite{nazari2021optimizing} conducted a survey study examining optimization efforts in serverless computing. The survey examines new strategies such as isolation techniques and intelligent algorithms to solve the cold start problem. In addition, solutions that increase the processing efficiency of functions executed in a container are also examined.}

\subsection{\textcolor{black}{Critical Analysis}}
\textcolor{black}{Table \ref{table:comparison_table} compares our work with existing surveys is given. The '*' symbol indicates just an overview, the '+' symbol indicates comprehensive discussion, and the '-' indicates the absence of the feature. They are also abbreviations for CSR: Cold Start Relevance, FRD: Future Research Directions, and CFCS: Comprehensive Analysis for Cold Start Solutions. While some of these articles cover the cold start issue superficially, to our knowledge, no previous work has focused on the cold start issue in detail and an in-depth analysis of its solutions. This survey provides a comprehensive analysis of the studies found in the literature to solve cold start problem. It also provides an overview of the cold start issue in serverless and the factors affecting cold start. This paper brings different technical approaches into a clear framework by categorizing them into a detailed taxonomy.}

\begin{table*}[t]
\caption{\small \textcolor{black}{Comparison of our Systematic Literature Review  with Existing Surveys}}
\label{table:comparison_table}

\begin{center}

\footnotesize
\begin{tabular}{@{}ccccccccc@{}}
\toprule
\textbf{Work} &
  \textbf{Topic} &
  \textbf{Year} &
  \textbf{Type} &
  \textbf{Publisher} &
  \textbf{CSR} &
  \textbf{FRD} &
  \textbf{CFCS} \\ \midrule
  \cite{vahidinia2020cold} &
  \begin{tabular}[c]{@{}c@{}}Cold Start\end{tabular} &
  2020 &
  Review &
  IEEE &
  + &
  - &-
   \\
\cite{mampage2022holistic} &
  \begin{tabular}[c]{@{}c@{}}Resource \\ Management\end{tabular} &
  2022 &
  Taxonomy &
  ACM &
  * &
  + &-
   \\
\cite{li2022serverless2}  & Architecture   & 2022 & Survey   & ACM             & * & + & - \\
\cite{shafiei2022serverless}  & Applications   & 2022 & Survey   & ACM             & * & + & - \\
\cite{eismann2021state}  & Applications   & 2022 & Survey   & IEEE            & * & * & - \\
\cite{jawaddi2023autoscaling}  & Autoscaling    & 2023 & Taxonomy & Research Square & * & + &-  \\
\cite{hassan2021survey}  & General        & 2021 & Survey   & Springer        & * & * & - \\
\cite{li2022serverless}  & Infrastructure & 2023 & Survey   & IEEE            & * & * & - \\
\cite{cassel2022serverless}  & IoT            & 2022 & Review   & Elsevier        & * & * & - \\
\cite{eismann2008review}  & General        & 2021 & Review   & arXiv           &   & * & - \\
\cite{wen2023rise} & General        & 2023 & Review   & ACM             & * & + & - \\
\cite{marin2022serverless} & Security       & 2022 & Review   & Springer        & -  &  - & - \\
\cite{dittakavi2023cold} & Cold Start       & 2023 & Review   & Eduzone        & +  &  - & - \\
\cite{mallick2024securing} & Security       & 2024 & Review   & World Scientific News  & *  &  - & - \\
\cite{ahmadi2024challenges} & Security       & 2024 & Review   &  IJCSRR  & *  &  + & - \\
\cite{donta2023exploring} &  General      & 2023 & Review   &  MDPI Computers  & *  &  - & - \\
\cite{donta2023towards} &  Data Protocol      & 2023 & Review   &  IEEE EDGE  & -  &  + & - \\
\cite{huang2018machine} &  Mobile Communication & 2018 & Survey   &  Springer MNA  & -  &  - & - \\
\cite{nazari2021optimizing} &  Serverless Platforms Optimization & 2021 & Survey   & IEEE  & *  &  + & * \\

\textbf{This Work} &
  Cold Start &
  2024 &
  Review, Taxonomy &
  ACM &
  + &
  + &
  + \\ \bottomrule
\end{tabular}
\end{center}

\end{table*}

\section{BACKGROUND}\label{sec:background}
\begin{figure}[t]
	\centering
	\includegraphics[scale=1]{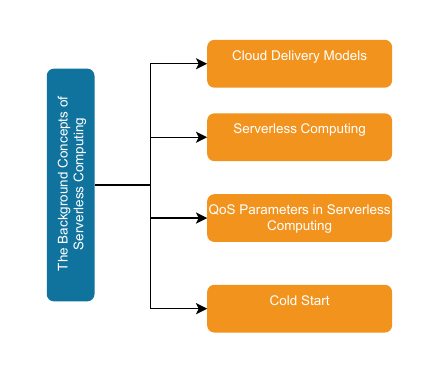}
	\Description{This figure illustrates the background of serverless computing and the cold start problem.}
	\caption{The Background of Serverless Computing and Cold Start}
	\label{fig:fig2}
\end{figure}

\textcolor{black}{In this section, we explain the background concepts related to serverless computing and cold start problem, as shown in Figure \ref{fig:fig2}.}

\subsection{\textcolor{black}{Cloud Models}}

\textcolor{black}{Cloud Models are examined under two subsections. The first subsection is cloud computing models according to deployment. The second subsection is cloud computing models according to delivery.} 

\subsubsection{Cloud Deployment Models}
\begin{figure}[t]
	\centering
	\includegraphics[scale=1]{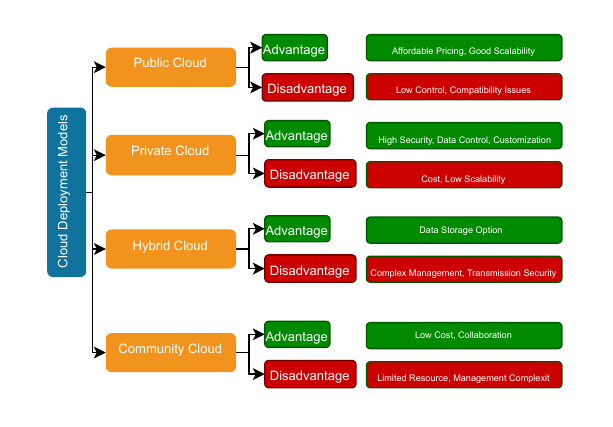}
	\Description{This figure shows various cloud deployment models.}
	\caption{Cloud Deployment Models}
	\label{fig:fig3}
\end{figure}

There are four cloud deployment models shown in Figure \ref{fig:fig3} based on their different features and characteristics, such as public cloud, private cloud, hybrid cloud, and community cloud. Furthermore, \textit{Section A} of \textbf{\textit{Appendix A}} provides a detailed explanation of these models.

\subsubsection{Cloud Delivery Models}
\begin{figure}[t]
	\centering
	\includegraphics[scale=0.65]{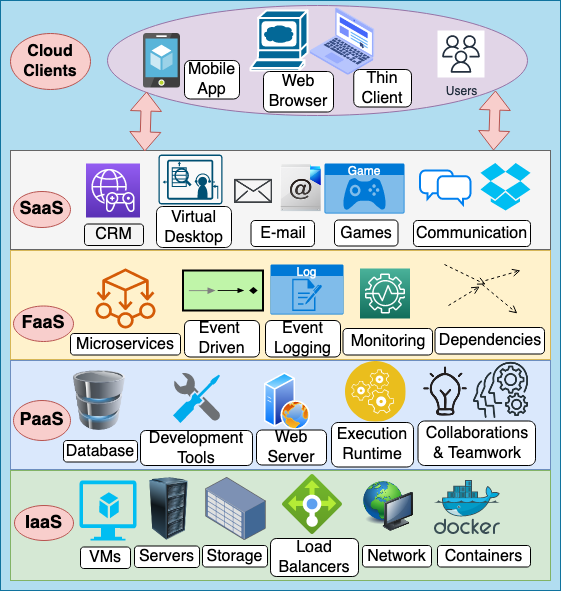}
	\Description{This figure presents cloud delivery models and their associated services.}
	\caption{Cloud Delivery Models and their Services}
	\label{fig:fig4}
\end{figure}

There are four cloud Delivery models shown in Figure \ref{fig:fig4} based on their different features and characteristics, such as Infrastructure as a Service (IaaS), Platform as a Service (PaaS), Software as a Service (SaaS), and Function as a Service (FaaS). Furthermore, \textit{Section B} of \textbf{\textit{Appendix A}} provides a detailed explanation of these models.

\begin{figure}[t]
	\centering
	\includegraphics[scale=0.55]{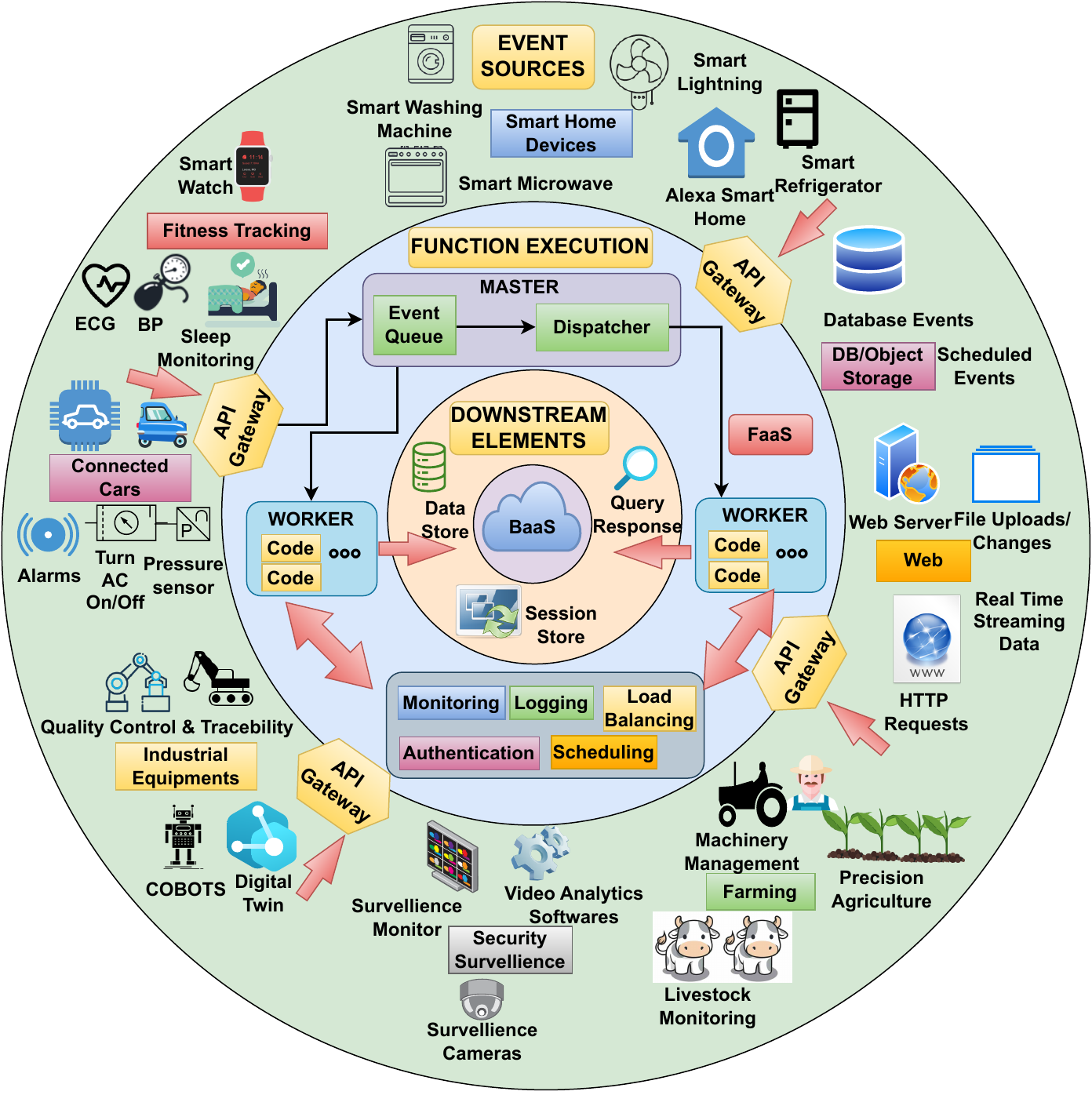}
	\Description{This figure displays the architecture and workflow in serverless computing.}
	\caption{Architecture and Workflow in Serverless Computing}
	\label{fig:fig5}
\end{figure}

\subsubsection{Serverless Computing Architecture} 
A general architecture of serverless computing is shown in Figure \ref{fig:fig5}. The functionality of this paradigm can be demonstrated via its following components:

\begin{itemize}

\item Event Sources: These include sources such as database change, file upload, HTTP requests, scheduled timers, etc. Although this paradigm was initially proposed for the cloud, however, it has positioned itself in IoT use cases by bringing functions closer to the devices \cite{edgeai}. It comprises a wide range of application scenarios such as ubiquitous phones, smart home devices (amazon echo, ring doorbell, etc.), fitness tracking (monitoring blood sugar levels, heart rate, recording quality sleep), industrial equipment (Collaborative Robots known as COBOTS), connected cars (turning on alarms on driver sleep detection, applying brakes, accelerating), farming (Livestock monitoring, precision agriculture) and nevertheless streaming application data which yields time-series data as depicted in Fig \ref{fig:fig5}. All these applications thrive for an apt paradigm that can provide real-time responsiveness and agility in order to incorporate an escalating number of IoT equipment and the corresponding surge in data volume. 

\item API Gateway: It acts as an interface between the event sources which can be called as frontend and the Function Execution layer providing FaaS \cite{EdgeBus}. It is responsible for routing the incoming request to specific serverless functions. Apart from this, it performs some pre-processing operations such as request validation, aggregating requests into batches, protocol transformation, and response formatting before forwarding it back to the consumer/requesting device. 
\item Function Execution: This component defines the core of serverless computing architecture, with the advent of which the responsibility of application deployment, management, scalability, and data security can be offloaded to cloud servers \cite{kumar2024comprehensive}. On the occurrence of an event, the request is forwarded to the event queue via API gateway, which further aims to invoke pertinent functions as per set defined rules. Then, the scheduler is responsible for dispatching the incoming request to the appropriate worker node, which provides a suitable environment comprising required resource requirements. In order to complete the function execution, the required application code is loaded from the repository. Finally, the user receives the response once execution is complete after which created environment is destroyed and corresponding resources are released \cite{mampage2022holistic}.

\item Downstream Elements: It comprises the elements such as data store, session store, query response, etc. which are triggered in response to the execution of a serverless function \cite{chen2018edge}. These elements are generally invoked as a result of the completion or successful execution of a serverless function. 

\item Backend-as-a-Service (BaaS): Although the cloud remains an indispensable part of serverless computing, however, the details of infrastructural functionality, scalability and management remain a black box for the serverless computing consumer \cite{nandhakumar2023edgeaisim}. These characteristics of serverless enables offloading of the database server and application logic to the cloud thus enabling the professionals to hone in on Business Logics (BL). 

\end{itemize}

\subsubsection{Serverless Computing Platforms}

\textcolor{black}{ Serverless platforms are generally examined under two main headings: open source and commercial. The classification of serverless platforms is given in Figure \ref{fig:fig7}, and a detailed review of these platforms is given in \textit{Section C} of \textbf{\textit{Appendix A}} .}

\subsubsection{Serverless Computing Applications}
\begin{figure}[t]
	\centering
	\includegraphics[scale=0.17]{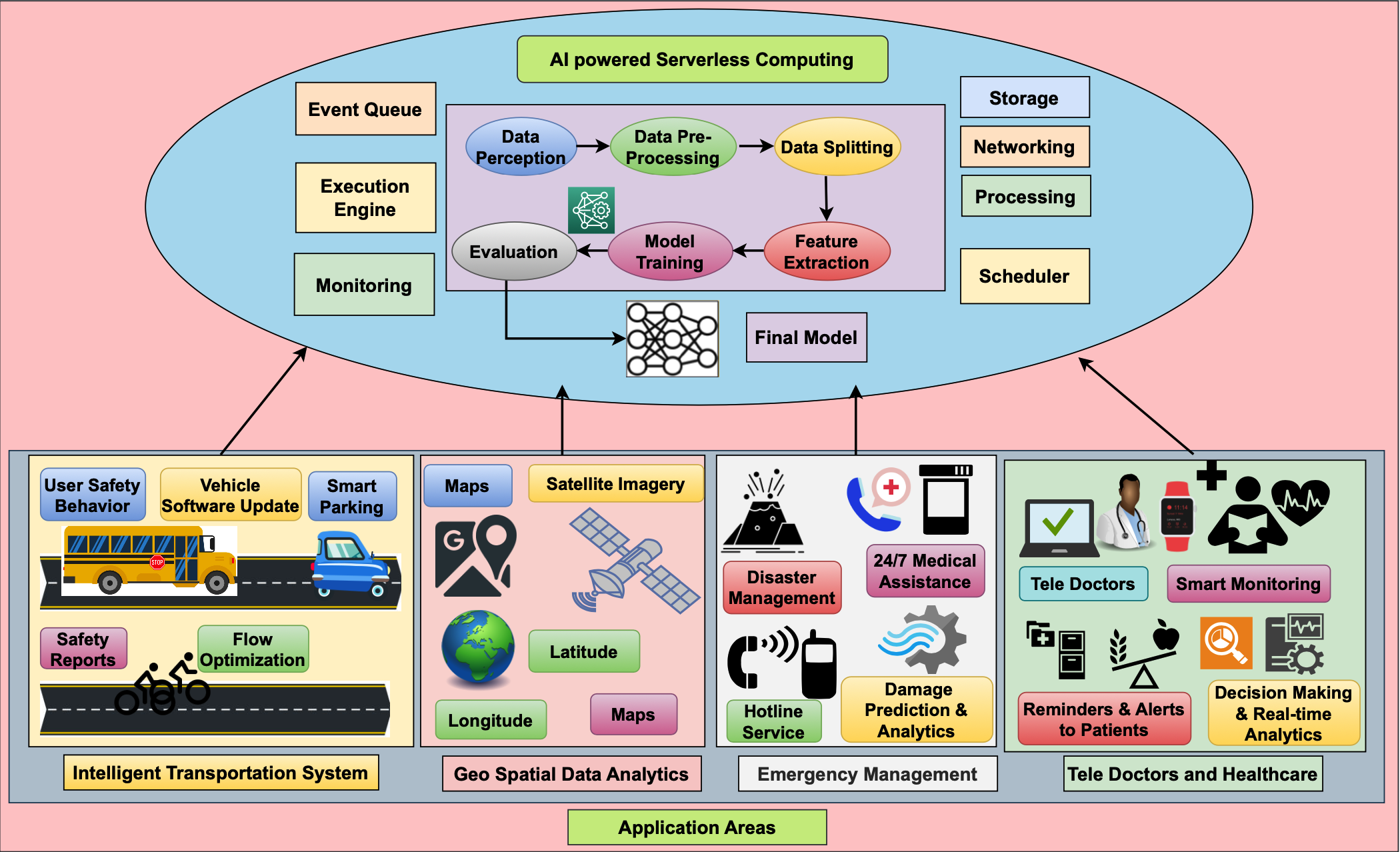}
	\Description{This figure shows AI-powered serverless computing applications.}
	\caption{AI-powered Serverless Computing Applications}
	\label{fig:fig6}
\end{figure}

\begin{figure}[t]
\centering
\begin{minipage}{0.8\textwidth}
\centering
\includegraphics[scale=1]{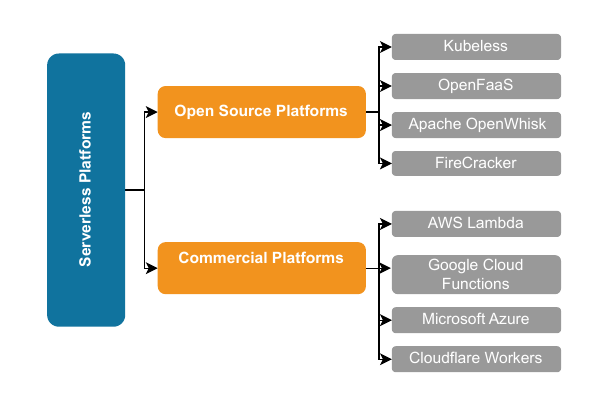}
\Description{This figure presents the taxonomy of serverless platforms.}
\caption{The Taxonomy of Serverless Platforms}
\label{fig:fig7}
\end{minipage}\hfill
\begin{minipage}{0.8\textwidth}
\centering
\includegraphics[scale=1]{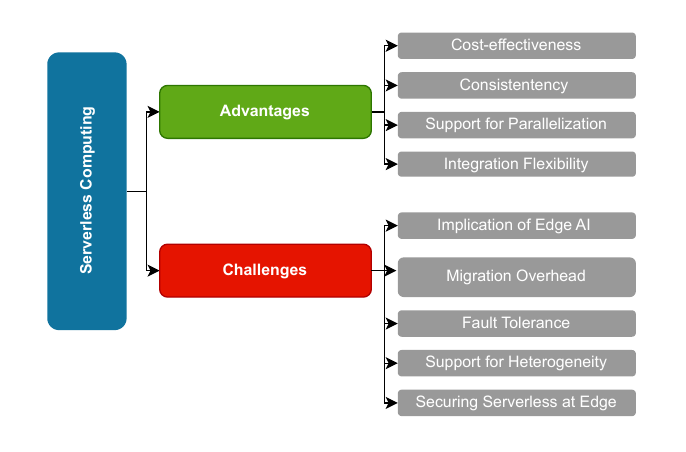}
\Description{This figure summarizes the advantages and challenges of serverless computing.}
\caption{Summary of Advantage \& Challenge in Serverless Computing}
\label{fig:fig8}
\end{minipage}
\end{figure}

The use cases for this architectural paradigm span various domains ranging from web applications and APIs, providing effortless scaling of business to IoT applications, requiring real-time data processing as shown in Figure \ref{fig:fig6}, which shows how these applications are utilizing Artificial Intelligence (AI) to optimize the serverelss computing based service.

\par \textbf{Taxonomy of Serverless Applications:} Further, we have proposed a detailed taxonomy of serverless applications. These applications are discussed below:

\begin{itemize}
\item \textbf{Real-time analysis of Geospatial data:} This data comprises addresses, maps, Satellite Imagery (SI), coordinates (longitudes, latitudes), etc. relating to different geographical and topographic phenomena. Such data constitutes an aggregation of physical features, which requires a framework for the perception and cognition of features. Apart from optimizing latency, Serverless Computing (SC) improves the processing in the case of high-dimensional hyper-spectral data \cite{bebortta2020geospatial}. \textcolor{black}{Mete et al. \cite{mete2021implementation} deployed Geographic Information Systems (GIS) applications to the serverless platform AWS Lambda. As a result of the tests, it was seen that GIS applications on serverless platforms respond faster under high workloads.}

\item \textbf{Emergency Management:} Some use-cases such as disaster management, identifying damage extent, handling pandemic situations such as delivering essentials to affected zones, and emergency hotlines require scalable resources, and Business Intelligence (BI) capabilities in order to coordinate relief operations. Serverless extends all cloud-based functionalities, streamlines deployment, and handles dynamic workloads, thus making data processing much more convenient \cite{shafiei2022serverless}. \textcolor{black} {
Franz et al. \cite{franz2018reunifying} proposed a system that aims to quickly reunite children who were lost after a disaster with their families, with the IoT-based system called ReuniFam. Families and rescue teams use Raspberry, an IoT device located in shelters, to upload images of missing children to a website. If there is any match between the uploaded images and images from sources such as drones, the location of the missing children is updated in the system. The authors used AWS Lambda, a serverless platform, for fast face recognition time.}

\item \textbf{Intelligent Transportation System (ITS):} The increasing trend of autonomous vehicles thrives for capabilities such as vehicle route planning, smart parking, emergency braking, weather incorporation, fleet management, etc. Hence, leveraging SC capabilities can result in a more agile and responsive ITS. For instance, this computing paradigm has been implicated in the calculation of the Origin/Destination (O/D) matrix for Bus Rapid Transit (BRT) utilizing Bluetooth signals \cite{herrera2018smart}.

\item \textbf{Telemedicine and Healthcare:} SC enables organizations to deliver healthcare services in a more personalized manner, empowering appointment scheduling, remote diagnosis, virtual health assistants, bots, and most importantly tele doctors, enabling patients to get medical assistance regardless of their physical whereabouts. For example, \textcolor{black}{Golec et al. \cite{golec2021ifaasbus} proposed a serverless-based framework that provides early Covid diagnosis for an increasing number of patients.}

\item \textbf{AI and ML-based Enterprise Data Intelligence:} It provides an ideal platform for executing AI models optimizing resource allocation, and leveraging event-driven architecture \cite{golec2024computing}. It allows the execution of AI algorithms cost-effectively, as resources are provisioned only when required. \textcolor{black}{Jefferson et al. \cite{jefferson2022resource} recommended a system sentiment analysis using Natural Language Processing (NLP) techniques. To take advantage of the processing power and affordability of serverless computing, they deployed the Naïve Bayes model to AWS Lambda, a serverless platform.}

\item \textcolor{black}{\textbf{Blockchain-based Serverless Applications:} Blockchain-based applications have problems such as high network congestion due to the number of users and transaction volume \cite{dehghani2020blockchain}. Serverless could be a potential solution to this problem with dynamic scalability property. In addition, blockchain applications (smart contracts, etc.) work with trigger logic based on events such as transactions and data changes. Serverless's event-drive architecture is suitable for these applications. Golec et al. \cite{golec2022aiblock} proposed a serverless-based framework that enables early COVID detection. They have integrated a blockchain module to enable the integration of patients' sensitive data such as health information and biometric data.}
 
\item \textcolor{black}{\textbf{IoT-based Serverless Applications:} Since serverless platforms can scale with increasing demand, they are suited for IoT applications with varying workloads and data amounts \cite{prokube}. In addition, it can provide an economical model for IoT applications with sudden workloads thanks to the pay-as-you-go model. When the literature is examined, many IoT-based serverless applications can be seen. Benedict  \cite{benedict2020serverless} proposed a new framework using IoT-based serverless and a societal application that monitors air quality for smart cities.}

\end{itemize}

\textcolor{black}{Table \ref{table:taxapp} shows a comparison of existing works based on the proposed taxonomy of serverless applications and platforms.}

\begin{table*}[t]
\caption{\small \textcolor{black}{Comparison of Existing Works based on the Taxonomy of Serverless Applications}}
\label{table:taxapp}
\begin{center}
\resizebox{.6\textwidth}{!}{
\footnotesize
\begin{tabular}{@{}ccc@{}}
\toprule
\textbf{Work} & \textbf{Application Type}                    & \textbf{Platform} \\ \midrule
\cite{mete2021implementation}             & Real-time Analysis of Geospatial Data        & AWS Lambda        \\
\cite{franz2018reunifying}             & Emergency Management                         & AWS Lambda        \\
\cite{herrera2018smart}             & Intelligent Transportation System (ITS)      & AWS Lambda, GCP   \\
\cite{golec2021ifaasbus}             & Telemedicine and Healthcare                  & GCP               \\
\cite{jefferson2022resource}             & AI and ML-based Enterprise Data Intelligence & AWS Lambda        \\
\cite{golec2022aiblock}             & Blockchain-based Serverless Applications     & GCP               \\
\cite{benedict2020serverless}             & IoT based Serverless Applications            & NA                \\ \bottomrule
\end{tabular}
}
\end{center}
\end{table*}

\subsection{Quality of Service (QoS) Parameters in Serverless Computing} 
The following are discussions of these QoS parameters: 
\begin{itemize}
    \item \emph{Latency}: It is defined as the time delay between consumer request submission and receiving the response from the corresponding serverless function \cite{golec2023qos,sedlak2024equilibrium}. This time is resultant of request queuing time, request-to-resource mapping time, and finally the serverless function execution time. Hence, reducing the cold start delay in resource set-up significantly optimizes this parameter making it suitable for real-life use cases.
    \item \emph{Throughput}: It is measured by the number of requests processed in unit time by the serverless system. This metric is considered effective as it helps in determining the overall system efficiency along with the underlying system's capability to handle high arrival rate workload \cite{mahmoudi2019optimizing,pujol2024causality}. Consequently, due to the negative correlation between mean latency and system throughput, techniques used to reduce the latency, ultimately ameliorates the system's overall throughput. 
   \item \emph{User Experience (UX)}: The agile infrastructure of serverless computing not only reduces cost but also eases the hassle of administrative overhead \cite{shin2015beyond}. The Service Providers (SPs) aim at delivering positive UX via optimizing warm-up strategies, periodic monitoring and analyzing the performance metrics, 
    \item \emph{Cost}: The billing and pricing model offered by a particular serverless platform decides the function execution cost \cite{golec2023qos}. Due to its pay-as-you-go pricing model, the cost is significantly determined by the function execution time. Apart from the service provider context, the cost is determined by the number of functions requested and resources utilized for the function execution time span. Hence, this parameter becomes of crucial consideration by both service providers and service consumers. 
     \item \emph{Scalability}: Serverless paradigm possesses the capability of automatically provisioning resources as and when desired by the incoming workloads. Scalability can be vertical or horizontal. The horizontal one is characterized by the addition and elimination of existing containers, whereas the vertical scaling aims at elevating (adding more cores ) or reducing (removing cores) capabilities of existing containers \cite{somma2020less}. 
      \item  Resource Utilization: The autonomous resource handling capabilities of serverless make it a suitable choice for serving geographically distributed edge/ fog nodes. After application deployment on the serverless platform, the service provider becomes accountable for managing the whole chain of resource management (RM) activities such as resource provisioning, scheduling, and final allocation along with monitoring and resource scaling \cite{mampage2022holistic,gill2024edge}. 
\end{itemize}

\subsubsection{Serverless Computing Advantages \& Challenges} 

In this section, we explain the advantages and challenges of serverless computing, which are summarised in Figure \ref{fig:fig8}.

The advantages of serverless computing are discussed below:
\begin{itemize}

\item \emph{Cost-effectiveness}: This computing offers lower cost as compared to traditional hosting, as the consumer needs to pay for specific amounts of time, hence there is no upkeep cost during idle times. In contrast to conventional VM (which come with a hefty expense of duplicating data along with OS and underlying application environment), serverless offer lightweight abstractions in the form of unikernels such as microVMs, which automatically scale the resources as per incoming requests \cite{morabito2018consolidate}\cite{krishnamurthi2023serverless}. 
\item \emph{Consistent for IoT Applications}: Serverless platform provides fine-grained scalability and resource provisioning which suits well for spiky workloads of IoT use cases \cite{eismann2020serverless, singh2022machine}. 
\item \emph{Support for Parallelized Execution}: it equips applications such as Defence IoT (DIoT) with the capability to carry on computations in a parallelizable fashion. For instance, whenever any hazard occurs, prompt and independent task processing capabilities are required for equipment to be traced, so as to ensure timely response and at the same time not affecting other mission-critical operations. 
\item \emph{Flexibility to be integrated with Cloud-Edge Computing}: The latest research is trending towards creating a collaborative cloud-fog-edge environment in order to cater to real-world application scenarios. Although serverless was originally designed for the cloud, however, to ameliorate edge capabilities, Serverless-powered edge platforms are emerging such as AWS IoT Greengrass \cite{el2021platforms}.
\end{itemize}

The challenges of serverless computing can be listed as follows:
\begin{itemize}
    \item \emph{Implication of Edge AI}: Despite its numerous advantages, the practical implication of Edge AI in serverless voices serious challenges, as training and testing ML models at the network edge in the computation finite resources environment of serverless computing becomes I/O intensive \cite{edgeai}. It contradicts the fact that serverless is designated to work well with CPU-bound functions, which ultimately makes it cost-effective \cite{gill2024edge}. In fact, this computing aims to minimize resource usage, both in execution as well as in idle states.
    \item \emph{Migration Overhead}: The adoption of vendor-specific proprietary APIs, runtime environments, and other native services makes it challenging to migrate with another vendor. Migration becomes a tedious task as it involves differences in billing models, reconfiguring infrastructure, and significant changes in code.
    \item \emph{Fault Tolerance}: The feature of fault tolerance ensures that the running application remains available and responsive even in case of container fail over. 
    \item \emph{Support for Heterogeneous Hardware}: the platforms support for heterogeneous hardware such as Graphics Processing Unit (GPU), and Field Programmable Gate Arrays (FPGAs) remains a challenging task. 
    \item \emph{Securing Serverless at Edge}: Although serverless is backed by firewalls and a Trusted Execution Environment (TEE) at the cloud layer, security gets a backseat when serverless computing is bought at the edge \cite{EdgeBus}. It occurs because of distributed nature of microservices and multi-tenancy amongst services, further, the deployment of functions in a containerized manner makes it lesser isolated in comparison to VMs \cite{aslanpour2021serverless}. 

\end{itemize}

\subsection{Cold Start Problem}

\textcolor{black}{In serverless computing, each function is assigned to a separate container for the execution of functions \cite{baldini2017serverless}. If the container is ready, the function is directly assigned to this container, and the function is executed. However, if there is no ready container, which is mostly the case, it is necessary to prepare a new container to execute the function. In the serverless paradigm, containers are released after a specific time ($\tau$) after executing functions to avoid wasting resources. This process is known as the scale to zero \cite{li2022serverless}. After the containers are scaled to zero, it is necessary to start the container again for the requests coming to the server. Starting a new container and preparing the function for execution causes a specific latency. In the serverless paradigm, this latency period is called a cold start \cite{kim2019practical}. The main causes of cold start are as follows:}

\begin{itemize} 
\item \textcolor{black}{Since there is no ready container when the first request comes to the server, it is necessary to start a new container. Preparatory processes will take time, such as starting a new container and loading the runtimes and libraries needed to execute the function. This causes a cold start.}

\item \textcolor{black}{Serverless paradigm automatically scales resources if needed thanks to its dynamic scalability feature. Sudden increases in requests can cause scalability delays. This delay is another reason why cold start occurs.}
 
\item \textcolor{black}{In serverless platforms, resources (CPU, RAM, etc.) are shared between functions. Moreover, resource problems such as resource contention caused by spikes in demand can cause cold start latency \cite{golec2023healthfaas}.}
\end{itemize}

\textcolor{black}{Figure \ref{fig:fig10} shows the steps from creating a container to executing a function to illustrate the cold start process. These processes are generally the same on all platforms.}

\begin{itemize}
    
\item \textcolor{black}{\textit{Loading and Preparation Phase}: When a function request is sent to the platform, the container should be prepared to execute this function. The platform first parses the headers and input parameters from the incoming data. It then allocates the necessary resources, such as CPU and RAM, to execute the function. To execute the function, a container is started, and the necessary dependencies, such as libraries and runtime, are loaded into the container. Function code is deployed to the container and made ready for execution.}

\item  \textcolor{black}{\textit{Function Execution}: After the container loading and preparation phase is completed, the function assigned to the container is executed. In running the function, the platform monitors and manages the necessary resource management. External service interactions of the function with the network, such as the API, occur in this process. After the execution process is completed, the response is prepared and sent to the client. Resources are kept warm for $\tau$ to respond to new requests and prevent a cold start; this process is known as a warm start. If there is no request during this time, the containers are scaled to zero, saving energy.}

\end{itemize}

\section{REVIEW METHODOLOGY} \label{sec:Methodology} 
In this section, we discuss the review methodology used to conduct this systematic review of cold start latency in serverless computing.

\subsection{Cold Start Papers Collection}

\textcolor{black}{In this section, the evaluation process followed to collect the studies on the cold start problem in serverless computing is mentioned. Serverless computing is still a new field of research considering it was first introduced in 2014. Therefore, there may be literature publications that are meaningless or beyond the scope of this paper. For this, the paper collection process should be carried out with great care. To find the papers examined in this survey; We have benefited from leading digital databases such as IEEE, Springer,  ACM, ScienceDirect, and Wiley. Since serverless computing was introduced in 2014, naturally all articles are for after 2014. Keywords used to find related articles are as follows:}
\begin{equation}
[(Serverless Computing) \left |  \right | (Cold Start)] \&
[(Serverless Computing)  \left |  \right |  (Cold Start)]
\label{eq:1}
\end{equation}
\begin{equation}
[(Resource Management) \left |  \right | (Serverless Computing)] \& 
 [(Cold Start)  \left |  \right |  (FaaS)]
\label{eq:2}
\end{equation}
\begin{equation}
[(Cold Start Mitigation)] \& [(Serverless Computing)
\label{eq:3}
\end{equation}
\begin{equation}
 [(Serverless Computing)] | |[(Cold Start)]
\label{eq:4}
\end{equation}

\par
\begin{figure}[t]
	\centering
	\includegraphics[scale=0.45]{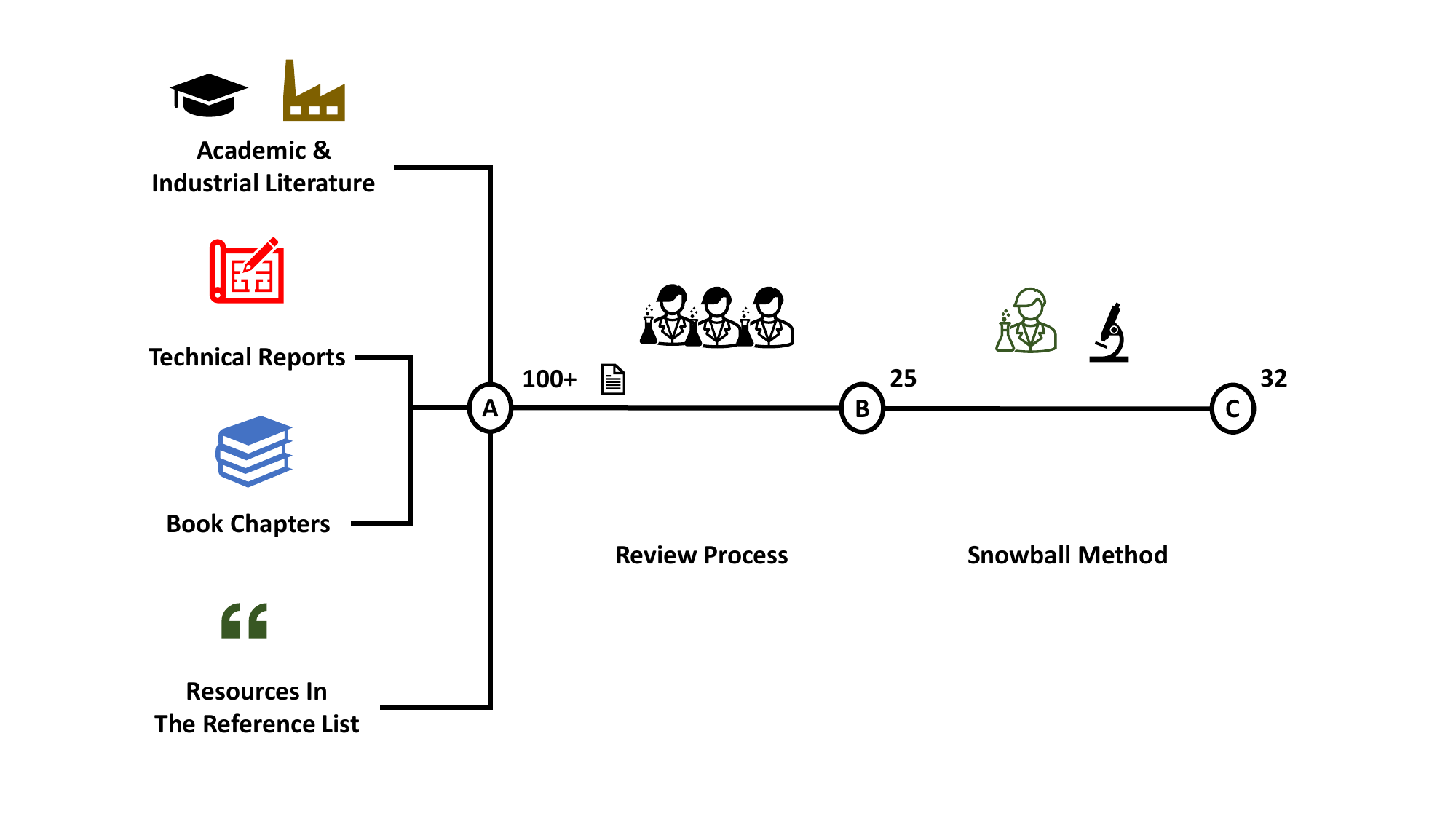}
	\Description{This figure illustrates the process of collecting papers related to cold start issues.}
	\caption{The Paper Collection Process}
	\label{fig:fig9}
\end{figure}

 \textcolor{black}{Using these keywords, a total of over 100 serverless computing and cold start-related materials were obtained from Academic \& Industrial Literature, Technical Reports, Book Chapters, and Resources in the Reference List (A). In the second stage (B), the abstract and conclusion parts of the obtained papers were systematically examined. Further, unrelated papers were excluded from the research. To remove potential bias in paper selection and screening, the entire review process (A-B) was carried out jointly among the authors participating in this survey. The 25 papers obtained are directly related to solving cold start in serverless computing. In the last stage, the backward snowball method was applied by an expert author to expand the paper set \cite{voicu1997using}. Snowball Method is a sampling technique used to collect hard-to-reach data \cite{iftikhar2022ai,kitchenham2004procedures}. The expert writer first analyzes a paper about cold start. It then scans all citations in the subsections and reference section of this paper. In this way, papers that do not contain the keywords 'Cold Start' or 'Serverless' but focus on solving cold start in serverless are identified. As a result, the number of papers obtained reached 32 (C). Figure \ref{fig:fig9} shows the paper collection process for this survey.}

\begin{figure}[t]
\centering
\begin{minipage}{0.9\textwidth}
\centering
\includegraphics[scale=1]{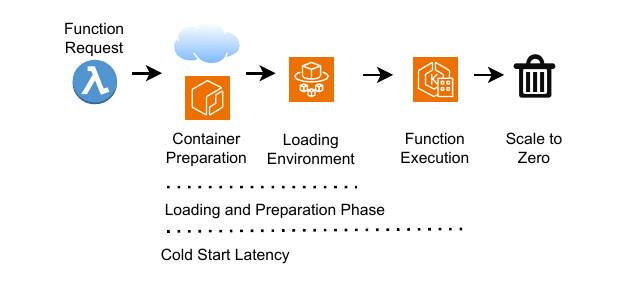}
\Description{This figure shows the cold start latency process in serverless computing.}
\caption{The Cold Start Latency Process}
\label{fig:fig10}
\end{minipage}\hfill
\begin{minipage}{0.9\textwidth}
\centering
\includegraphics[scale=1]{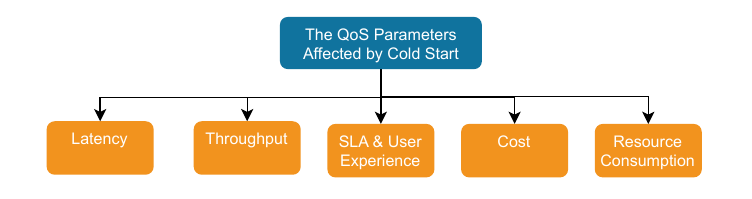}
\Description{This figure presents the QoS parameters that are affected by cold start latency in serverless computing.}
\caption{The QoS Parameters Affected by Cold Start}
\label{fig:fig11}
\end{minipage}
\end{figure}

\subsection{Research Questions} \label{sec:RQ}
\textcolor{black}{One of the most important aspects of a systematic review is to identify research questions \cite{keele}. In this section, we define research questions to examine and analyze current literature on the cold start problem in serverless computing. Research questions, motivation, and relevant section information are given in Table \ref{table:researchquest}.}

\begin{table*}[t]
\caption{\small \textcolor{black}{The Research Questions of the Survey }}
\label{table:researchquest}
\begin{center}
\resizebox{.7\textwidth}{!}{
\footnotesize
\begin{tabular}{|c|c|c|c|}
\hline
\textbf{No} &
  \textbf{\begin{tabular}[c]{@{}c@{}}Research \\ Question\end{tabular}} &
  \textbf{Motivation} &
  \textbf{Section} \\ \hline
1 &
  \begin{tabular}[c]{@{}c@{}}What is the effect of \\ Cold Start on QoS?\end{tabular} &
  \begin{tabular}[c]{@{}c@{}}The purpose of this research question \\ is to investigate how cold start affects \\ which QoS parameters.\end{tabular} &
  5.1 \\ \hline
2 &
  \begin{tabular}[c]{@{}c@{}}What factors affect \\ the Cold Start latency time?\end{tabular} &
  \begin{tabular}[c]{@{}c@{}}The purpose of this research question \\ is to determine the factors \\ affecting cold start latency.\end{tabular} &
  5.2 \\ \hline
3 &
  \begin{tabular}[c]{@{}c@{}}What are the current \\ studies used to mitigate \\ the Cold Start issue?\end{tabular} &
  \begin{tabular}[c]{@{}c@{}}The purpose of this research question \\ is to identify current studies for cold start \\ and to search for current solutions for cold start.\end{tabular} &
  5.3 \\ \hline
4 &  
  \begin{tabular}[c]{@{}c@{}}Which platforms are the\\  current studies been implemented?\end{tabular} &
  \begin{tabular}[c]{@{}c@{}}The purpose of this research question \\ is to identify the platforms where current studies\\  for the cold start are applied and evaluated.\end{tabular} &
  5.4 \\ \hline
5 &
  \begin{tabular}[c]{@{}c@{}}Where are existing studies \\ published and do they share \\ the dataset or code?\end{tabular} &
  \begin{tabular}[c]{@{}c@{}}The purpose of this research question \\ is to identify publishers that publish \\ current studies for cold start. \\ It is also aimed to explore \\ the reproducibility of the current work.\end{tabular} &
  5.5 \\ \hline
\end{tabular}
}
\end{center}
\end{table*}

\section{ANALYSIS}\label{sec:analysis}
In this section, we discuss how we addressed the above-mentioned questions in this study.
\subsection{RQ1: What is the effect of Cold Start on QoS?} \label{sec:RQ1}

\textcolor{black}{The purpose of this research question is to investigate how cold start affects which QoS parameters. The QoS parameters that cold start can affect are given in Figure \ref{fig:fig11}. The cold start problem caused by the serverless paradigm adversely affects various QoS parameters such as latency, throughput, SLA, and User Experience \cite{lee2015location,chen2017your}. Developers need to consider these effects when designing their applications. In this section, we examine how a cold start affects these parameters;}

\begin{itemize}

\item \textbf{Latency}: \textcolor{black}{It is the time taken until the data from the source reaches the destination and comes back, and it is usually measured in milliseconds (ms) \cite{golec2023qos}. It is one of the most frequently used QoS parameters to measure the performance of an application. Golec et al. \cite{golec2022aiblock} found high delays due to cold start in a serverless-based application that followed patients instantaneously. \textcolor{black}{They sent an increasing number of concurrent requests to the server to represent an increasing number of users. The results showed that 100 concurrent requests had more latency than 200 concurrent requests (5 ms). The reason for this is the cold start situation that occurs when the first requests (100 simultaneous requests) arrive on the server.} Especially in applications such as time-sensitive and instant patient follow-up, cold start may cause unwanted latency.}

\item \textbf{Throughput}: \textcolor{black}{It is a QoS parameter defined as the data transmitted per second over a network or communication channel \cite{golec2023qos}. In order to achieve SLA, the use of resources, i.e. Throughput, is important. Sudden increases in server traffic due to cold start cause resource contention and negatively affect throughput. Golec et al. \cite{golec2023healthfaas} found in their study that there were significant decreases in throughput due to resource contention.\textcolor{black}{ They sent varying numbers of workloads (100, 200, 300, 500, and 1000) to the server. The resource contention that occurred due to the cold start in the system (500 requests) reduced the throughput value (from 470 to 430 (P/Sec)).}}

\item \textbf{SLA and User Experience}: \textcolor{black}{It covers all experiences such as accessibility and functionality experienced by users using serverless systems while interacting with the platform \cite{golec2023qos}. It is one of the important concepts that service providers should pay attention to for customer continuity \cite{ristov2012new}. Cold start negatively affects User Experience as it causes unwanted latency \cite{bardsley2018serverless}\cite{ustiugov2021benchmarking}.}

\item \textbf{Cost}: \textcolor{black}{Serverless computing, with a pay-as-you-go model, charges customers only for the time resources are used. However, a cold start may mean more charges as it will increase the execution time of the function \cite{baldini2017serverless}. Also, some platforms like Google Cloud Functions (GCF) and AWS Lambda \cite{vahidinia2022mitigating} implement keep-warm strategies that require containers to run idle to mitigate the effects of a cold start. This strategy causes additional costs for customers.}

\item \textbf{Scalability}: \textcolor{black}{Serverless computing can automatically scale resources with increasing demand \cite{mampage2022holistic}. The prolongation of the delay times due to cold start will also affect the scalability time. This causes performance loss in serverless computing.}

\item \textbf{Resource Consumption}: \textcolor{black}{High cold start frequency may cause non-optimal allocation in case of creating new instances as it will make busy the system's resources such as CPU and RAM.}

\end{itemize}

\subsubsection{\textcolor{black}{Optimized Caching Mechanisms}}
\textcolor{black}{Optimized caching mechanisms are currently used in almost all serverless platforms to reduce latency and increase efficiency.}

\begin{itemize}
    \item \textbf{Impact of optimized caching mechanisms on latency}: \textcolor{black}{Caching data close to where the calculation is performed eliminates the need to retrieve data from remote databases and enables faster processing of functions. AWS Lambda implements a caching strategy with Amazon ElastiCache service, and experimental results show latency reduction \cite{pelle2019towards}. This service appears as the Memorystore (Redis) service in GCP and it is observed that the latency decreases in the same way with ElastiCache \cite{chen2016towards}. Azure, on the other hand, reduces latency by performing caching operations using Azure Cache \cite{gunarathne2011iterative}.}

    \item \textbf{Impact of optimized caching mechanisms on throughput}: \textcolor{black}{Thanks to the optimized caching strategy, more simultaneous requests can be responded to and this increases throughput as this means more data. Likewise, it has been reported that ElastiCache, Memorystore and Azure Cache increase throughput because they allow more data processing \cite{kavitha2020task,kulkarni2020achieving, agarwal2012azurebench}.}
\end{itemize}

\subsubsection{\textcolor{black}{Performance Metrics Used To Evaluate Cold Start Solutions}}

\textcolor{black}{This subsection explains the basic metrics used to evaluate the effectiveness of cold start frequency reduction and cold start prevention efforts in serverless computing. These metrics are:}

\begin{itemize}

\item \textcolor{black}{\textbf{Resource Consumption}: This metric, usually expressed in units such as RAM consumption (MB) and CPU usage amount (GHz), represents the amount of resources used in the cold start prevention solution (such as starting a new container usage) \cite{golec2023qos}.}

\item \textcolor{black}{\textbf{Cost}: This metric expresses the additional costs that the cold start solutions bring to the system depending on the methods they use \cite{baldini2017serverless}. This metric is usually used when evaluating solutions that require additional resource consumption and is represented in USD currency based on the execution cost.}

\item \textcolor{black}{\textbf{Scalability}: This metric expresses the horizontal and vertical scaling performance to evaluate solutions to the cold start problem \cite{vahidinia2022mitigating}. The number of containers launched per second can be used as a unit.}

\end{itemize}

\subsection{RQ2: What factors affect the Cold Start latency?}\label{sec:RQ2}

\begin{figure}[t]
	\centering
	\includegraphics[scale=0.6]{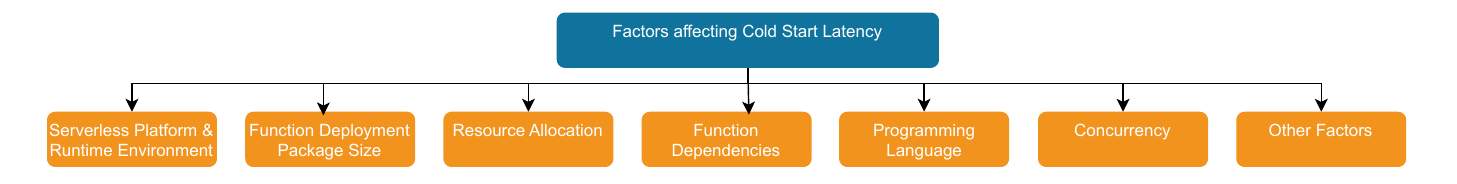}
	\Description{This figure illustrates the various factors that influence cold start latency in serverless computing.}
	\caption{Factors influencing Cold Start Latency}
	\label{fig:fig12}
\end{figure}

\textcolor{black}{The purpose of this research question is to investigate the factors affecting the cold start latency that occurs in serverless computing. Figure \ref{fig:fig12} shows these factors. Although the factors affecting cold start can be a very wide range, they are generally as follows:}

\begin{itemize}
    \item  \textbf{Serverless Platform and Runtime Environment}: \textcolor{black}{Each commercial and open-source platform uses different isolation and container runtime techniques specific to its architecture \cite{agarwal2021reinforcement}. Therefore, cold start latency times differ on each platform \cite{mampage2022holistic}. Li et al.\cite{li2022serverless2} compared cold start latency in various isolation and container runtime environments. The results showed that HyperContainer had the highest cold start latency while Process-based Docker had the lowest cold start latency. In another study \cite{jackson2018investigation}, the authors compared cold start latency for AWS Lambda and Microsoft Azure. The results showed that the Microsoft Azure platform has a better performance compared to AWS Lambda. A similar comparison is made in \cite{manner2018cold} for AWS Lambda and Microsoft Azure platforms. \textcolor{black}{Factors such as the software language used on serverless platforms, function dependencies (file size), CPU, and RAM affect the cold start latency period. Interpreted languages such as Python, JavaScript, and Node.js have shorter cold start latency than compiled languages such as JAVA, C\#, and .NET \cite{pawlik2018performance}. Each platform has optimization options and design and engineering features for different languages \cite{wang2018peeking}. That's why factors such as programming language and function dependencies cause different performance on each platform. For example, AWS Lambda runs the Python language faster than other platforms \cite{gimenez2019framework}. In their experiments, Wang et al compared AWS Lambda, GCF, and Azure for Nodejs \cite{wang2018peeking}. The results confirm that Nodejs creates smaller cold start latency on AWS. Experiments by Lee et al show similar results \cite{lee2018evaluation}. AWS has a faster runtime in Python, Nodejs, C\#, and Java languages than, Azure, IBM, and Google. Another factor that affects cold start latency is function dependencies (file size). Golec et al. In their experiments, they calculated the cold start latency of functions with function dependencies of 1 Kb, 16 MB, and 37 MB in size to GCF. The results show that cold start increases with the size of function dependencies. There are new trends, such as Pre-Warming and Edge Computing, which will be explained in later sections, that can potentially mitigate these factors more effectively, and programming language-specific optimizations offered by the previously described platforms.}}
    
    \item \textbf{Function Deployment Package Size:} \textcolor{black}{With the increase of the Function deployment package, cold start latency is expected to increase \cite{li2022serverless2} \cite{kumari2023acpm}. Because it will take longer to load large-size packages into existing containers than small-size packages. To test this, Manner et al. \cite{manner2018cold} uploaded deployment packages of various sizes to the AWS Lambda platform and calculated cold start amounts. Experiment results showed that the function deployment package had a negative effect on cold start latency.}
    
    \item \textbf{Resource Allocation:} \textcolor{black}{On serverless platforms, resources such as RAM and CPU can be adjusted manually in the environment configuration. Depending on the amount of resources (RAM and CPU), the deployment time of the containers also varies and this time is expected to affect the cold start \cite{hassan2021survey}. The authors in \cite{li2022serverless2}, \cite{manner2018cold}, and \cite{golec2023healthfaas} conducted experiments to examine the relationship between RAM amount and cold start on serverless platforms. The results showed that the cold start time decreased in proportion to the increased amount of RAM. It should be noted that on most serverless platforms the CPU scales linearly with the amount of RAM \cite{kumari2023acpm}.}
    
    \item \textbf{Function Dependencies: } \textcolor{black}{Before a function can be deployed, dependencies such as libraries required for the function's execution must be loaded into the containers \cite{kumari2023acpm}. Since this loading process takes time, it has a direct effect on cold start \cite{akkus2018sand}\cite{kaffes2019centralized}. To pre-provisioning dependencies for commonly used functions can reduce the number of cold starts \cite{wen2023rise}.}
    
    \item \textbf{Programming Language:} \textcolor{black}{Serverless platforms come with various programming language options such as Python, Java, and C\# \cite{eismann2021state}. Since the software languages have different environment overheads and architectures, the amount of cold start they have will also be different. High-level languages such as Python and Java have additional startup delays because they contain language runtimes such as JVM \cite{du2020catalyzer}. In \cite{jackson2018investigation}, \cite{manner2018cold}\cite{golec2023healthfaas}\cite{li2022serverless2} experiments showed that interpretative languages (Python, Java ) have higher cold start than compiled languages.}

    \item \textbf{Concurrency:} \textcolor{black}{With the scaling feature in serverless computing, a new container is started for each concurrent request. As a result, peak loads occur due to excessive use of resources and peak loads trigger cold start \cite{manner2018cold}. Mohan et al. \cite{mohan2019agile} and Ustiugov et al. \cite{ustiugov2021benchmarking} conducted experiments on two different platforms to observe the effect of concurrent on cold start. The results show that cold start increases with increasing concurrency.}

    \item  \textbf{Other Factors:} \textcolor{black}{The other factors affecting cold start can be:}
            \begin{itemize} 
            \item \textcolor{black}{If the function is dependent on the data in the storage, the storage access process will increase the cold start time \cite{vahidinia2022mitigating},}
            \item \textcolor{black}{Designing a function to run in a Virtual Private Cloud (VPC) increases cold start latency due to network isolation functions \cite{kumari2023acpm},}
            \item \textcolor{black}{Depending on the geographical region infrastructure where the function is distributed, cold start latency may be affected \cite{qureshi2010power}.}
            \end{itemize}
\end{itemize}

\subsection{RQ3: What are the current studies used to mitigate the Cold Start Problem?} \label{sec:RQ3}

\begin{figure}[t]
	\centering
	\includegraphics[scale=0.41]{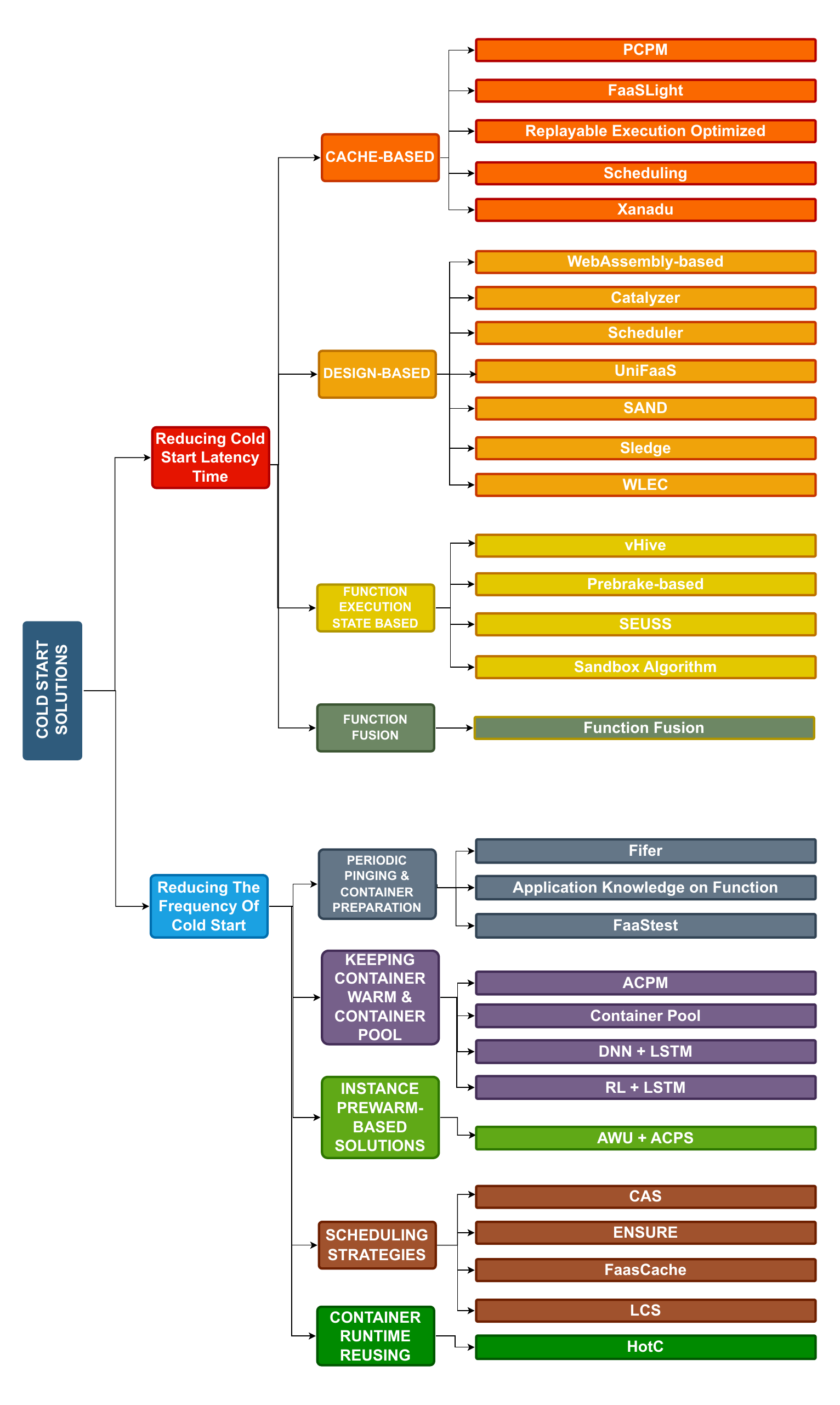}
	\Description{This figure presents the taxonomy of techniques for reducing cold start latency time and cold start frequency.}
	\caption{\textcolor{black}{The Taxonomy of Cold Start Latency Time Reduction Techniques and Cold Start Frequency Reduction Techniques.}}
	\label{fig:fig13}
\end{figure}

\textcolor{black}{The purpose of this research question is to identify and search current studies for cold start. To answer RQ3, the first two authors carefully examine the contents of the 32 papers available and then obtain the research summary. Then, the solutions obtained from the papers are classified under the necessary subheadings. Figure \ref{fig:fig13} shows the taxonomy of the current techniques used to mitigate the cold start problem.}

\textcolor{black}{It is important to solve the cold start problem in order to increase the performance of QoS parameters and serverless applications. One of the biggest problems caused by cold start issues is latency and it is seen that some studies have been done in the literature to solve this problem. Literature studies are generally examined under two main headings. These; (i) studies to reduce the cold start latency period and (ii) studies to reduce the frequency of cold start.}

\subsubsection{Studies to reduce cold start latency time}

\textcolor{black}{The studies to reduce cold start latency are generally aimed at speeding up container preparation and loading function libraries. The studies can be examined under 4 subheadings. Table \ref{table:cl} shows the studies and methods used to reduce cold start latency time. It should be noted that these subheadings may vary according to the review methodology:}
\begin{itemize}
     \item \textbf{Cache-based Solutions:} \textcolor{black}{Serverless computing uses application delivery technology containers that provide isolated environments as runtime environments. Cache-based solutions rely on preloading required or commonly used libraries into containers. In this way, it is aimed to speed up the container preparation process and shorten the cold start latency period. Mohan et al. \cite{mohan2019agile} proposed a solution based on pre-rendering network creation to speed up container provisioning. They drastically reduced cold start latency by loading networks into paused containers using a system they call a Pause-Container Pool Manager (PCPM). In another study, LIU et al. \cite{liu2023faaslight} recommended FaaSLight, a new approach based on app optimization. FaaSLight reduces cold start latency by loading only the core part of the application code, which is one of the root causes of cold start. In the study suggested by Wang et al. \cite{wang2019replayable}, with the replayable execution optimized method, the image of the applications can be saved and shared between containers. In this way, applications are restored if necessary, thus shortening the application start-up time and thus the cold start latency period. In \cite{zuk2020scheduling}, new scheduling methods are proposed, an approach that reduces latency for chain functions (functions that call each other). After the first function is called, the runtime environment for the next function to be called is predicted and prepared in advance. In this way, the cold start latency time can be reduced. In another similar study, \cite{daw2020xanadu} proposed a tool called Xanadu, which provides resources when needed to reduce the cold start that occurs in chained functions. Cache-based solutions can be inspiring to reduce cold start latency time, but determining an effective cache policy may not be as easy as it seems. Also, most of the cache-based solutions are those that require high resource consumption for the runtime environment \cite{wen2023rise}.} \textcolor{black}{Cache-based solutions vary on each platform in terms of implementation and effectiveness. This is due to differences in the architecture and optimization techniques of each serverless platform. Since each platform has its own strengths and trade-offs, it is most logical to decide on the platform to be used according to the project. Amazon, Microsoft, and Google serverless services use Amazon ElastiCache, Memorystore, and Azure Cache for Redis to speed up access to applications \cite{zhang2022infinistore,vergadia2022visualizing,kumar2022using}.} \textcolor{black}{The cache-based solutions interact with the underlying architecture of serverless computing, its differences, and tradeoffs in terms of implementation and effectiveness were examined in detail in \textit{Section A} of \textbf{\textit{Appendix B}}.}

     \item \textbf{Design-based Solutions:} \textcolor{black}{Studies on improving isolation and runtime architectural designs to reduce cold start latency in serverless computing. Hall et al. \cite{hall2019execution} introduced a new terminology study that leverages WebAssembly as an alternative to containers for serverless functions. First, serverless access models are characterized to determine runtime requirements, and then performance comparisons are made with the Openwhisk platform to show that WebAssembly is a good alternative. It has been found that the proposed run has less cold start latency than containers. The authors in \cite{du2020catalyzer} proposed Catalyzer, a new sandboxing architecture with strong isolation and fast function initialization. Catalyzer restores the function instance from the checkpoint image to reduce the initialization cost. Thus, it can reduce the cold start latency time. The study suggested by Vedaei et al. \cite{kaffes2019centralized} proposed a new scheduler in serverless computing that eliminates queue imbalances and also reduces interference. Using The scheduler, only function code and its dependencies (libraries, etc.) can be processed in the sandbox without the need for a full operating system image. Thus, the cold start latency time can be reduced. In \cite{mistry2020demonstrating}, the authors introduced UniFaaS as a serverless platform for edge devices. They suggested using unikernels against the container technology used as runtime on serverless platforms. And they reported that UniFaaS has much lower latency for both cold and warm starts. Akkus et. al. \cite{akkus2018sand} introduced a new serverless platform, SAND. By introducing a new sandboxing approach, SAND was able to reduce the container preparation process and thus the cold start latency time. The authors in \cite{gadepalli2020sledge} presented a new serverless framework Sledge for edge based on WebAssembly. Sledge has fast startup times by supporting high-density computation, so it shows great promise in reducing cold start latency. In \cite{solaiman2020wlec}, a new container architecture WLEC has been proposed. WLEC uses S2LRU++ Cache replacement policies, thereby speeding up the container preparation process and reducing cold start latency. \textcolor{black}{Oakes et al. \cite{oakes2018sock} They proposed a new container system called SOCK. SOCK works by optimizing workloads (Avoiding Kernel scalability bottleneck, Implementing Zygote provisioning strategy) and experiments have shown that SOCK is 2.8x faster than AWS Lambda. Li et al. \cite{li2022rund} proposed a new lightweight container runtime called RunD. RunD uses MicroVM to provide high-density deployment and high-concurrency startup. The authors adopted RunD as an AliBaba serverless container and reported in their experiments that RunD had the performance of starting over 200 containers per second. Shillaker et al. \cite{shillaker2020faasm} have introduced a new isolation abstraction called Faaslet. Faaslet aims to reduce resource load by supporting memory sharing between functions. The runtime in which Faaslet is implemented is Function-as-a-Service Microservice (FAASM). FAASM is a working environment that increases the performance of serverless functions. FAASM can reduce startup time and therefore cold start latency by restoring Faaslet from snapshots. These three studies examined use different runtime environments: container, MicroVM, and WASM, respectively. For this reason, each runtime has its own characteristics and performs differently in different optimizations. Among the runtime environments examined, WASM has the fastest startup time, MicroVM has the strongest isolation, and containers have the best performance for lightweight applications. Containers; It has (i) resource management optimization \cite{vhatkar2020particle}, (ii) image optimization to reduce container image size. MicroVMs; It has optimizations such as (i) snapshotting to speed up booting, (ii) memory ballooning for memory management \cite{hu2018hub}. WASM; It uses strategies such as (i) eliminating dead code for code optimization, (ii) caching WASM modules to speed up future executions \cite{ylenius2020mitigating}. As a result, each isolation and runtime architectural design has different characteristics. And design-based solutions can be extended in future work, not only for cold start solutions but also for safety and performance studies.} \textcolor{black}{Design-based Solutions vary depending on the architecture and design of the serverless platform on which they are applied.} \textcolor{black}{Design-based solutions interact with the underlying architecture of serverless computing, its differences, and trade-offs in terms of implementation and effectiveness were examined in detail in \textit{Section B} of \textbf{\textit{Appendix B}}.}} 

      \item \textbf{Function Execution State-based Solutions:} \textcolor{black}{ In this method, the execution state for a function is stored locally, and if it is called again, the platform starts the function quickly using the stored state. In this way, it is aimed to reduce the cold start latency time. Ustiugov et.al. \cite{ustiugov2021benchmarking} have proposed a new framework vHive that proposes to run an image of a function stored on disk instead of starting a function from scratch. The authors identified the main reason for the cold boot delay as restoring an image of a function populating the guest memory. For this reason, they succeeded in reducing cold start latency by 3.7 times by pre-fetching the working set of pages created when a function is called for the first time to the guest memory. The authors in \cite{silva2020prebaking} suggested restoring functions from a previously saved image using a prebrake-based method. In this way, it is aimed to reduce the cold start latency time by reducing the function process start-up time. The study suggested by Cadden et al. \cite{cadden2020seuss} introduced a new system-level method called SEUSS that enables the fast distribution of functions. Functions for this are deployed from unikernel snapshots. In this way, it is expected that the cold start latency time will decrease. In \cite{kumari2023acpm}, authors propose an approach to reduce cold start latency by assigning similar functions to the same containers using the sandbox algorithm. While function execution state-based solutions are successful in reducing cold start latency, they may not be practical for situations such as when the input data payload is very different. Because in such cases, the runtime has to wait for the input data and the function execution state to be loaded.\textcolor{black}{Function Execution State-based Solutions vary depending on the architecture and design of the serverless platform on which they are applied.} \textcolor{black}{ Function execution state-based solutions interact with the underlying architecture of serverless computing, its differences, and tradeoffs in terms of implementation and effectiveness were examined in detail in \textit{Section C} of \textbf{\textit{Appendix B}}.}}
    
      \item \textbf{Function Fusion-based solutions:} \textcolor{black}{This solution method combines sequential functions and aims to remove the cold start of the second function. Lee et.al. \cite{lee2021mitigating} suggested a Function Fusion-based method to reduce cold start latency. If two sequential functions are combined into one function, the cold start from the second function is eliminated and there is only one cold start delay. In fusion-based operations, even if the cold start of the second function is eliminated when parallel functions are combined, in serverless computing, the processing time may increase because the functions are run sequentially. Therefore, a new fusion-based method for parallel functions is still an area of research.\textcolor{black}{Function Fusion-based Solutions vary depending on the architecture and design of the serverless platform on which they are applied.} \textcolor{black}{ Function fusion-based solutions interact with the underlying architecture of serverless computing, its differences, and tradeoffs in terms of implementation and effectiveness were examined in detail in \textit{Section D} of \textbf{\textit{Appendix B}}}}.

\end{itemize}

\begin{table*}[t]
\caption{\small \textcolor{black}{Classification of Cold Start Latency (CSL) Time Reduction Techniques }}
\label{table:cl}

\begin{center}
\footnotesize
\begin{tabular}{@{}cccc@{}}
\toprule
\textbf{Solution Method} &
  \textbf{Objective} &
  \textbf{Work} &
  \textbf{Technique Used} \\ \midrule
Cache-based &
  CSL &
  \begin{tabular}[c]{@{}c@{}}\cite{mohan2019agile}\\ \cite{liu2023faaslight}\\ \cite{wang2019replayable}\\ \cite{zuk2020scheduling}\\ \cite{daw2020xanadu}\end{tabular} &
  \begin{tabular}[c]{@{}c@{}}PCPM \\ FaaSLight \\ Replayable Execution Optimized \\ Scheduling \\ Xanadu\end{tabular} \\
Design-based &
  CSL &
  \begin{tabular}[c]{@{}c@{}}\cite{hall2019execution}\\ \cite{du2020catalyzer}\\  \cite{kaffes2019centralized}\\  \cite{mistry2020demonstrating}\\  \cite{akkus2018sand}\\  \cite{gadepalli2020sledge}\\  \cite{solaiman2020wlec}\end{tabular} &
  \begin{tabular}[c]{@{}c@{}}WebAssembly-based \\  Catalyzer \\ Scheduler \\ UniFaaS \\ SAND \\ Sledge \\ WLEC\end{tabular} \\
Function Execution State-based &
  CSL &
  \begin{tabular}[c]{@{}c@{}}\cite{ustiugov2021benchmarking}\\ \cite{silva2020prebaking}\\ \cite{cadden2020seuss}\\ \cite{kumari2023acpm}\end{tabular} &
  \begin{tabular}[c]{@{}c@{}}vHive\\ Prebrake-based \\ SEUSS \\ Sandbox Algorithm\end{tabular} \\
Function Fusion &
  CSL &
  \cite{lee2021mitigating} &
  Function Fusion \\ \bottomrule
\end{tabular}
\end{center}

\end{table*}

\subsubsection{Studies to reduce the frequency of cold start occurrence}
\textcolor{black}{Studies to reduce the cold start frequency can be examined under 5 subheadings. Table \ref{table:cf} shows studies and methods used to reduce the frequency of cold start occurrence. It should be noted that these subheadings may vary according to the review methodology.}

\begin{figure}[t]
    \centering
    \includegraphics[width=0.75\linewidth]{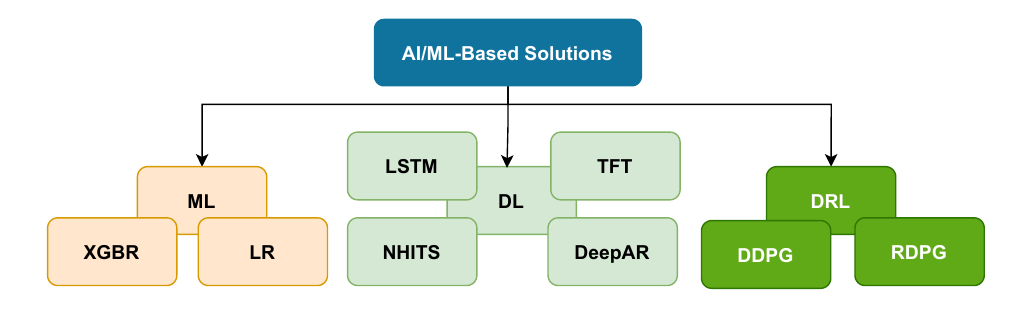}
    \Description{This figure illustrates AI-based solutions for addressing issues in serverless computing.}
    \caption{AI-based solutions}
    \label{fig:fig14}
\end{figure}

\begin{itemize}
    \item \textbf{Periodic Pinging and Container Preparation:} \textcolor{black}{In this method, containers are prepared in advance to meet new requests on the serverless platform by following a specific period or workload, and cold start formation is prevented. Gunasekaran et.al. \cite{gunasekaran2020fifer} recommended Fifer, a new resource management framework that provides cluster efficiency and efficient container utilization. Load forecasting is done using the Long short-term memory (LSTM) model in Fifer and containers are proactively prebuilt. The authors in \cite{bermbach2020using} suggested a system that prepares containers in advance using application knowledge on function. In \cite{horovitz2019faastest}, an ML-based cost and performance optimization strategy called FaaStest is proposed. FaaStest learns the behavioral models of the service provider and recommends the most suitable model to the user in terms of cost and performance. It also pre-provisions containers for functions by making a time series-based prediction to prevent cold start, predicting when requests are called. The period to follow for pinging can be obtained by analyzing factors such as function CPU usage and request time. However, analyzing these factors and training the AI models used for ping requires much effort. Also, unnecessary pinging can result in unnecessary resource wastage as containers run idle. \textcolor{black}{Further, AI-based solutions are developed, such as the ATOM and MASTER frameworks \cite{golec2023atom, golec2024master}, in which the request patterns coming to the server are predicted using time-series models as shown in Figure \ref{fig:fig14}. These models predict the occurrence of cold start using Deep Reinforcement Learning (DRL) and Deep Learning (DL) methods. Thus, they aim to lay the foundation for future cold start prevention studies that do not waste resources (resource-sensitive). The authors chose DRL and DL methods because these two models are successful in solving complex and nonlinear problems. However, there are still challenges to be overcome in the integration of solutions based on AI/ML models into serverless computing environments: (i) Due to the nature of the serverless computing paradigm, a cold start delay occurs when functions are launched for the first time. AI models with high startup requirements (RAM, CPU, large amount of data, etc.) can further increase this latency. (ii) Serverless platforms do not contain large amounts of resources because they work using functions that trigger each other. For this reason, AI models that require high processing power may lead to problems such as execution errors and reduced scaling performance in serverless environments. (iii) One of the most important features of serverless computing is the pay-as-you-go pricing model. Since AI models will consume a high amount of resources, they may cause the user not to benefit from the pay-as-you-go advantage. In this solution method, ping can be sent to the server in cases where a cold start does not occur. This may cause unnecessary waste of resources. For this reason, it is very important to increase the prediction accuracy of the AI models used (such as using new models). The DRL models examined within the ATOM framework are Deep Deterministic Policy Gradient (DDPG) and Recurrent Deterministic Policy Gradient (RDPG). Experiments have shown that these models have some advantages and disadvantages compared to each other in terms of prediction performances, computation time, and $CO_2$ emission amount. The RDPG model is 6 times slower than DDPG. This is because Recurrent Neural Networks (RNNs) add complexity to the model to increase the learning performance of RDPG. Likewise, it is seen that DDPG is more successful in cold start prediction performances. This is because DDPG can more successfully predict complex problems such as cold start with its discovery mechanism. In the MASTER framework, DDPG, LSTM, XGB Regressor,  linear regression (LR), DeepAr, Neural Hierarchical Interpolation for Time Series Forecasting (NHITS), and Teamfight Tactics (TFT) models are compared. The results show that the XGB Regressor model has higher success than all other models. This is because ML models perform better on small data sets such as cold start. Additionally, DL models perform poorly on noisy datasets compared to ML.}}
     \item \textbf{Keeping Container Warm and Container Pool:} {Most commercial platforms like AWS, and GCP keep containers alive for a certain amount of time after function execution is complete \cite{wen2023rise}. During this period, requests that may come to the server are assigned to these containers that are kept alive, and thus it is planned to prevent the occurrence of a cold start. However, this policy wastes resources as it cannot detect the exact time of cold start occurrence. Another similar method is the container pool used in the Fission and Knative platforms. In this method, constantly running containers are used to respond to requests. This method causes a waste of resources as it will require resources to work in vain. Kumari et.al. \cite{kumari2023acpm} proposed a new model called ACPM that provides the runtime provisioning of containers. In the two-stage model, first, the number of containers that need to be preheated is determined using the LSTM model. An integrated docker module is then used to deliver the heated containers faster and reduce the cold start frequency. In \cite{lin2019mitigating} has been suggested to create a pool of hot containers to handle requests to the server. However, this solution is a resource-intensive process. Kumari et.al \cite{kumari2022mitigating} proposed a new method based on hot containers. Using deep neural networks and LSTM models, the number of hot containers and the keeping times of the containers are determined. So they try to suggest a resource-aware based approach to prevent a cold start. In another similar study, \cite{vahidinia2022mitigating} uses a reinforcement learning method and LSTM models to determine the number of hot containers and the time they are kept warm. One of the biggest limitations of this solution method is unnecessary resource consumption. Because, even in cases where a cold start does not occur, certain system resources are used to keep containers warm. To prevent this, AI-based container warm-up strategies can be applied.}
      \item \textbf{Instance Prewarm-based Solutions:} \textcolor{black}{It is aimed to prevent the occurrence of cold start by pre-starting the function instances required to execute the requests coming to the server. Xu et al. \cite{xu2019adaptive} introduced a new resource-sensitive optimization strategy to reduce the cold start frequency. First, with the Adaptive Warm-Up (AWU) strategy, the call times of functions are estimated with a time series model. And according to this prediction result, the functions are warmed. However, these approaches may require resources to run idle, as they cannot show the same performance in sudden demand increases or determine the cold start time precisely. In this solution method, it is difficult to estimate the time that the instances need to be heated. With the increase in the number of functions, the management of pre-heated samples becomes complex. AI-based models can be leveraged to overcome these issues. Thus, for example, the heating time will be estimated more accurately.}
      \item \textbf{Scheduling Strategies:} \textcolor{black}{It is the study of container lifecycle and scaling policy to reduce cold start frequency in serverless computing. Wu et al. \cite{wu2022container} proposed a new container lifecycle-aware scheduling strategy called CAS. CAS controls the request distribution and builds each container according to the required lifecycle stage. When scheduling a request, CAS prefers the workers with containers. And it uses the first available container for the worker of choice. In this way, it is aimed to reduce the cold start frequency. The authors in \cite{suresh2020ensure} introduced ENSURE, a new efficient scheduler and resource manager for serverless platforms. ENSURE is designed to eliminate cold start and application resource contention using queuing concepts. ENSURE classifies function requests and flexibly scales resources based on varying workload traffic to prevent cold start. The authors in \cite{fuerst2021faascache} proposed a new system called FaasCache, which follows a GReedy-Dual-based policy to prevent cold start. GReedy-Dual is a size and frequency-sensitive caching policy. Using an analogy, the authors suggested that the keep-alive and object-caching methods could be adapted to each other. Therefore, they showed that the GReedy-Dual cache policy can be adapted to an effective keep-alive policy. The authors proposed a new container management scheme based on container sharing called Pagurus. According to this scheme, instead of starting a new container for incoming requests to the server, they appoint idle hot containers. In this way, it is aimed to prevent cold start, the time it takes to start a new container. In \cite{sethi2023lcs}, a scheduling approach called LCS is proposed to keep containers hot for longer. In this affinity-based approach, the last container used to execute incoming requests is detected. In this solution method, there are problems such as algorithm complexity that may arise when determining resource contention and scheduling strategies. To overcome these problems, adaptive scheduling algorithms and strategies to distribute the load equally to the instances can be used.}
      \item \textbf{Container Runtime Reusing:} \textcolor{black}{This method, which is similar to the container pool method to reduce the cold start frequency, uses the runtime environment instead of warm containers. Suo et.al. \cite{suo2021tackling} introduced the container-based runtime management framework HotC. Demand is forecasted using the exponential smoothing model and Markov chain method, and the necessary runtime environment is provided from the container runtime pool. In this way, it is tried to prevent cold start by using containers efficiently. The container remaining from previous executions should be carefully cleaned. Otherwise, it will cause security vulnerabilities. For this reason, safe container-cleaning processes must be implemented.}
\end{itemize}

\begin{table*}[t]
\caption{\small \textcolor{black}{Classification of Cold Start Frequency (CSF) Reduction Techniques }}
\label{table:cf}
\begin{center}
\resizebox{.65\textwidth}{!}{
\footnotesize
\begin{tabular}{@{}cccc@{}}
\toprule
\textbf{Solution Method}         & \textbf{Objective} & \textbf{Work}           & \textbf{Technique Used} \\ \midrule
\begin{tabular}[c]{@{}c@{}}Periodic Pinging and \\ Container Preparation\end{tabular} &
  CSF &
  \begin{tabular}[c]{@{}c@{}}\cite{gunasekaran2020fifer}\\ \cite{bermbach2020using}\\ \cite{horovitz2019faastest}\end{tabular} &
  \begin{tabular}[c]{@{}c@{}}Fifer \\ Application Knowledge on Function \\ FaaStest\end{tabular} \\
\begin{tabular}[c]{@{}c@{}}Keeping Container Warm and \\ Container Pool\end{tabular} &
  CsF &
  \begin{tabular}[c]{@{}c@{}}\cite{kumari2023acpm} \\ \cite{lin2019mitigating} \\ \cite{kumari2022mitigating} \\ \cite{vahidinia2022mitigating}\end{tabular} &
  \begin{tabular}[c]{@{}c@{}}ACPM \\ Container Pool \\ DNN + LSTM \\ RL + LSTM\end{tabular} \\
Instance Prewarm-based Solutions & CSF                 & \cite{xu2019adaptive}  & AWU + ACPS              \\
Scheduling Strategies &
  CSF &
  \begin{tabular}[c]{@{}c@{}}\cite{wu2022container} \\ \cite{suresh2020ensure} \\ \cite{fuerst2021faascache} \\ \cite{sethi2023lcs}\end{tabular} &
  \begin{tabular}[c]{@{}c@{}}CAS \\ ENSURE \\ FaasCache \\ LCS\end{tabular} \\
Container Runtime Reusing        & CSF                 & \cite{suo2021tackling} & HotC                    \\ \bottomrule
\end{tabular}
}
\end{center}
\end{table*}

\subsection{RQ4: Which platforms have existing studies been implemented?}\label{sec:RQ4}

\textcolor{black}{The purpose of this research question is to investigate which serverless platforms the existing solutions for the cold start problem are applied to and evaluated. The first two authors of this paper identified the platforms used by carefully examining the Abstract, Conclusion, and Implementation sections of the existing studies, respectively. Fig \ref{fig:fig15a} shows the percentage distribution of serverless platforms on which the existing solutions reviewed in this paper are implemented. As can be seen from the figure, the current studies are commercial, open-source, and simulation-based platforms, including OpenWhisk, AWS Lambda, Knative, Kubernetes, OpenFaaS, Microsoft Azure, Simulation-based, GCF, Nuclio, Firecracker, Huawei Cloud Functions, and Kubeless. The majority of existing studies (\cite{hall2019execution}, \cite{mohan2019agile}, \cite{wu2022container}, \cite{suresh2020ensure}, \cite{fuerst2021faascache}, \cite{vahidinia2022mitigating},  \cite{akkus2018sand}, \cite{cadden2020seuss}, \cite{bermbach2020using},\cite{daw2020xanadu}) use Openwhisk, an open-source serverless platform of 34.37\%. The reasons for this are; (i) Openwhisk allows users to design their own custom runtime environments. (ii) Since Openwhisk is open source, users can easily distribute their own solutions. The second most used platform is AWS Lambda with 29.03\% (\cite{solaiman2020wlec},\cite{bermbach2020using}, \cite{kumari2022mitigating}, \cite{lee2021mitigating}, \cite{sethi2023lcs}, \cite{liu2023faaslight},\cite{kumari2023acpm}). Also, AWS Lambda is the most used commercial platform compared to other commercial platforms like Microsoft and Google. This may be because AWS Lambda was the first commercial serverless platform, and hence greater adoption. Also, it may be that the infrastructure support it offers for researchers in deploying cold start solutions to the platform is more mature \cite{wen2023rise}. The third most preferred platforms are Knative (\cite{daw2020xanadu}, \cite{lin2019mitigating},\cite{ustiugov2021benchmarking}), Kubernetes(\cite{horovitz2019faastest}, \cite{gunasekaran2020fifer}, \cite{xu2019adaptive}), and OpenFaaS (\cite{suo2021tackling},\cite{silva2020prebaking},\cite{kumari2022mitigating}) with 9.67\% rates. All three technologies have flexible and efficient management and efficient resource options for developers. Microsoft Azure and simulation-based platforms account for 6.45\% of platforms that cold start mitigation works implemented (\cite{kumari2022mitigating}, \cite{li2022help}, \cite{zuk2020scheduling}, \cite{gias2020cocoa}). Cloud Functions, Nuclio, Firecracker, Huawei Cloud Functions, and Kubeless are the least preferred platforms in the studies reviewed.}

\textcolor{black}{The scalability of solutions for cold start issues varies across different serverless platforms \cite{silva2019pure}. An example is a warm start. In AWS Lambda, containers continue to run for a while even after the function has completed its execution. In case of a new function call, these containers are reused to reduce the cold start latency period. Similar systems also exist in Azure and GCP. However, the performance of this solution method varies due to reasons such as cross-platform system architecture, and container startup speed \cite{mcgrath2017serverless}. Other reasons for the difference in scalability performance of solution methods across platforms are (i) Isolation Techniques and (ii) Runtime Environment \cite{lee2018evaluation}. Each platform uses different container isolation techniques depending on its specific architecture and this creates differences in the scalability of solutions for cold start issues \cite{perez2018serverless}. AWS Lambda uses container-based solutions like "container reuse" and "keep-warm", GCP, VPC (Virtual Private Cloud), Azure, and "Durable Functions" \cite{chen2023hyscaler}. Runtime may vary on each platform. For example, languages like Go and Node.js run faster on AWS Lambda than on Java or .NET.}

\subsection{RQ5: Where are existing studies published and do they share the dataset or code?}\label{sec:RQ5}

\begin{figure}[t]
    \centering
    \begin{subfigure}[b]{1\textwidth}
        \centering
        \includegraphics[scale=0.5]{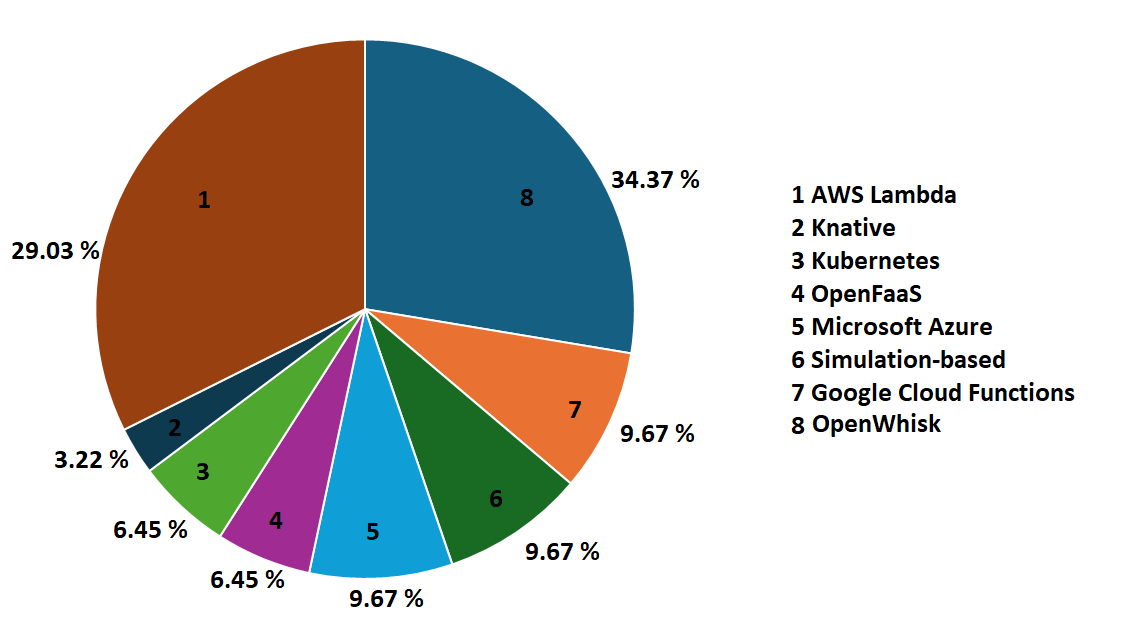}
        \Description{This subfigure shows the percentage distribution of platforms used in research studies.}
        \caption{}
        \label{fig:fig15a}
    \end{subfigure}
    \hspace{0.21\textwidth}
    \begin{subfigure}[b]{1\textwidth}
        \centering
        \includegraphics[scale=0.5]{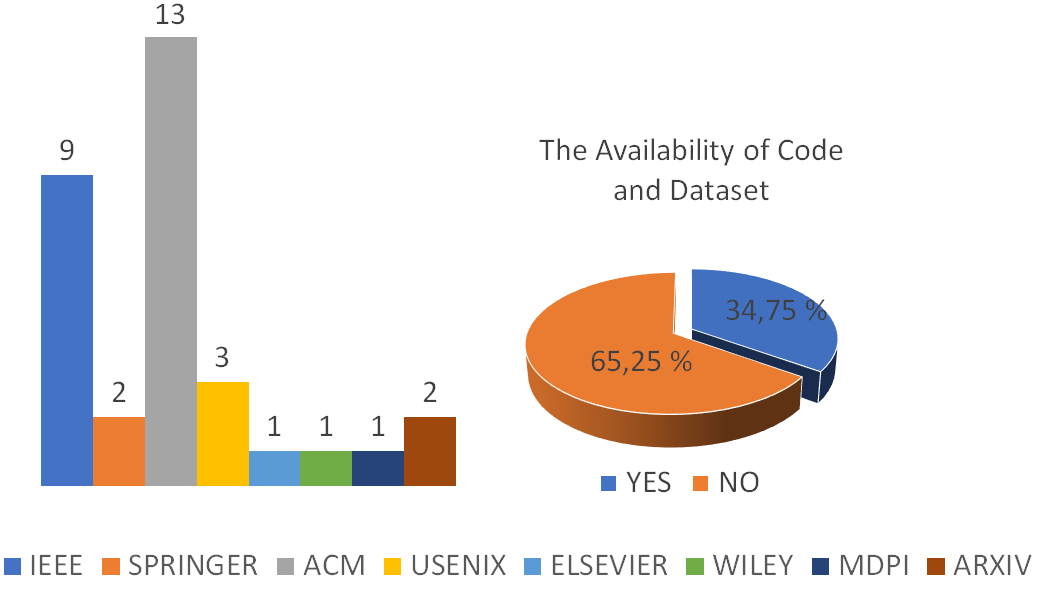}
        \Description{This subfigure shows the distribution of research venues and the availability of code and datasets.}
        \caption{}
        \label{fig:fig15b}
    \end{subfigure}
    \Description{This figure consists of two subfigures: (a) the percentage distribution of platforms and (b) the distribution of research venues along with the availability of code and datasets.}
    \caption{(a) The Percentage Distribution of Platforms (b) Distribution of Venues of Research Works \& Availability of Code and Dataset}
    \label{fig:combined}
\end{figure}

\textcolor{black}{ With this research question, it investigates where the studies are published and whether the dataset used in the current studies is shared with researchers.} \textcolor{black}{Additionally, in order to provide a more comprehensive evaluation, we also add evaluation information about the publication year of the article, the number of citations, and the quality of the conference/journal in \textbf{\textit{Appendix C}}.} \textcolor{black}{The first two authors identified where they were published by typing the names of the articles from Google Scholar. Figure \ref{fig:fig15b} shows the organizations where current studies are published. As can be seen, most of the works about the cold start solution in serverless computing have been published in ACM.} \textcolor{black}{Accessibility of the dataset is very important in terms of reproducing solutions and opening the door to new research. To answer this question, the paper's third author first scanned the Dataset subsection. If he could not find a definitive answer to this question, he carefully studied the entire paper and especially the reference section. Figure \ref{fig:fig15b} shows the percentage distribution showing the accessibility of the dataset and code for the cold start papers reviewed in this paper. Datasets or codes are available for only 34.75\% of studies. This indicates that most of the authors focused only on design and implementation descriptions rather than the reproducibility of studies.}

\section{OPEN CHALLENGES AND FUTURE DIRECTIONS} \label{sec:challengesandfuture}

\begin{figure}[t]
    \centering
    \includegraphics[width=1\linewidth]{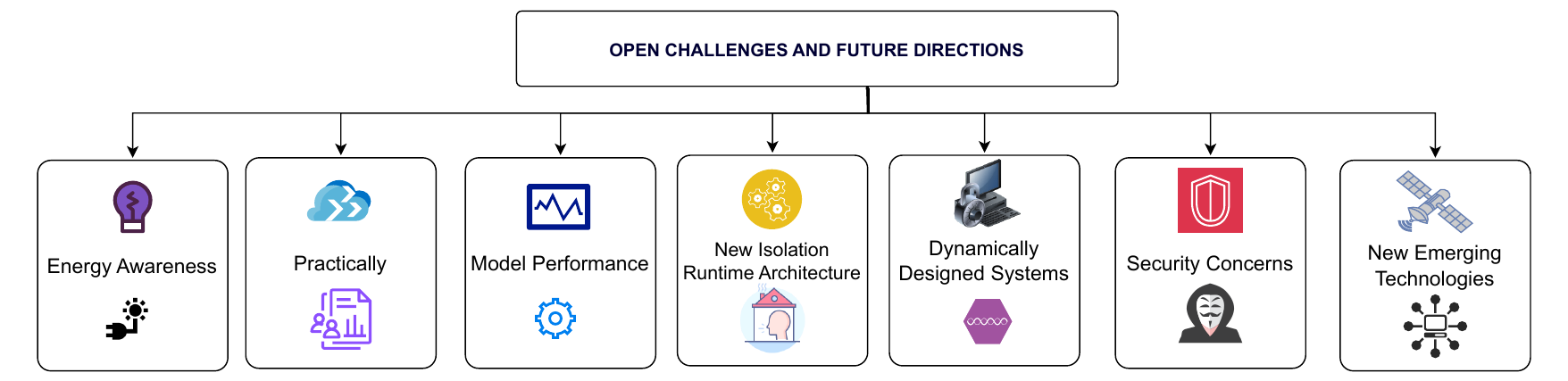}
    \Description{This figure presents the taxonomy of open challenges and future directions in serverless computing.}
    \caption{\textcolor{black}{Taxonomy of Open Challenges And Future Directions}}
    \label{fig:fig16}
\end{figure}

\textcolor{black}{In this systematic research paper, over 100 papers were scanned to examine the cold start problem in serverless computing and categorize current solutions. Keele's Systematic Literature Reviews method \cite{keele} was followed and the snowballing technique was applied to make a systematic review, and the 32 most meaningful and up-to-date cold start research papers were identified and classified. Based on the taxonomy study, existing studies in the literature were reviewed and compared for 5 Research Questions. In this section, we identify open challenges and future directions for the cold start issue still waiting to be resolved in serverless computing. A taxonomy for open challenges and future directions is provided in Figure \ref{fig:fig16}. We have examined the techniques used to reduce cold start frequency and latency under two general headings in subsection \ref{sec:RQ3}. The techniques generally get trade-offs such as Energy-awareness, Practicality, and Model Performance. New solutions are needed that reduce or eliminate these trade-offs.} As an independent research gap from these, also energy-sensitive simulators are needed to simulate the cold start situation in serverless and serverless edge environments \cite{nandhakumar2023edgeaisim}.

\subsection{Energy Awareness} 
\textcolor{black}{ Some of the preferred methods for reducing cold start frequency and cold start latency require wasting resources. These methods can be listed as follows:}

\begin{itemize}
    \item \textcolor{black}{\textbf{Periodic pinging-based studies.} These studies \cite{gunasekaran2020fifer}, \cite{bermbach2020using}, \cite{horovitz2019faastest}, in which periodically repetitive workloads are followed and containers are made ready on the server, often require containers to run in vain. In this solution method, new energy-aware studies can be done by using effective and high-performance AI models.}

    \item \textcolor{black}{\textbf{Studies based on Keeping Container Warm and Container Pool.} These studies \cite{kumari2023acpm}, \cite{lin2019mitigating}, \cite{kumari2022mitigating}, \cite{vahidinia2022mitigating}, which are based on the fact that the containers whose function execution is completed work for a certain period of time without scale to zero, or the use of constantly running containers, require resources to be wasted.}

    \item \textcolor{black}{\textbf{Instance Prewarm-based studies. }In these studies \cite{xu2019adaptive}, where the function instances make been ready before the request arrives, sudden demand increases cannot be coped with, and the accuracy rates of the AI forecasting models used are not at a satisfactory level.}

    \item \textcolor{black}{\textbf{Container Runtime Reusing studies.} In these studies \cite{suo2021tackling} based on the container runtime pool method, runtime environments are kept running continuously, similar to the container pool method. Keeping these runtime environments working will require resources to be wasted. There is a need for studies that provide energy optimization in the environment pool by making high-accuracy demand forecasting.}

    \item \textcolor{black}{\textbf{Cache-based studies.} Cache-based studies are based on the principle of preparing the necessary libraries in advance and quickly installing them in containers to reduce cold start latency and increase the performance of the system \cite{mohan2019agile}, \cite{liu2023faaslight}, \cite{wang2019replayable}, \cite{zuk2020scheduling}, \ cite{daw2020xanadu}. However, in addition to the advantages they offer, since they are solutions that require high resource consumption, studies can be carried out that take into account various optimization aspects in this field. For example, tailored caching methods can be designed in scenarios such as applications requiring high volumes of data, taking into account applicable scenarios and load characteristics. Additionally, dynamic caching strategies can be developed to adapt to scenarios with changing workloads \cite{souri2024tournament}.}

    \item \textbf{Dynamically Designed Solutions}: \textcolor{black}{These are solution methods that can be dynamically adjusted to minimize the cold start latency. These methods; (i) edge-based solutions that reduce the amount of latency by moving the computing power closer to the edge of the network (\cite{koch2021empirical} such as AWS Lambda, Lambda@Edge), (ii) systems that aim to reduce the amount of cold start latency by configuring the scaling speed \cite{srinivasan2018google}, and (iii) solutions where functions are kept warm (Azure, Premium Plan \cite{olariu2023challenges}). All these solutions bring with them the disadvantage of unnecessary waste of resources.}
    
\end{itemize}

\subsection{Practicality} 

\textcolor{black}{Some of the preferred methods to reduce the cold start frequency and latency still bring impractical issues such as long processing time. New practical solutions can be offered by using new optimization techniques.}

\begin{itemize}
    \item \textcolor{black}{\textbf{Function Execution State-based studies.} These studies \cite{ustiugov2021benchmarking}, \cite{silva2020prebaking}, \cite{cadden2020seuss}, \cite{kumari2023acpm}, which are based on the logic of storing function execution states and using them for fast startup if needed, will take a long time to load runtime input data and execution states in scenarios where the input data payload differs. Therefore, there is a need for new studies that include scenarios where the input data payload differs.}

    \item  \textcolor{black}{\textbf{Function Fusion-based studies.} In these studies \cite{lee2021mitigating}, which aim to remove the cold start of the second function by combining sequential functions, the case of combining parallel functions still awaits to be investigated as a big challenge.}
    
\end{itemize}
 
\subsection{Model Performance} 

\textcolor{black}{Some of the preferred methods to reduce cold start frequency and latency try to predict workloads and call times of functions using AI-based models \cite{gunasekaran2020fifer},
\cite{bermbach2020using},
\cite{horovitz2019faastest}, \cite{xu2019adaptive}. However, the AI prediction models' performances still have not reached satisfactory levels. Transfer Learning techniques and cold start datasets can be used to improve the performance of AI models. Transfer learning is a technique used in environments with limited computational capabilities and little data \cite{chowdhury2023covidetector}. Reusing previously trained models for another task can increase model performance. When the literature is examined, it is seen that there is a need for an open-source dataset regarding the cold start. Models trained with datasets created from a real environment workload will give more accurate prediction results for cold start prediction. Additionally, in the future, ML models with high prediction accuracy may inspire solutions to improve cold start. ML models are expected to have higher prediction performance than DRL and DL models in small and noisy datasets such as cold start datasets.}

\subsection{ New Isolation and Runtime Architecture}
\textcolor{black}{Isolation and runtime architectures are promising in solving various challenges (security etc.), especially the cold start problem in serverless computing \cite{Transforming2024}. An example is the lightweight and secure virtualization architecture AWS Firecracker. AWS Firecracker is a microVM-based technology and offers advantages such as minimal overhead and faster operation for serverless instances compared to traditional VMs \cite{agache2020firecracker}. Thanks to minimal overhead, microVMs are started faster and cold start latency is directly reduced. Another example is Google's proposed gVisor technology. gVisor acts as an additional layer between the application and the processor, blocking system calls and reducing cold start latency as it provides flexibility by working compatible with existing containers \cite{young2019true}.} \textcolor{black}{Future researchers may combine lightweight microVMs and system call interception techniques to further reduce cold start latency. In addition, serious improvements can be made on the cold start time and response time of serverless computing with new techniques to be developed on container life cycle and scalability policies. In order to understand the limitations of the new Isolation and runtime architecture studies carried out to reduce cold start latency, the effects of resource consumption and scalabiity on performance can be examined.}

\subsection{\textcolor{black}{Dynamically Designed Serverless Computing Systems}}

\textcolor{black}{In this subsection, the difficulties of designing serverless computing systems that can be dynamically adjusted to minimize the cold start problem in the serverless paradigm will be discussed. Almost all serverless platforms offer dynamic adjustment options to reduce cold start. The options and challenges offered by AWS Lambda, GCF, and Microsoft Azure are as follows:}

\begin{itemize}
    
\item \textcolor{black}{ AWS Lambda: AWS aims to minimize the latency effect caused by cold start by processing data at the edge with Lambda@Edge technology \cite{koch2021empirical}. Since the data is processed close to the data source, the response delay is reduced relative to the communication delay between the user and the server. The other dynamic solution is to optimize the VPCs used for the lambda function. In addition to the advantages these solutions offer, they also bring additional costs and scalability problems since edge devices have limited processing power and resources.}

\item \textcolor{black}{ Google Cloud Functions (GCF): GCF allows the scaling rate to be configured so that the amount of cold start latency can be reduced when new instances are started \cite{srinivasan2018google}. Additionally, GCF provides the option to adjust the number of ready instances that can meet new requests. In this way, it is aimed to prevent a cold start that may occur, as there are instances that can meet the requests in a working state. These dynamic solutions offered by GCF also have disadvantages such as cost and configuration complexity.}

\item \textcolor{black}{ Microsoft Azure: Azure tries to prevent cold start by keeping functions warm with the Premium Plan \cite{olariu2023challenges}. This dynamic solution has disadvantages such as unnecessary waste of resources and additional costs.}

\end{itemize}

\textcolor{black}{Apart from these developments, some of the promising research areas in the solutions offered regarding cold start are as follows:}

\begin{itemize}

\item \textcolor{black}{ Network optimization: Promising developments that can reduce cold start in the field of network optimization can be listed as follows: (i) Edge computing: Edge computing is an emerging technology that aims to reduce latency and unnecessary bandwidth usage by bringing computing devices closer to the data source. It can be used to reduce latency caused by a cold start. The serverless edge computing paradigm, which has recently emerged by combining the serverless and edge computing paradigms, is a new research field that attracts attention in the academy \cite{golec2024master}. (ii) Software-Defined Networking (SDN): It is a network management that configures network management dynamically, unlike traditional methods \cite{banaie2022serverless}. It can be used to dynamically route function calls in serverless computing.}

\item \textcolor{black}{ Compiler Technology: Ahead-of-time compilation (AOT), the premature compilation of code deployed for compilation in serverless computing, can be a dynamic solution to reduce cold start latency. Additionally, faster-running runtime architectures can be used to reduce cold start, as reviewed in subsection 6.4.}

\item \textcolor{black}{ Resource Prediction: In addition to the ML performance improvements that predict function calls mentioned in Subsection 6.3, anomaly detection methods can also be used for request patterns that may require scaling.}
    
\end{itemize}

\subsection{\textcolor{black}{Security Concerns}}

\textcolor{black}{While existing cold start solutions hold great promise in optimizing latency, which is a fundamental problem in serverless computing, they still have fundamental limitations in addressing security concerns \cite{kumar2024comprehensive}. These limitations and examples are as follows;}

\begin{itemize}

\item \textcolor{black}{Lack of Inter-Function Isolation: Techniques such as reusing containers whose function execution has been completed and reusing function images are some of the existing solutions. However, these solutions carry security risks by leaving the door open for hackers to obtain unsanitized data from previous functions.}

\item  \textcolor{black}{Extending the Lifecycle of Vulnerable Containers: Some cold start solution methods may increase the risk of attack for hackers as they extend the lifecycle of vulnerable containers.}

\item  \textcolor{black}{Security is the Responsibility of Cloud Providers: In a serverless platform, security is assumed to be handled by the service provider, and therefore cold start solutions do not guarantee additional protection.}
\end{itemize}

\textcolor{black}{Future researchers can focus on stronger isolation mechanisms and high-security cold start solutions to address security concerns such as cold start attacks in serverless computing.}

\subsection{New Emerging Technologies}

\textcolor{black}{New technologies such as edge computing and 5G can be used to reduce the negative effects of cold start issues such as latency in serverless computing. Edge computing is based on the basic principle of processing data close to the data source, reducing unnecessary bandwidth usage and latency compared to remote servers \cite{golec2024computing}. In this way, the concept of serverless edge computing, which emerged from the combination of serverless computing and edge computing, has attracted great attention \cite{golec2023atom}. On the other hand, with the widespread use of high-speed data transfer technologies such as 5G and 6G, communication will accelerate and latency periods will shorten. This will contribute greatly to the tolerance of latency periods that occur due to cold starts.}

\section{CONCLUSIONS} \label{sec:conclusions}

\textcolor{black}{In this paper, a systematic review and taxonomy study is presented for the cold start problem, which is still waiting for a solution in the serverless paradigm. First, over 100 most recent literature studies on cold start in serverless computing were collected by the authors to create a taxonomy. A systematic study was carried out by two authors to identify relevant studies that focused only on the cold start problem. And at the last stage, a paper set consisting of 32 papers was obtained by the author, who is an expert on the subject, using the snowball method. Using the paper set, the authors sought answers to five research questions regarding cold start solutions. With the first research question, the relationship between cold start and QoS was questioned. In the second research question, the factors affecting cold start latency were investigated. With the third research question, solution studies on cold start and cold start latency were classified and examined. In the fourth research question, which platforms the relevant solution methods were applied to were investigated and statistical information was obtained. In the last research question, the distribution of the journals in which the relevant research articles were published and whether the dataset/codes were shared were investigated.} \textcolor{black}{
A classification is made of various approaches offered by academic and industrial organizations to reduce cold start frequency and shorten cold start latency in serverless computing. These approaches mainly include AI, ML, DL, and RL-based approaches as well as popular techniques such as caching and application-level optimization. Existing methods are classified according to their common points as a result of the analysis. The final section discusses the open challenges of cold start in serverless computing and identifies future directions, providing valuable information for researchers. Future studies can further investigate new approaches in the future, such as integrating AI-based prediction models and caching mechanisms. Additionally, studies can be conducted to evaluate the performance of solutions proposed for different serverless platforms to improve this article.}

\section*{Appendix A: Cloud Models}
Cloud Models are examined under two subsections. The first subsection is cloud computing models according to deployment. The second subsection is cloud computing models according to delivery.
\subsection*{\textbf{A.1 }Cloud Deployment Models}
Here's an explanation of the cloud deployment models:

\begin{itemize}

\item \textcolor{black}{\textbf{Public Cloud}: It is a cloud model where services are provided by third-party providers such as Google, Amazon and Microsoft \cite{manvi2014resource}. It brings advantages such as an affordable pricing model, better scalability, and intangible infrastructure while control over the data is less as data is controlled by the third party.}

\item \textcolor{black}{\textbf{Private Cloud}: The infrastructure is allocated to a single company/individual that offers advantages higher security, more control over data, and customization in line with needs compared to other models \cite{beimborn2011platform}.}

\item \textcolor{black}{\textbf{Hybrid Cloud}: It offers to combine public and private cloud and benefit from the advantages of both \cite{cusumano2010cloud}. In this paradigm, two different infrastructures are integrated, management is a bit complicated and data transmission security between two clouds may pose a risk.}

\item \textcolor{black}{\textbf{Community Cloud}: It is a type of cloud that is shared between institutions for a common goal such as security \cite{shahrad2019architectural}.  Its resources are limited compared to the public cloud. Since it involves a large number of members, management can be a bit complicated.}

\end{itemize}

\subsection*{A.2 Cloud Delivery Models}
Below is an explanation of the cloud delivery models, including Infrastructure as a Service (IaaS), Platform as a Service (PaaS), Software as a Service (SaaS), and Function as a Service (FaaS):

\begin{itemize}

\item \textbf{IaaS:}
This service model provides the capability to provision processing, storage, networks, and other underlying resources, where the responsibility of managing all hardware, host operating system, virtualization software, guest operating system, and the applications deployed becomes the liability of the user \cite{manvi2014resource}.  Some prominent IaaS Cloud Service Providers (CSPs) are Amazon Elastic Compute (EC2), Amazon S3, Elastic Scale, etc. 

\item \textbf{PaaS:}
The offerings of this model include the capability to provision an APT platform and environment for developers in order to deploy their applications created using programming languages, libraries, and various tools supported by SPs \cite{beimborn2011platform}. AWS Elastic Beanstalk, Azure WebApps, Compute App Engine are few PaaS modules.

\item \textbf{SaaS:}
It implies direct access to on-demand applications from the internet via some user credentials. Software applications might include sharing emails, online office suites, Customer Relationship Management (CRM) software, social networking, and photo editing software which are accessed from any web browser or thin client \cite{cusumano2010cloud}. A few examples include Dropbox, GMail, Salesforce.com, and Zoom. 

\item \textbf{FaaS:}
It constitutes a software architecture where an application is partitioned into events and functions. The functions are event-driven and are triggered either by users' HTTP requests or in response to other events, occurring within the SP platform \cite{shahrad2019architectural}. A few characteristics of functions include lightweight, short-lived, written in any programming language and finally hosted in deeply virtualized manner in Virtual Machines (VMs).

\end{itemize}

\subsection{Serverless Computing Platforms}
Serverless platforms are generally examined under two main sections: open source and commercial.

\textbf{Open Source Platforms:} These platforms are discussed below:

\begin{itemize} 

\item \emph{Kubeless}: It is built on top of Kubernetes, it enables to deployment of functions utilizing a streamlined development workflow where locally developed functions can be packed into containers \cite{balla2020open}. It uses Kubernetes resources to manage function execution and scaling. In addition, it implements stateless execution using storage units called Kubernetes Persistent Volumes, along with interaction with other resources including ConfigMaps, Secrets which enable storage of sensitive information (passwords, tokens, ssh keys) etc. 
\item  \emph{OpenFaaS}: This is the most preferred open-source serverless platform which eases the deployment of event-driven functions along with microservices on Kubernetes without requiring boiler-plate coding \cite{le2022openfaas}. Furthermore, it enables to pack the code in the form of docker image/ Open Container Initiative (OCI) containers via Kubernetes.
\item \emph{Apache OpenWhisk}: It accepts the source code as input which originates from various events such as timers, message queues, and even websites, and delivers the services to consumers via REST APIs \cite{quevedo2019evaluating}. It majorly comprises 3 components: (1) Controller: this component manages entities, handles triggered events, and routes action invocations (2) Invoker: it launches the containers in order to execute actions (3) Action Containers: this component is responsible for executing actions.  
\item \emph{FireCracker}: This platform provides lightweight VMs known as micro VMs \cite{agache2020firecracker}. It consists of a Worker manager module that routes incoming queries to worker nodes. Multiple slots are offered by Lambda, in order to serve functional units. In case of unavailability of a slot, time-based leasing is used by the Placement Service component. 

\end{itemize}

\textbf{Commercial Platforms:} The following are discussions of these platforms:  

\begin{itemize} 
    \item	\emph{AWS Lambda}: It is the most leading and powerful computational platform for hosting serverless applications introduced in 2014 \cite{villamizar2016infrastructure}. In contrast to the traditional approaches based upon physical or virtual servers, it enables its service consumers to focus on business logic. Here, the functions denote the primary unit of Lambda, which can represent a set of specific functionalities such as backup, querying, etc. which are created, updated, modified, and finally deployed. The use cases of this platform involve API hosting, event processing, and managing ad-hoc timer-based jobs. Its workflow involves: (1) Function creation: Console or Command Line Interface (CLI) (2) Defining and uploading code in AWS via multiple programming languages such as Python, JavaScript, Ruby, etc. (3) Deployment and Function Run (direct invocation or API). Apart from this, its global availability and built-in security mechanism via Virtual Private Cloud (VPC), make it one of the prominent platforms for building APIs and web applications.   
    \item	\emph{Google Cloud Functions}: The popular approach of FaaS enables developers to write a piece of code that is triggered in response to any event which either sends the response back or further spawns events \cite{malawski2020serverless}. Apart from providing multiple programming language support, it provides a large set of libraries and frameworks. In addition, it is equipped with CloudRun service which enables to deploy of a docker image instead of application source code. As CloudRun is built with Knative, hence it can be used with Kubernetes cluster too. 
    \item \emph{Microsoft Azure}: This platform provides continually updated infrastructure and resources needed to cater the consumer applications \cite{chappell2008introducing}. Functions enable serverless computing which can be utilized to build Web APIs, send responses in lieu of database change, process IoT data streams, manage message queues, etc. Here, the functions metadata is stored in Azure storage tables whereas the function code resides in Azure storage blobs. 
    \item \emph{Cloudflare Workers (CFW)}: It provides a platform that enables developers to build and deploy javascript functions on Cloudflare edge networks, which are referred to as Cloudflare workers \cite{schwarzl2022robust}. Executing the functions in close proximity to end-users helps to minimize the latency. In addition, this serverless platform enables its developers to run code at multiple places in order to reap the benefits of both CFW and serverless computing paradigms.
\end{itemize}

\section{Appendix B: Interaction Of Cold Start Solutions With The Underlying Architecture Of Serverless Computing}

This section examines how some cold start solution methods differ in terms of implementation and effectiveness across platforms, and what the trade-offs involved.

\subsection{Cache-based Solutions:}

     \begin{itemize}
         \item \textbf{Implementation Differences}
         \begin{itemize}
             \item \textcolor{black}{ElastiCache \cite{hafeez2018elmem}: AWS Lambda functions to provide low latency by accessing ElastiCache instances through the VPC where it is installed.}
             \item \textcolor{black}{Memorystore \cite{tulving1971retroactive}: Google Cloud Functions (GCF) access memory store caches through VPC connectors. GCF requires additional setup to facilitate integration for VPC access.}
             \item \textcolor{black}{Azure Cache for Redis \cite{kumar2022using}: Azure functions can connect to the Redis service via a Virtual network (VNet). Software Development Kits (SDKs) are used to facilitate this process.}
         \end{itemize}
     \end{itemize}

        \begin{itemize}
         \item \textbf{Effectiveness and Trade-offs}
         \begin{itemize}
             \item \textcolor{black}{ElastiCache \cite{hafeez2018elmem}: Due to its VPC feature and automatic scaling feature, it has lower latency and better scaling than cache-based solutions of other platforms. Although it is cost-effective, it becomes costly under excessive load (high request rate, large data, etc.). It has a certain complexity, although lower than other models.} 
             \item \textcolor{black}{Memorystore \cite{tulving1971retroactive}: Scaling performance and latency may be negatively affected under excessive load as it may require additional installation and is sensitive to configuration settings. Although it is affordable, the cost increases as the VPC egress fee increases. It is highly complex due to VPC connectors.}
             \item \textcolor{black}{Azure Cache for Redis \cite{kumar2022using}: Although the latency is relatively good, the initial configuration settings are complex. Additionally, scaling performance may be lower in high-scale situations. VNet peering increases cost and installation complexity.}
         \end{itemize}
     \end{itemize}

\textcolor{black}{These solution methods have disadvantages such as cache consistency and additional resource load on the system. A dynamic cache management strategy can be applied to optimize resource usage.}

\subsection{Design-based Solutions:}

\begin{itemize}
         \item \textbf{Implementation Differences}
         \begin{itemize}
             \item \textcolor{black}{AWS Lambda, GCF, and Azure have a wide range of applications with multi-language support. It provides tools for deploying serverless applications, such as AWS Lambda Serverless Application Model (SAM) and CloudFormation, GCF Google Cloud Deployment Manager, and Azure Resource Manager \cite{karthikeyan2017azure}.}
         \end{itemize}
         \item \textbf{Effectiveness and Trade-offs}
          \begin{itemize}
             \item \textcolor{black}{Design-based Solutions have different effects on different platforms. Although all platforms provide automatic scaling when compared in terms of scalability, AWS Lambda works more effectively in operations requiring high requests and large amounts of data. When compared in terms of latency and especially cold start latency, it was seen that AWS Lambda and GCF had lower latency than other platforms \cite{kiener2021towards}.}
             \item \textcolor{black}{In addition to the advantages offered by Design-based Solutions, it also includes trade-offs such as increased vendor dependency, more complex debugging processes, and unpredictable cost increases \cite{martens2012decision}.}
         \end{itemize}
     \end{itemize}
     
\textcolor{black}{Design-based solutions applied to reduce cold start increase complexity in application architecture. Additionally, the fact that the design is platform-specific means that it cannot be used for other platforms. To overcome these problems, the solutions should be designed in a platform-independent manner. Additionally, when designing the solution, the complexity it will create in the system should also be taken into consideration}

\subsection{Function Execution State-based Solutions:}
      \begin{itemize}
         \item \textbf{Implementation Differences}
         \begin{itemize}
             \item \textcolor{black}{AWS Lambda uses AWS Step Functions to coordinate multiple functions \cite{figiela2018performance}. AWS Step Functions coordinate state transitions by transferring state information between functions. The application that performs this operation in GCF is Google Cloud Workflows \cite{malawski2020serverless}. Google Cloud Workflows manages state by processing function calls. A similar application is Durable Functions, available in Microsoft Azure. Durable Functions manages function state and checkpoints \cite{cao2015checkpointing}.}
         \end{itemize}
         \item \textbf{Effectiveness and Trade-offs}
          \begin{itemize}
             \item \textcolor{black}{In complex scenarios, AWS Step Functions and Azure Durable Functions are more effective than Google Cloud Workflows with their powerful orchestration capabilities \cite{malawski2020serverless}. Likewise, when managing stateful workflows, AWS Step Functions and Azure Durable Functions are better at auto-scaling \cite{chowhan2018hands}.}
             \item \textcolor{black}{All platforms come with trade-offs such as higher costs and overhead in vendor lock-in and stateful workflows.}
         \end{itemize}
     \end{itemize}
\textcolor{black}{In these solution methods, a resource contention situation may arise when more than one function uses the same resources. Additionally, scaling is more difficult if these solutions contain stateful functions. Applications such as AWS Step Functions and Azure Durable Functions can be used to overcome all these problems.}

\subsection{Function Fusion-based Solutions:}
\begin{itemize}
         \item \textbf{Implementation Differences}
         \begin{itemize}
             \item \textcolor{black}{AWS Lambda combines multiple functions into a single function (AWS SAM) as a solution specifically for mitigating cold starts. Multiple functions can be incorporated into larger functions using GCF, and Google Cloud Console. Azure allows you to bypass cold start by combining multiple functions through applications such as the Azure Portal \cite{maenpaa2009cloud}.}
         \end{itemize}
         \item \textbf{Effectiveness and Trade-offs}
          \begin{itemize}
             \item \textcolor{black}{By combining functions on all three platforms, function call overhead is reduced and cold start latency is reduced. Although these have advantages, they result in more complex code and debugging. Additionally, large functions reduce autoscaling performance \cite{ferraris2012evaluating}.} 
         \end{itemize}
     \end{itemize}

\textcolor{black}{In these solution methods, since more than one function is combined, complexity in resource consumption and execution increases. As a
solution, the modular strategy, which involves the fusion of logically related functions,
can be used.}

\section{Appendix c: Quality Assessment and Publication Venues (Journals and Conferences)}

\textcolor{black}{After conducting a literature review, we carried out quality assessment procedures, as well as the methods in subsection 4.1, to select the most relevant ones from the 110 articles obtained. For this, we evaluated the quality of the studies using criteria such as objectivity and bias [68] and created a form as in Table \ref{table:examining}. The questions in the form were used to find research publications on the problem of serverless computing and cold start and to select articles within these publications that focused on solving the problem of cold start in serverless computing. Table \ref{table:examining2} contains evaluation information (SCI Rank) regarding the publication year of the articles obtained based on the answers to these questions, the number of citations, and the quality of the conference/journal.}

 \begin{table*}[ht]
\caption{\small \textcolor{black}{The Research Examining Questions }}
\label{table:examining}
\begin{center}
\footnotesize
\begin{tabular}{|l|c|c|}
\hline
\textbf{Question} &
  \textbf{Yes} &
  \textbf{No} \\ \hline
\begin{tabular}[c]{@{}l@{}}Q1) Is the cold start problem in serverless computing discussed in \\ the reviewed article?\end{tabular} &
   &
   \\ \hline
\begin{tabular}[c]{@{}l@{}}Q2) Is the main focus of this article on solving a problem \\ related to resource management in serverless computing?\end{tabular} &
   &
   \\ \hline
\begin{tabular}[c]{@{}l@{}}Q3) Does this article provide a solution method for the \\ cold start problem in serverless computing? (strategy, framework, AI model, etc.)\end{tabular} &
   &
   \\ \hline
\begin{tabular}[c]{@{}l@{}}Q4) Under which subheading is the presented solution \\ method examined? (reducing cold start latency time or \\ reducing the frequency of cold start occurrence etc.)\end{tabular} &
   &
   \\ \hline
\begin{tabular}[c]{@{}l@{}}Q5) Which solution method and which technique does \\ the presented solution method use? (Cache-based and FaaSLight etc.)\end{tabular} &
   &
   \\ \hline
\begin{tabular}[c]{@{}l@{}}Q6) Are any simulated environments or real platforms\\  used to test the offered solution performance? If yes, which platform is used?\end{tabular} &
   &
   \\ \hline
\begin{tabular}[c]{@{}l@{}}Q7) What is the published place, year of publication, \\ and number of citations of the article under review?\end{tabular} &
   &
   \\ \hline
\end{tabular}
\end{center}
\end{table*}

\begin{table*}[ht]
\caption{\small \textcolor{black}{Distribution of Journals and Conferences Containing Research on Serverless Computing}}
\label{table:examining2}
\begin{center}
\footnotesize
\begin{tabular}{|c|c|c|c|}
\hline
\textbf{Publication Venue and Year} & \textbf{Type} & \textbf{Citations} & \textbf{SCI Rank (H Index)} \\ \hline
2021 IEEE CCGRID                & Conferance & 52  & 30  \\ \hline
2024 Springer Cluster Computing & Journal    & 18  & 69  \\ \hline
2019 IEEE ICPADS                & Conferance & 43  & 46  \\ \hline
2019 USENIX HotCloud            & Journal    & 184 & 5   \\ \hline
2022 IEEE CCGRID                & Conferance & 13  & 30  \\ \hline
2019 ACM PICITDI                & Conferance & 172 & 12  \\ \hline
2018 IEEE/ACM UCC Companion     & Conferance & 77  & NA  \\ \hline
2021 ACM ASPLOS                 & Conference & 157 & 106 \\ \hline
2020 ACM ASPLOS                 & Conference & 248 & 106 \\ \hline
2020 Elsevier FGCS              & Journal    & 24  & 164 \\ \hline
2019 ACM SOCC                   & Conference & 102 & 17  \\ \hline
2020 IEEE MASCOTS               & Conference & 29  & 39  \\ \hline
2018 IEEE/ACM                   & Conference & 150 & 9   \\ \hline
2022 ACM ApPLIED                & Conference & 15  & NA  \\ \hline
2022 Wiley SPE                  & Journal    & 36  & 77  \\ \hline
2020 IEEE CloudCom              & Conference & 8   & 16  \\ \hline
2016 USENIX OSDI                & Conference & 100 & 8   \\ \hline
2020 ACM SOCC                   & Conference & 98  & 19  \\ \hline
2020 IEEE ACSOS                 & Conference & 73  & 4   \\ \hline
2021 ACM ASPLOS                 & Conference & 149 & 106 \\ \hline
2020 ACM DEBS                   & Conference & 90  & 14  \\ \hline
2023 ACM TSEM                   & Journal    & 31  & 88  \\ \hline
2019 Springer GECON             & Conferance & 23  & 3   \\ \hline
2020 Middleware                 & Conferance & 75  & 2   \\ \hline
2023 IEEE IoT Journal           & Journal    & 16  & 179 \\ \hline
2022 USENIX ATC                 & Conferance & 43  & 15  \\ \hline
2023 ACM ICDCN                  & Conferance & 8   & 11  \\ \hline
2021 MDPI Sensors               & Conferance & 30  & NA  \\ \hline
2022 IEEE INDICON               & Conferance & 6   & 6   \\ \hline
2022 IEEE IoT Journal           & Journal    & 43  & 179 \\ \hline
2020 ACM PICMC                  & Conferance & 98  & 11  \\ \hline
2019 ACM PFEC                   & Conferance & 77  & 22  \\ \hline
2018 USENUX ATC                 & Conferance & 468 & 15  \\ \hline
2020 IEEE ISCA                  & Conferance & 15  & 133 \\ \hline
2018 IEEE SmartCloud            & Conferance & 45  & 11  \\ \hline
2020 ACM PFECC                  & Conferance & 142 & 21  \\ \hline
2020 IEEE PIMC                  & Conferance & 79  & 11  \\ \hline
2018 USENIX ATC                 & Conferance & 376 & 15  \\ \hline
2021 IEEE CLUSTER               & Conferance & 20  & 51  \\ \hline
2020 ACM PSAC                   & Conferance & 84  & 76  \\ \hline
2020 IEEE IC2E                  & Conferance & 35  & 17  \\ \hline
2020 ACM PIMC                   & Conferance & 80  & 4   \\ \hline
\end{tabular}
\end{center}
\end{table*}

\bibliography{bibliography.bib}

\begin{thebibliography}{155}
\providecommand{\natexlab}[1]{#1}
\providecommand{\url}[1]{\texttt{#1}}
\expandafter\ifx\csname urlstyle\endcsname\relax
  \providecommand{\doi}[1]{doi: #1}\else
  \providecommand{\doi}{doi: \begingroup \urlstyle{rm}\Url}\fi

\bibitem[Gill et~al.(2022)Gill, Xu, Ottaviani, et~al.]{gill2022ai}
Sukhpal~Singh Gill, Minxian Xu, Carlo Ottaviani, et~al.
\newblock Ai for next generation computing: Emerging trends and future directions.
\newblock \emph{Internet of Things}, 19:\penalty0 100514, 2022.

\bibitem[Jonas et~al.(2019)Jonas, Schleier-Smith, Sreekanti, et~al.]{jonas2019cloud}
Eric Jonas, Johann Schleier-Smith, Vikram Sreekanti, et~al.
\newblock Cloud programming simplified: A berkeley view on serverless computing.
\newblock \emph{arXiv preprint arXiv:1902.03383}, 2019.

\bibitem[Villamizar et~al.(2016)Villamizar, Garces, Ochoa, et~al.]{villamizar2016infrastructure}
Mario Villamizar, Oscar Garces, Lina Ochoa, et~al.
\newblock Infrastructure cost comparison of running web applications in the cloud using aws lambda and monolithic and microservice architectures.
\newblock In \emph{2016 16th IEEE/ACM International Symposium on Cluster, Cloud and Grid Computing (CCGrid)}, pages 179--182. IEEE, 2016.

\bibitem[Golec et~al.(2023{\natexlab{a}})Golec, Gill, Parlikad, and Uhlig]{golec2023healthfaas}
Muhammed Golec, Sukhpal~Singh Gill, Ajith~Kumar Parlikad, and Steve Uhlig.
\newblock Healthfaas: Ai based smart healthcare system for heart patients using serverless computing.
\newblock \emph{IEEE Internet of Things Journal}, 2023{\natexlab{a}}.

\bibitem[Wen et~al.(2023)Wen, Chen, Jin, and Liu]{wen2023rise}
Jinfeng Wen, Zhenpeng Chen, Xin Jin, and Xuanzhe Liu.
\newblock Rise of the planet of serverless computing: A systematic review.
\newblock \emph{ACM Transactions on Software Engineering and Methodology}, 2023.

\bibitem[Chen et~al.(2018)Chen, Wan, Celesti, Li, Abbas, and Zhang]{chen2018edge}
Baotong Chen, Jiafu Wan, Antonio Celesti, Di~Li, Haider Abbas, and Qin Zhang.
\newblock Edge computing in iot-based manufacturing.
\newblock \emph{IEEE Communications Magazine}, 56\penalty0 (9):\penalty0 103--109, 2018.

\bibitem[Xie et~al.(2021)Xie, Tang, Qiao, Zhu, Yu, and Huang]{xie2021serverless}
Renchao Xie, Qinqin Tang, Shi Qiao, Han Zhu, F~Richard Yu, and Tao Huang.
\newblock When serverless computing meets edge computing: Architecture, challenges, and open issues.
\newblock \emph{IEEE Wireless Communications}, 28\penalty0 (5):\penalty0 126--133, 2021.

\bibitem[Baresi and Mendon{\c{c}}a(2019)]{baresi2019towards}
Luciano Baresi and Danilo~Filgueira Mendon{\c{c}}a.
\newblock Towards a serverless platform for edge computing.
\newblock In \emph{2019 IEEE International Conference on Fog Computing (ICFC)}, pages 1--10. IEEE, 2019.

\bibitem[Mar(2024)]{MarketsandMarkets}
2024.
\newblock URL \url{https://www.marketsandmarkets.com/Market-Reports/edge-computing-market-133384090.html}.

\bibitem[Golec et~al.(2021)Golec, Ozturac, Pooranian, Gill, and Buyya]{golec2021ifaasbus}
Muhammed Golec, Ridvan Ozturac, Zahra Pooranian, Sukhpal~Singh Gill, and Rajkumar Buyya.
\newblock Ifaasbus: A security-and privacy-based lightweight framework for serverless computing using iot and machine learning.
\newblock \emph{IEEE Transactions on Industrial Informatics}, 18\penalty0 (5):\penalty0 3522--3529, 2021.

\bibitem[Castro et~al.(2019)Castro, Ishakian, Muthusamy, and Slominski]{castro2019rise}
Paul Castro, Vatche Ishakian, Vinod Muthusamy, and Aleksander Slominski.
\newblock The rise of serverless computing.
\newblock \emph{Communications of the ACM}, 62\penalty0 (12):\penalty0 44--54, 2019.

\bibitem[Baldini et~al.(2017)Baldini, Castro, Chang, et~al.]{baldini2017serverless}
Ioana Baldini, Paul Castro, Kerry Chang, et~al.
\newblock Serverless computing: Current trends and open problems.
\newblock \emph{Research advances in cloud computing}, pages 1--20, 2017.

\bibitem[Kim and Lee(2019)]{kim2019practical}
Jeongchul Kim and Kyungyong Lee.
\newblock Practical cloud workloads for serverless faas.
\newblock In \emph{Proceedings of the ACM Symposium on Cloud Computing}, pages 477--477, 2019.

\bibitem[Golec et~al.(2023{\natexlab{b}})Golec, Iftikhar, Prabhakaran, Gill, and Uhlig]{golec2023qos}
Muhammed Golec, Sundas Iftikhar, Pratibha Prabhakaran, Sukhpal~Singh Gill, and Steve Uhlig.
\newblock Qos analysis for serverless computing using machine learning.
\newblock In \emph{Serverless Computing: Principles and Paradigms}, pages 175--192. Springer, 2023{\natexlab{b}}.

\bibitem[Keele(2007)]{keele}
Staffs Keele.
\newblock Guidelines for performing systematic literature reviews in software engineering.
\newblock \emph{Researchgate}, 2007.

\bibitem[Vahidinia et~al.(2020)Vahidinia, Farahani, and Aliee]{vahidinia2020cold}
Parichehr Vahidinia, Bahar Farahani, and Fereidoon~Shams Aliee.
\newblock Cold start in serverless computing: Current trends and mitigation strategies.
\newblock In \emph{International Conference on Omni-layer Intelligent Systems}, pages 1--7. IEEE, 2020.

\bibitem[Mampage et~al.(2022)Mampage, Karunasekera, and Buyya]{mampage2022holistic}
Anupama Mampage, Shanika Karunasekera, and Rajkumar Buyya.
\newblock A holistic view on resource management in serverless computing environments: Taxonomy and future directions.
\newblock \emph{ACM Computing Surveys}, 54\penalty0 (11s):\penalty0 1--36, 2022.

\bibitem[Li et~al.(2022{\natexlab{a}})Li, Guo, Cheng, Chen, He, and Guo]{li2022serverless2}
Zijun Li, Linsong Guo, Jiagan Cheng, Quan Chen, BingSheng He, and Minyi Guo.
\newblock The serverless computing survey: A technical primer for design architecture.
\newblock \emph{ACM Computing Surveys (CSUR)}, 54\penalty0 (10s):\penalty0 1--34, 2022{\natexlab{a}}.

\bibitem[Shafiei et~al.(2022)Shafiei, Khonsari, and Mousavi]{shafiei2022serverless}
Hossein Shafiei, Ahmad Khonsari, and Payam Mousavi.
\newblock Serverless computing: a survey of opportunities, challenges, and applications.
\newblock \emph{ACM Computing Surveys}, 54\penalty0 (11s):\penalty0 1--32, 2022.

\bibitem[Eismann et~al.(2021)Eismann, Scheuner, et~al.]{eismann2021state}
Simon Eismann, Joel Scheuner, et~al.
\newblock The state of serverless applications: Collection, characterization, and community consensus.
\newblock \emph{IEEE Transactions on Software Engineering}, 48\penalty0 (10):\penalty0 4152--4166, 2021.

\bibitem[Jawaddi and Ismail(2023)]{jawaddi2023autoscaling}
Siti Nuraishah~Agos Jawaddi and Azlan Ismail.
\newblock Autoscaling in serverless computing: Taxonomy and openchallenges.
\newblock \emph{europepmc}, 2023.

\bibitem[Hassan et~al.(2021)Hassan, Barakat, and Sarhan]{hassan2021survey}
Hassan~B Hassan, Saman~A Barakat, and Qusay~I Sarhan.
\newblock Survey on serverless computing.
\newblock \emph{Journal of Cloud Computing}, 10\penalty0 (1):\penalty0 1--29, 2021.

\bibitem[Li et~al.(2022{\natexlab{b}})Li, Lin, Wang, Ye, and Xu]{li2022serverless}
Yongkang Li, Yanying Lin, Yang Wang, Kejiang Ye, and Chengzhong Xu.
\newblock Serverless computing: state-of-the-art, challenges and opportunities.
\newblock \emph{IEEE Transactions on Services Computing}, 16\penalty0 (2):\penalty0 1522--1539, 2022{\natexlab{b}}.

\bibitem[Eismann et~al.(2024)Eismann, Scheuner, Eyk, Schwinger, Grohmann, Herbst, Abad, and Iosup]{eismann2008review}
S~Eismann, J~Scheuner, EV~Eyk, M~Schwinger, J~Grohmann, NR~Herbst, CL~Abad, and A~Iosup.
\newblock A review of serverless use cases and their characteristics. arxiv 2020.
\newblock \emph{arXiv preprint arXiv:2008.11110}, 2024.

\bibitem[Cassel et~al.(2022)Cassel, Rodrigues, da~Rosa~Righi, Bez, Nepomuceno, and da~Costa]{cassel2022serverless}
Gustavo Andr{\'e}~Setti Cassel, Vinicius~Facco Rodrigues, Rodrigo da~Rosa~Righi, Marta~Rosecler Bez, Andressa~Cruz Nepomuceno, and Cristiano~Andr{\'e} da~Costa.
\newblock Serverless computing for internet of things: A systematic literature review.
\newblock \emph{Future Generation Computer Systems}, 128:\penalty0 299--316, 2022.

\bibitem[Marin et~al.(2022)Marin, Perino, and Di~Pietro]{marin2022serverless}
Eduard Marin, Diego Perino, and Roberto Di~Pietro.
\newblock Serverless computing: a security perspective.
\newblock \emph{Journal of Cloud Computing}, 11\penalty0 (1):\penalty0 1--12, 2022.

\bibitem[Dittakavi(2023)]{dittakavi2023cold}
Raghava Satya~Saikrishna Dittakavi.
\newblock Cold start latency in serverless computing: Current trends and mitigation techniques.
\newblock \emph{Eduzone: International Peer Reviewed/Refereed Multidisciplinary Journal}, 12\penalty0 (2):\penalty0 135--139, 2023.

\bibitem[Mallick and Nath(2024)]{mallick2024securing}
Md~Abu~Imran Mallick and Rishab Nath.
\newblock Securing the server-less frontier: Challenges and innovative solutions in network security for server-less computing.
\newblock \emph{Reading Time}, 2024:\penalty0 04--15, 2024.

\bibitem[Ahmadi(2024)]{ahmadi2024challenges}
Sina Ahmadi.
\newblock Challenges and solutions in network security for serverless computing.
\newblock \emph{International Journal of Current Science Research and Review}, 7\penalty0 (01):\penalty0 218--229, 2024.

\bibitem[Donta et~al.(2023)Donta, Murturi, Casamayor~Pujol, Sedlak, and Dustdar]{donta2023exploring}
Praveen~Kumar Donta, Ilir Murturi, Victor Casamayor~Pujol, Boris Sedlak, and Schahram Dustdar.
\newblock Exploring the potential of distributed computing continuum systems.
\newblock \emph{Computers}, 12\penalty0 (10):\penalty0 198, 2023.

\bibitem[Donta and Dustdar(2023)]{donta2023towards}
Praveen~Kumar Donta and Schahram Dustdar.
\newblock Towards intelligent data protocols for the edge.
\newblock In \emph{2023 IEEE International Conference on Edge Computing and Communications (EDGE)}, pages 372--380. IEEE, 2023.

\bibitem[Huang et~al.(2018)Huang, Ma, and Hu]{huang2018machine}
Xin-Lin Huang, Xiaomin Ma, and Fei Hu.
\newblock Machine learning and intelligent communications.
\newblock \emph{Mobile Networks and Applications}, 23:\penalty0 68--70, 2018.

\bibitem[Nazari et~al.(2021)Nazari, Goodarzy, Keller, Rozner, and Mishra]{nazari2021optimizing}
Maziyar Nazari, Sepideh Goodarzy, Eric Keller, Eric Rozner, and Shivakant Mishra.
\newblock Optimizing and extending serverless platforms: A survey.
\newblock In \emph{2021 Eighth International Conference on Software Defined Systems (SDS)}, pages 1--8. IEEE, 2021.

\bibitem[Singh and Gill(2023)]{edgeai}
Raghubir Singh and Sukhpal~Singh Gill.
\newblock Edge ai: a survey.
\newblock \emph{Internet of Things and Cyber-Physical Systems}, 3:\penalty0 71--92, 2023.

\bibitem[Ali et~al.(2024{\natexlab{a}})Ali, Golec, Gill, Wu, Cuadrado, and Uhlig]{EdgeBus}
Babar Ali, Muhammed Golec, Sukhpal~Singh Gill, Huaming Wu, Felix Cuadrado, and Steve Uhlig.
\newblock Edgebus: Co-simulation based resource management for heterogeneous mobile edge computing environments.
\newblock \emph{Internet of Things}, 28:\penalty0 101368, 2024{\natexlab{a}}.
\newblock ISSN 2542-6605.

\bibitem[Kumar et~al.(2024)Kumar, Dwivedi, et~al.]{kumar2024comprehensive}
Surendra Kumar, Mridula Dwivedi, et~al.
\newblock A comprehensive review of vulnerabilities and ai-enabled defense against ddos attacks for securing cloud services.
\newblock \emph{Computer Science Review}, 53:\penalty0 100661, 2024.

\bibitem[Nandhakumar et~al.(2024)Nandhakumar, Baranwal, et~al.]{nandhakumar2023edgeaisim}
Aadharsh~Roshan Nandhakumar, Ayush Baranwal, et~al.
\newblock Edgeaisim: A toolkit for simulation and modelling of ai models in edge computing environments.
\newblock \emph{Measurement: Sensors}, 31:\penalty0 100939, 2024.

\bibitem[Bebortta et~al.(2020)Bebortta, Das, Kandpal, et~al.]{bebortta2020geospatial}
Sujit Bebortta, Saneev~Kumar Das, Meenakshi Kandpal, et~al.
\newblock Geospatial serverless computing: Architectures, tools and future directions.
\newblock \emph{ISPRS International Journal of Geo-Information}, 9\penalty0 (5):\penalty0 311, 2020.

\bibitem[Mete and Yomralioglu(2021)]{mete2021implementation}
Muhammed~Oguzhan Mete and Tahsin Yomralioglu.
\newblock Implementation of serverless cloud gis platform for land valuation.
\newblock \emph{International Journal of Digital Earth}, 14\penalty0 (7):\penalty0 836--850, 2021.

\bibitem[Franz et~al.(2018)Franz, Nagasuri, Wartman, Ventrella, and Esposito]{franz2018reunifying}
Justin Franz, Tanmayi Nagasuri, Andrew Wartman, Agnese~V Ventrella, and Flavio Esposito.
\newblock Reunifying families after a disaster via serverless computing and raspberry pis.
\newblock In \emph{2018 IEEE International Symposium on Local and Metropolitan Area Networks (LANMAN)}, pages 131--132. IEEE, 2018.

\bibitem[Herrera-Quintero et~al.(2018)Herrera-Quintero, Vega-Alfonso, Banse, and Zambrano]{herrera2018smart}
Luis~Felipe Herrera-Quintero, Julian~Camilo Vega-Alfonso, Klaus Bodo~Albert Banse, and Eduardo~Carrillo Zambrano.
\newblock Smart its sensor for the transportation planning based on iot approaches using serverless and microservices architecture.
\newblock \emph{IEEE Intelligent Transportation Systems Magazine}, 10\penalty0 (2):\penalty0 17--27, 2018.

\bibitem[Golec and Gill(2024)]{golec2024computing}
Muhammed Golec and Sukhpal~Singh Gill.
\newblock Computing: Looking back and moving forward.
\newblock \emph{21st International Conference on Smart Business Technologies (ICSBT 2024)}, pages 7--14, 2024.

\bibitem[Jefferson et~al.(2022)Jefferson, Chelliah, and Surianarayanan]{jefferson2022resource}
Steve Jefferson, Pethuru Chelliah, and Chellammal Surianarayanan.
\newblock A resource-optimized and accelerated sentiment analysis method using serverless computing.
\newblock \emph{Procedia Computer Science}, 215:\penalty0 33--43, 2022.

\bibitem[Dehghani et~al.(2020)Dehghani, Ghiasi, Niknam, Kavousi-Fard, Shasadeghi, Ghadimi, and Taghizadeh-Hesary]{dehghani2020blockchain}
Moslem Dehghani, Mohammad Ghiasi, Taher Niknam, Abdollah Kavousi-Fard, Mokhtar Shasadeghi, Noradin Ghadimi, and Farhad Taghizadeh-Hesary.
\newblock Blockchain-based securing of data exchange in a power transmission system considering congestion management and social welfare.
\newblock \emph{Sustainability}, 13\penalty0 (1):\penalty0 90, 2020.

\bibitem[Golec et~al.(2022)Golec, Chowdhury, Jaglan, Gill, and Uhlig]{golec2022aiblock}
Muhammed Golec, Deepraj Chowdhury, Shivam Jaglan, Sukhpal~Singh Gill, and Steve Uhlig.
\newblock Aiblock: Blockchain based lightweight framework for serverless computing using ai.
\newblock In \emph{2022 22nd IEEE International Symposium on Cluster, Cloud and Internet Computing (CCGrid)}, pages 886--892. IEEE, 2022.

\bibitem[Ali et~al.(2024{\natexlab{b}})Ali, Golec, Singh~Gill, Cuadrado, and Uhlig]{prokube}
Babar Ali, Muhammed Golec, Sukhpal Singh~Gill, Felix Cuadrado, and Steve Uhlig.
\newblock Prokube: Proactive kubernetes orchestrator for inference in heterogeneous edge computing.
\newblock \emph{International Journal of Network Management}, page e2298, 2024{\natexlab{b}}.

\bibitem[Benedict(2020)]{benedict2020serverless}
Shajulin Benedict.
\newblock Serverless blockchain-enabled architecture for iot societal applications.
\newblock \emph{IEEE Transactions on Computational Social Systems}, 7\penalty0 (5):\penalty0 1146--1158, 2020.

\bibitem[Sedlak et~al.(2024)Sedlak, Pujol, Donta, and Dustdar]{sedlak2024equilibrium}
Boris Sedlak, Victor~Casamayor Pujol, Praveen~Kumar Donta, and Schahram Dustdar.
\newblock Equilibrium in the computing continuum through active inference.
\newblock \emph{Future Generation Computer Systems}, 2024.

\bibitem[Mahmoudi et~al.(2019)Mahmoudi, Lin, Khazaei, and Litoiu]{mahmoudi2019optimizing}
Nima Mahmoudi, Changyuan Lin, Hamzeh Khazaei, and Marin Litoiu.
\newblock Optimizing serverless computing: Introducing an adaptive function placement algorithm.
\newblock In \emph{Proceedings of the 29th Annual International Conference on Computer Science and Software Engineering}, pages 203--213, 2019.

\bibitem[Pujol et~al.(2024)Pujol, Sedlak, Donta, and Dustdar]{pujol2024causality}
V{\'\i}ctor~Casamayor Pujol, Boris Sedlak, Praveen~Kumar Donta, and Schahram Dustdar.
\newblock On causality in distributed continuum systems.
\newblock \emph{IEEE Internet Computing}, 28\penalty0 (2):\penalty0 57--64, 2024.

\bibitem[Shin(2015)]{shin2015beyond}
Donghee Shin.
\newblock Beyond user experience of cloud service: Implication for value sensitive approach.
\newblock \emph{Telematics and Informatics}, 32\penalty0 (1):\penalty0 33--44, 2015.

\bibitem[Somma et~al.(2020)Somma, Ayimba, Casari, Romano, and Mancuso]{somma2020less}
Gaetano Somma, Constantine Ayimba, Paolo Casari, Simon~Pietro Romano, and Vincenzo Mancuso.
\newblock When less is more: Core-restricted container provisioning for serverless computing.
\newblock In \emph{IEEE INFOCOM 2020-IEEE Conference on Computer Communications Workshops (INFOCOM WKSHPS)}, pages 1153--1159. IEEE, 2020.

\bibitem[Gill et~al.(2024)Gill, Golec, Hu, et~al.]{gill2024edge}
Sukhpal~Singh Gill, Muhammed Golec, Jianmin Hu, et~al.
\newblock Edge ai: A taxonomy, systematic review and future directions.
\newblock \emph{Technical Report, arXiv preprint arXiv:2407.04053}, 2024.

\bibitem[Morabito et~al.(2018)Morabito, Cozzolino, Ding, Beijar, and Ott]{morabito2018consolidate}
Roberto Morabito, Vittorio Cozzolino, Aaron~Yi Ding, Nicklas Beijar, and Jorg Ott.
\newblock Consolidate iot edge computing with lightweight virtualization.
\newblock \emph{IEEE network}, 32\penalty0 (1):\penalty0 102--111, 2018.

\bibitem[Krishnamurthi et~al.(2023)]{krishnamurthi2023serverless}
Rajalakshmi Krishnamurthi et~al.
\newblock Serverless computing: New trends and research directions.
\newblock \emph{Serverless Computing: Principles and Paradigms}, pages 1--13, 2023.

\bibitem[Eismann et~al.(2020)Eismann, Scheuner, Van~Eyk, Schwinger, Grohmann, Herbst, Abad, and Iosup]{eismann2020serverless}
Simon Eismann, Joel Scheuner, Erwin Van~Eyk, Maximilian Schwinger, Johannes Grohmann, Nikolas Herbst, Cristina~L Abad, and Alexandru Iosup.
\newblock Serverless applications: Why, when, and how?
\newblock \emph{IEEE Software}, 38\penalty0 (1):\penalty0 32--39, 2020.

\bibitem[Singh et~al.(2022)]{singh2022machine}
Parminder Singh et~al.
\newblock Machine learning for cloud, fog, edge and serverless computing environments: comparisons, performance evaluation benchmark and future directions.
\newblock \emph{International Journal of Grid and Utility Computing}, 13\penalty0 (4):\penalty0 447--457, 2022.

\bibitem[El~Ioini et~al.(2021)El~Ioini, H{\"a}stbacka, Pahl, and Taibi]{el2021platforms}
Nabil El~Ioini, David H{\"a}stbacka, Claus Pahl, and Davide Taibi.
\newblock Platforms for serverless at the edge: a review.
\newblock In \emph{Advances in Service-Oriented and Cloud Computing: International Workshops of ESOCC 2020, Heraklion, Crete, Greece, September 28--30, 2020, Revised Selected Papers 8}, pages 29--40. Springer, 2021.

\bibitem[Aslanpour et~al.(2021)Aslanpour, Toosi, Cicconetti, et~al.]{aslanpour2021serverless}
Mohammad~S Aslanpour, Adel~N Toosi, Claudio Cicconetti, et~al.
\newblock Serverless edge computing: vision and challenges.
\newblock In \emph{Proceedings of the 2021 Australasian Computer Science Week Multiconference}, pages 1--10, 2021.

\bibitem[Voicu and Babonea(1997)]{voicu1997using}
Mirela-Cristina Voicu and Alina-Mihaela Babonea.
\newblock Using the snowball method in marketing research on hidden populations.
\newblock \emph{Social Problems}, 44\penalty0 (2):\penalty0 1341--1351, 1997.

\bibitem[Iftikhar et~al.(2022)Iftikhar, Gill, Song, et~al.]{iftikhar2022ai}
Sundas Iftikhar, Sukhpal~Singh Gill, Chenghao Song, et~al.
\newblock Ai-based fog and edge computing: A systematic review, taxonomy and future directions.
\newblock \emph{Internet of Things}, page 100674, 2022.

\bibitem[Kitchenham(2004)]{kitchenham2004procedures}
Barbara Kitchenham.
\newblock Procedures for performing systematic reviews.
\newblock \emph{Keele, UK, Keele University}, 33\penalty0 (2004):\penalty0 1--26, 2004.

\bibitem[Lee et~al.(2015)Lee, Park, and Baik]{lee2015location}
Kwangkyu Lee, Jinhee Park, and Jongmoon Baik.
\newblock Location-based web service qos prediction via preference propagation for improving cold start problem.
\newblock In \emph{2015 IEEE International Conference on Web Services}, pages 177--184. IEEE, 2015.

\bibitem[Chen et~al.(2017)Chen, Shen, Li, and You]{chen2017your}
Zhen Chen, Limin Shen, Feng Li, and Dianlong You.
\newblock Your neighbors alleviate cold-start: On geographical neighborhood influence to collaborative web service qos prediction.
\newblock \emph{Knowledge-Based Systems}, 138:\penalty0 188--201, 2017.

\bibitem[Ristov et~al.(2012)Ristov, Gusev, and Kostoska]{ristov2012new}
Sasko Ristov, Marjan Gusev, and Magdalena Kostoska.
\newblock A new methodology for security evaluation in cloud computing.
\newblock In \emph{2012 Proceedings of the 35th International Convention MIPRO}, pages 1484--1489. IEEE, 2012.

\bibitem[Bardsley et~al.(2018)Bardsley, Ryan, and Howard]{bardsley2018serverless}
Daniel Bardsley, Larry Ryan, and John Howard.
\newblock Serverless performance and optimization strategies.
\newblock In \emph{2018 IEEE International Conference on Smart Cloud (SmartCloud)}, pages 19--26. IEEE, 2018.

\bibitem[Ustiugov et~al.(2021)Ustiugov, Petrov, Kogias, Bugnion, and Grot]{ustiugov2021benchmarking}
Dmitrii Ustiugov, Plamen Petrov, Marios Kogias, Edouard Bugnion, and Boris Grot.
\newblock Benchmarking, analysis, and optimization of serverless function snapshots.
\newblock In \emph{Proceedings of the 26th ACM International Conference on Architectural Support for Programming Languages and Operating Systems}, pages 559--572, 2021.

\bibitem[Vahidinia et~al.(2022)Vahidinia, Farahani, and Aliee]{vahidinia2022mitigating}
Parichehr Vahidinia, Bahar Farahani, and Fereidoon~Shams Aliee.
\newblock Mitigating cold start problem in serverless computing: a reinforcement learning approach.
\newblock \emph{IEEE Internet of Things Journal}, 10\penalty0 (5):\penalty0 3917--3927, 2022.

\bibitem[Pelle et~al.(2019)Pelle, Czentye, D{\'o}ka, and Sonkoly]{pelle2019towards}
Istv{\'a}n Pelle, J{\'a}nos Czentye, J{\'a}nos D{\'o}ka, and Bal{\'a}zs Sonkoly.
\newblock Towards latency sensitive cloud native applications: A performance study on aws.
\newblock In \emph{12th International Conference on Cloud Computing}, pages 272--280. IEEE, 2019.

\bibitem[Chen et~al.(2016)Chen, Tang, Wang, Zhao, and Guo]{chen2016towards}
Shanshan Chen, Xiaoxin Tang, Hongwei Wang, Han Zhao, and Minyi Guo.
\newblock Towards scalable and reliable in-memory storage system: A case study with redis.
\newblock In \emph{2016 IEEE Trustcom/BigDataSE/ISPA}, pages 1660--1667. IEEE, 2016.

\bibitem[Gunarathne et~al.(2011)Gunarathne, Qiu, and Fox]{gunarathne2011iterative}
Thilina Gunarathne, Judy Qiu, and Geoffrey Fox.
\newblock Iterative mapreduce for azure cloud.
\newblock \emph{Proceedings of the CCA11 Cloud Computing and Its Applications}, pages 12--13, 2011.

\bibitem[Kavitha and Anita(2020)]{kavitha2020task}
C~Kavitha and X~Anita.
\newblock Task failure resilience technique for improving the performance of mapreduce in hadoop.
\newblock \emph{ETRI Journal}, 42\penalty0 (5):\penalty0 748--760, 2020.

\bibitem[Kulkarni et~al.(2020)Kulkarni, Chandramouli, and Stutsman]{kulkarni2020achieving}
Chinmay Kulkarni, Badrish Chandramouli, and Ryan Stutsman.
\newblock Achieving high throughput and elasticity in a larger-than-memory store.
\newblock \emph{arXiv preprint arXiv:2006.03206}, 2020.

\bibitem[Agarwal and Prasad(2012)]{agarwal2012azurebench}
Dinesh Agarwal and Sushil~K Prasad.
\newblock Azurebench: Benchmarking the storage services of the azure cloud platform.
\newblock In \emph{26th International Parallel and Distributed Processing Symposium Workshops \& PhD Forum}, pages 1048--1057, 2012.

\bibitem[Agarwal et~al.(2021)Agarwal, Rodriguez, and Buyya]{agarwal2021reinforcement}
Siddharth Agarwal, Maria~A Rodriguez, and Rajkumar Buyya.
\newblock A reinforcement learning approach to reduce serverless function cold start frequency.
\newblock In \emph{2021 IEEE/ACM 21st International Symposium on Cluster, Cloud and Internet Computing (CCGrid)}, pages 797--803. IEEE, 2021.

\bibitem[Jackson and Clynch(2018)]{jackson2018investigation}
David Jackson and Gary Clynch.
\newblock An investigation of the impact of language runtime on the performance and cost of serverless functions.
\newblock In \emph{2018 IEEE/ACM International Conference on Utility and Cloud Computing Companion (UCC Companion)}, pages 154--160. IEEE, 2018.

\bibitem[Manner et~al.(2018)Manner, Endre{\ss}, Heckel, and Wirtz]{manner2018cold}
Johannes Manner, Martin Endre{\ss}, Tobias Heckel, and Guido Wirtz.
\newblock Cold start influencing factors in function as a service.
\newblock In \emph{2018 IEEE/ACM International Conference on Utility and Cloud Computing Companion (UCC Companion)}, pages 181--188. IEEE, 2018.

\bibitem[Pawlik et~al.(2018)Pawlik, Figiela, and Malawski]{pawlik2018performance}
Maciej Pawlik, Kamil Figiela, and Maciej Malawski.
\newblock Performance evaluation of parallel cloud functions.
\newblock \emph{Poster presented at ICPP}, 2018.

\bibitem[Wang et~al.(2018)Wang, Li, Zhang, Ristenpart, and Swift]{wang2018peeking}
Liang Wang, Mengyuan Li, Yinqian Zhang, Thomas Ristenpart, and Michael Swift.
\newblock Peeking behind the curtains of serverless platforms.
\newblock In \emph{2018 USENIX Annual Technical Conference (USENIX ATC 18)}, pages 133--146, 2018.

\bibitem[Gim{\'e}nez-Alventosa et~al.(2019)Gim{\'e}nez-Alventosa, Molt{\'o}, and Caballer]{gimenez2019framework}
Vicent Gim{\'e}nez-Alventosa, Germ{\'a}n Molt{\'o}, and Miguel Caballer.
\newblock A framework and a performance assessment for serverless mapreduce on aws lambda.
\newblock \emph{Future Generation Computer Systems}, 97:\penalty0 259--274, 2019.

\bibitem[Lee et~al.(2018)Lee, Satyam, and Fox]{lee2018evaluation}
Hyungro Lee, Kumar Satyam, and Geoffrey Fox.
\newblock Evaluation of production serverless computing environments.
\newblock In \emph{2018 IEEE 11th International Conference on Cloud Computing (CLOUD)}, pages 442--450. IEEE, 2018.

\bibitem[Kumari and Sahoo(2023)]{kumari2023acpm}
Anisha Kumari and Bibhudatta Sahoo.
\newblock Acpm: adaptive container provisioning model to mitigate serverless cold-start.
\newblock \emph{Cluster Computing}, pages 1--28, 2023.

\bibitem[Akkus et~al.(2018)Akkus, Chen, Rimac, Stein, Satzke, Beck, Aditya, and Hilt]{akkus2018sand}
Istemi~Ekin Akkus, Ruichuan Chen, Ivica Rimac, Manuel Stein, Klaus Satzke, Andre Beck, Paarijaat Aditya, and Volker Hilt.
\newblock $\{$SAND$\}$: Towards $\{$High-Performance$\}$ serverless computing.
\newblock In \emph{2018 Usenix Annual Technical Conference (USENIX ATC 18)}, pages 923--935, 2018.

\bibitem[Kaffes et~al.(2019)Kaffes, Yadwadkar, and Kozyrakis]{kaffes2019centralized}
Kostis Kaffes, Neeraja~J Yadwadkar, and Christos Kozyrakis.
\newblock Centralized core-granular scheduling for serverless functions.
\newblock In \emph{Proceedings of the ACM symposium on cloud computing}, pages 158--164, 2019.

\bibitem[Du et~al.(2020)Du, Yu, Xia, et~al.]{du2020catalyzer}
Dong Du, Tianyi Yu, Yubin Xia, et~al.
\newblock Catalyzer: Sub-millisecond startup for serverless computing with initialization-less booting.
\newblock In \emph{Proceedings of the Twenty-Fifth International Conference on Architectural Support for Programming Languages and Operating Systems}, pages 467--481, 2020.

\bibitem[Mohan et~al.(2019)Mohan, Sane, Doshi, Edupuganti, Nayak, and Sukhomlinov]{mohan2019agile}
Anup Mohan, Harshad Sane, Kshitij Doshi, Saikrishna Edupuganti, Naren Nayak, and Vadim Sukhomlinov.
\newblock Agile cold starts for scalable serverless.
\newblock In \emph{11th USENIX Workshop on Hot Topics in Cloud Computing (HotCloud 19)}, 2019.

\bibitem[Qureshi(2010)]{qureshi2010power}
Asfandyar Qureshi.
\newblock \emph{Power-demand routing in massive geo-distributed systems}.
\newblock PhD thesis, Massachusetts Institute of Technology, 2010.

\bibitem[Liu et~al.(2023)Liu, Wen, Chen, Li, Chen, Liu, Wang, and Jin]{liu2023faaslight}
Xuanzhe Liu, Jinfeng Wen, Zhenpeng Chen, Ding Li, Junkai Chen, Yi~Liu, Haoyu Wang, and Xin Jin.
\newblock Faaslight: general application-level cold-start latency optimization for function-as-a-service in serverless computing.
\newblock \emph{ACM Transactions on Software Engineering and Methodology}, 2023.

\bibitem[Wang et~al.(2019)Wang, Ho, and Wu]{wang2019replayable}
Kai-Ting~Amy Wang, Rayson Ho, and Peng Wu.
\newblock Replayable execution optimized for page sharing for a managed runtime environment.
\newblock In \emph{Proceedings of the Fourteenth EuroSys Conference 2019}, pages 1--16, 2019.

\bibitem[Zuk and Rzadca(2020)]{zuk2020scheduling}
Pawel Zuk and Krzysztof Rzadca.
\newblock Scheduling methods to reduce response latency of function as a service.
\newblock In \emph{2020 IEEE 32nd International Symposium on Computer Architecture and High Performance Computing (SBAC-PAD)}, pages 132--140. IEEE, 2020.

\bibitem[Daw et~al.(2020)Daw, Bellur, and Kulkarni]{daw2020xanadu}
Nilanjan Daw, Umesh Bellur, and Purushottam Kulkarni.
\newblock Xanadu: Mitigating cascading cold starts in serverless function chain deployments.
\newblock In \emph{Proceedings of the 21st International Middleware Conference}, pages 356--370, 2020.

\bibitem[Zhang et~al.(2022)Zhang, Wang, Ma, Carver, et~al.]{zhang2022infinistore}
Jingyuan Zhang, Ao~Wang, Xiaolong Ma, Benjamin Carver, et~al.
\newblock Infinistore: Elastic serverless cloud storage.
\newblock \emph{arXiv preprint arXiv:2209.01496}, 2022.

\bibitem[Vergadia(2022)]{vergadia2022visualizing}
Priyanka Vergadia.
\newblock \emph{Visualizing Google Cloud: 101 Illustrated References for Cloud Engineers and Architects}.
\newblock John Wiley \& Sons, 2022.

\bibitem[Kumar(2022)]{kumar2022using}
Ankit Kumar.
\newblock \emph{Using Redis for persistent storage in serverless architecture to maintain state management}.
\newblock PhD thesis, Dublin, National College of Ireland, 2022.

\bibitem[Hall and Ramachandran(2019)]{hall2019execution}
Adam Hall and Umakishore Ramachandran.
\newblock An execution model for serverless functions at the edge.
\newblock In \emph{Proceedings of the International Conference on Internet of Things Design and Implementation}, pages 225--236, 2019.

\bibitem[Mistry et~al.(2020)Mistry, Stelea, Kumar, and Pasquier]{mistry2020demonstrating}
Chetankumar Mistry, Bogdan Stelea, Vijay Kumar, and Thomas Pasquier.
\newblock Demonstrating the practicality of unikernels to build a serverless platform at the edge.
\newblock In \emph{2020 IEEE International Conference on Cloud Computing Technology and Science (CloudCom)}, pages 25--32. IEEE, 2020.

\bibitem[Gadepalli et~al.(2020)Gadepalli, McBride, Peach, Cherkasova, and Parmer]{gadepalli2020sledge}
Phani~Kishore Gadepalli, Sean McBride, Gregor Peach, Ludmila Cherkasova, and Gabriel Parmer.
\newblock Sledge: A serverless-first, light-weight wasm runtime for the edge.
\newblock In \emph{Proceedings of the 21st International Middleware Conference}, pages 265--279, 2020.

\bibitem[Solaiman and Adnan(2020)]{solaiman2020wlec}
Khondokar Solaiman and Muhammad~Abdullah Adnan.
\newblock Wlec: A not so cold architecture to mitigate cold start problem in serverless computing.
\newblock In \emph{International Conference on Cloud Engineering}, pages 144--153. IEEE, 2020.

\bibitem[Oakes et~al.(2018)Oakes, Yang, Zhou, Houck, Harter, Arpaci-Dusseau, and Arpaci-Dusseau]{oakes2018sock}
Edward Oakes, Leon Yang, Dennis Zhou, Kevin Houck, Tyler Harter, Andrea Arpaci-Dusseau, and Remzi Arpaci-Dusseau.
\newblock $\{$SOCK$\}$: Rapid task provisioning with $\{$Serverless-Optimized$\}$ containers.
\newblock In \emph{2018 USENIX annual technical conference (USENIX ATC 18)}, pages 57--70, 2018.

\bibitem[Li et~al.(2022{\natexlab{c}})Li, Cheng, Chen, Guan, Bian, Tao, Zha, Wang, Han, and Guo]{li2022rund}
Zijun Li, Jiagan Cheng, Quan Chen, Eryu Guan, Zizheng Bian, Yi~Tao, Bin Zha, Qiang Wang, Weidong Han, and Minyi Guo.
\newblock $\{$RunD$\}$: A lightweight secure container runtime for high-density deployment and high-concurrency startup in serverless computing.
\newblock In \emph{2022 USENIX Annual Technical Conference (USENIX ATC 22)}, pages 53--68, 2022{\natexlab{c}}.

\bibitem[Shillaker and Pietzuch(2020)]{shillaker2020faasm}
Simon Shillaker and Peter Pietzuch.
\newblock Faasm: Lightweight isolation for efficient stateful serverless computing.
\newblock In \emph{2020 USENIX Annual Technical Conference (USENIX ATC 20)}, pages 419--433, 2020.

\bibitem[Vhatkar and Bhole(2020)]{vhatkar2020particle}
Kapil~Netaji Vhatkar and Girish~P Bhole.
\newblock Particle swarm optimisation with grey wolf optimisation for optimal container resource allocation in cloud.
\newblock \emph{IET Networks}, 9\penalty0 (4):\penalty0 189--199, 2020.

\bibitem[Hu et~al.(2018)Hu, Bai, Sha, Luo, Wang, and Wang]{hu2018hub}
Jingyuan Hu, Xiaokuang Bai, Sai Sha, Yingwei Luo, Xiaolin Wang, and Zhenlin Wang.
\newblock Hub: Hugepage ballooning in kernel-based virtual machines.
\newblock In \emph{Proceedings of the International Symposium on Memory Systems}, pages 31--37, 2018.

\bibitem[Ylenius(2020)]{ylenius2020mitigating}
Samuli Ylenius.
\newblock Mitigating javascript’s overhead with webassembly.
\newblock Master's thesis, University of Tampere, 2020.

\bibitem[Silva et~al.(2020)Silva, Fireman, and Pereira]{silva2020prebaking}
Paulo Silva, Daniel Fireman, and Thiago~Emmanuel Pereira.
\newblock Prebaking functions to warm the serverless cold start.
\newblock In \emph{Proceedings of the 21st International Middleware Conference}, pages 1--13, 2020.

\bibitem[Cadden et~al.(2020)Cadden, Unger, Awad, Dong, Krieger, and Appavoo]{cadden2020seuss}
James Cadden, Thomas Unger, Yara Awad, Han Dong, Orran Krieger, and Jonathan Appavoo.
\newblock Seuss: skip redundant paths to make serverless fast.
\newblock In \emph{Proceedings of the Fifteenth European Conference on Computer Systems}, pages 1--15, 2020.

\bibitem[Lee et~al.(2021)Lee, Yoon, Yeo, and Oh]{lee2021mitigating}
Seungjun Lee, Daegun Yoon, Sangho Yeo, and Sangyoon Oh.
\newblock Mitigating cold start problem in serverless computing with function fusion.
\newblock \emph{Sensors}, 21\penalty0 (24):\penalty0 8416, 2021.

\bibitem[Gunasekaran et~al.(2020)Gunasekaran, Thinakaran, Chidambaram, Kandemir, and Das]{gunasekaran2020fifer}
Jashwant~Raj Gunasekaran, Prashanth Thinakaran, Nachiappan Chidambaram, Mahmut~T Kandemir, and Chita~R Das.
\newblock Fifer: Tackling underutilization in the serverless era.
\newblock \emph{arXiv preprint arXiv:2008.12819}, 2020.

\bibitem[Bermbach et~al.(2020)Bermbach, Karakaya, and Buchholz]{bermbach2020using}
David Bermbach, Ahmet-Serdar Karakaya, and Simon Buchholz.
\newblock Using application knowledge to reduce cold starts in faas services.
\newblock In \emph{Proceedings of the 35th annual ACM symposium on applied computing}, pages 134--143, 2020.

\bibitem[Horovitz et~al.(2019)Horovitz, Amos, Baruch, Cohen, Oyar, and Deri]{horovitz2019faastest}
Shay Horovitz, Roei Amos, Ohad Baruch, Tomer Cohen, Tal Oyar, and Afik Deri.
\newblock Faastest-machine learning based cost and performance faas optimization.
\newblock In \emph{Economics of Grids, Clouds, Systems, and Services: 15th International Conference, GECON 2018, Pisa, Italy, September 18--20, 2018, Proceedings 15}, pages 171--186. Springer, 2019.

\bibitem[Golec et~al.(2023{\natexlab{c}})Golec, Gill, Cuadrado, Parlikad, Xu, Wu, and Uhlig]{golec2023atom}
Muhammed Golec, Sukhpal~Singh Gill, Felix Cuadrado, Ajith~Kumar Parlikad, Minxian Xu, Huaming Wu, and Steve Uhlig.
\newblock Atom: Ai-powered sustainable resource management for serverless edge computing environments.
\newblock \emph{IEEE Transactions on Sustainable Computing}, 2023{\natexlab{c}}.

\bibitem[Golec et~al.(2024)Golec, Gill, Wu, et~al.]{golec2024master}
Muhammed Golec, Sukhpal~Singh Gill, Huaming Wu, et~al.
\newblock Master: Machine learning-based cold start latency prediction framework in serverless edge computing environments for industry 4.0.
\newblock \emph{IEEE Journal of Selected Areas in Sensors}, 2024.

\bibitem[Lin and Glikson(2019)]{lin2019mitigating}
Ping-Min Lin and Alex Glikson.
\newblock Mitigating cold starts in serverless platforms: A pool-based approach.
\newblock \emph{arXiv preprint arXiv:1903.12221}, 2019.

\bibitem[Kumari et~al.(2022)Kumari, Sahoo, and Behera]{kumari2022mitigating}
Anisha Kumari, Bibhudatta Sahoo, and Ranjan~Kumar Behera.
\newblock Mitigating cold-start delay using warm-start containers in serverless platform.
\newblock In \emph{2022 IEEE 19th India Council International Conference (INDICON)}, pages 1--6. IEEE, 2022.

\bibitem[Xu et~al.(2019)Xu, Zhang, Geng, Wu, and Ma]{xu2019adaptive}
Zhengjun Xu, Haitao Zhang, Xin Geng, Qiong Wu, and Huadong Ma.
\newblock Adaptive function launching acceleration in serverless computing platforms.
\newblock In \emph{2019 IEEE 25th International Conference on Parallel and Distributed Systems (ICPADS)}, pages 9--16. IEEE, 2019.

\bibitem[Wu et~al.(2022)Wu, Tao, Fan, Huang, Zhang, Jin, Yu, and Cao]{wu2022container}
Song Wu, Zhiheng Tao, Hao Fan, Zhuo Huang, Xinmin Zhang, Hai Jin, Chen Yu, and Chun Cao.
\newblock Container lifecycle-aware scheduling for serverless computing.
\newblock \emph{Software: Practice and Experience}, 52\penalty0 (2):\penalty0 337--352, 2022.

\bibitem[Suresh et~al.(2020)Suresh, Somashekar, et~al.]{suresh2020ensure}
Amoghavarsha Suresh, Gagan Somashekar, et~al.
\newblock Ensure: Efficient scheduling and autonomous resource management in serverless environments.
\newblock In \emph{2020 IEEE International Conference on Autonomic Computing and Self-Organizing Systems (ACSOS)}, pages 1--10. IEEE, 2020.

\bibitem[Fuerst and Sharma(2021)]{fuerst2021faascache}
Alexander Fuerst and Prateek Sharma.
\newblock Faascache: keeping serverless computing alive with greedy-dual caching.
\newblock In \emph{Proceedings of the 26th ACM International Conference on Architectural Support for Programming Languages and Operating Systems}, pages 386--400, 2021.

\bibitem[Sethi et~al.(2023)Sethi, Addya, and Ghosh]{sethi2023lcs}
Biswajeet Sethi, Sourav~Kanti Addya, and Soumya~K Ghosh.
\newblock Lcs: alleviating total cold start latency in serverless applications with lru warm container approach.
\newblock In \emph{Proceedings of the 24th International Conference on Distributed Computing and Networking}, pages 197--206, 2023.

\bibitem[Suo et~al.(2021)Suo, Son, Cheng, Chen, and Baidya]{suo2021tackling}
Kun Suo, Junggab Son, Dazhao Cheng, Wei Chen, and Sabur Baidya.
\newblock Tackling cold start of serverless applications by efficient and adaptive container runtime reusing.
\newblock In \emph{2021 IEEE International Conference on Cluster Computing (CLUSTER)}, pages 433--443. IEEE, 2021.

\bibitem[Li et~al.(2022{\natexlab{d}})Li, Guo, Chen, Cheng, Xu, Zeng, Song, Ma, Yang, Li, et~al.]{li2022help}
Zijun Li, Linsong Guo, Quan Chen, Jiagan Cheng, Chuhao Xu, Deze Zeng, Zhuo Song, Tao Ma, Yong Yang, Chao Li, et~al.
\newblock Help rather than recycle: Alleviating cold startup in serverless computing through $\{$Inter-Function$\}$ container sharing.
\newblock In \emph{2022 USENIX Annual Technical Conference (USENIX ATC 22)}, pages 69--84, 2022{\natexlab{d}}.

\bibitem[Gias and Casale(2020)]{gias2020cocoa}
Alim~Ul Gias and Giuliano Casale.
\newblock Cocoa: Cold start aware capacity planning for function-as-a-service platforms.
\newblock In \emph{2020 28th International Symposium on Modeling, Analysis, and Simulation of Computer and Telecommunication Systems (MASCOTS)}, pages 1--8. IEEE, 2020.

\bibitem[Silva et~al.(2019)Silva, Carvalho, Pereira, Mour{\~a}o, and Rocha]{silva2019pure}
N{\'\i}collas Silva, Diego Carvalho, Adriano~CM Pereira, Fernando Mour{\~a}o, and Leonardo Rocha.
\newblock The pure cold-start problem: A deep study about how to conquer first-time users in recommendations domains.
\newblock \emph{Information Systems}, 80:\penalty0 1--12, 2019.

\bibitem[McGrath and Brenner(2017)]{mcgrath2017serverless}
Garrett McGrath and Paul~R Brenner.
\newblock Serverless computing: Design, implementation, and performance.
\newblock In \emph{2017 IEEE 37th International Conference on Distributed Computing Systems Workshops (ICDCSW)}, pages 405--410. IEEE, 2017.

\bibitem[P{\'e}rez et~al.(2018)P{\'e}rez, Molt{\'o}, Caballer, and Calatrava]{perez2018serverless}
Alfonso P{\'e}rez, Germ{\'a}n Molt{\'o}, Miguel Caballer, and Amanda Calatrava.
\newblock Serverless computing for container-based architectures.
\newblock \emph{Future Generation Computer Systems}, 83:\penalty0 50--59, 2018.

\bibitem[Chen et~al.(2023)Chen, Li, Ota, and Dong]{chen2023hyscaler}
Zhenke Chen, He~Li, Kaoru Ota, and Mianxiong Dong.
\newblock Hyscaler: a dynamic, hybrid vnf scaling system for building elastic service function chains across multiple servers.
\newblock \emph{IEEE Transactions on Network and Service Management}, 20\penalty0 (4):\penalty0 4803--4814, 2023.

\bibitem[Souri et~al.(2024)Souri, Mood, Gao, and Li]{souri2024tournament}
Alireza Souri, Sepehr~Ebrahimi Mood, Mingliang Gao, and Kuan-Ching Li.
\newblock Tournament based equilibrium optimization for minimizing energy consumption on dynamic task scheduling in cloud-edge computing.
\newblock \emph{Cluster Computing}, pages 1--13, 2024.

\bibitem[Koch and Hao(2021)]{koch2021empirical}
Jacob Koch and Wei Hao.
\newblock An empirical study in edge computing using aws.
\newblock In \emph{2021 IEEE 11th Annual Computing and Communication Workshop and Conference (CCWC)}, pages 0542--0549. IEEE, 2021.

\bibitem[Srinivasan et~al.(2018)Srinivasan, Ravi, and Raj]{srinivasan2018google}
Vitthal Srinivasan, Janani Ravi, and Judy Raj.
\newblock \emph{Google Cloud Platform for Architects: Design and manage powerful cloud solutions}.
\newblock Packt Publishing Ltd, 2018.

\bibitem[Olariu and Alboaie(2023)]{olariu2023challenges}
Florin Olariu and Lenuța Alboaie.
\newblock Challenges in optimizing migration costs from on-premises to microsoft azure.
\newblock \emph{Procedia Computer Science}, 225:\penalty0 3649--3659, 2023.

\bibitem[Chowdhury et~al.(2023)Chowdhury, Das, Dey, et~al.]{chowdhury2023covidetector}
Deepraj Chowdhury, Anik Das, Ajoy Dey, et~al.
\newblock Covidetector: A transfer learning-based semi supervised approach to detect covid-19 using cxr images.
\newblock \emph{BenchCouncil Transactions on Benchmarks, Standards and Evaluations}, 3\penalty0 (2):\penalty0 100119, 2023.

\bibitem[Gill and Buyya(2024)]{Transforming2024}
Sukhpal~Singh Gill and Rajkumar Buyya.
\newblock Transforming research with quantum computing.
\newblock \emph{Journal of Economy and Technology}, 2:\penalty0 1--8, 2024.
\newblock ISSN 2949-9488.

\bibitem[Agache et~al.(2020)Agache, Brooker, Iordache, Liguori, Neugebauer, Piwonka, and Popa]{agache2020firecracker}
Alexandru Agache, Marc Brooker, Alexandra Iordache, Anthony Liguori, Rolf Neugebauer, Phil Piwonka, and Diana-Maria Popa.
\newblock Firecracker: Lightweight virtualization for serverless applications.
\newblock In \emph{17th USENIX symposium on networked systems design and implementation (NSDI 20)}, pages 419--434, 2020.

\bibitem[Young et~al.(2019)Young, Zhu, Caraza-Harter, Arpaci-Dusseau, and Arpaci-Dusseau]{young2019true}
Ethan~G Young, Pengfei Zhu, Tyler Caraza-Harter, Andrea~C Arpaci-Dusseau, and Remzi~H Arpaci-Dusseau.
\newblock The true cost of containing: A $\{$gVisor$\}$ case study.
\newblock In \emph{11th USENIX Workshop on Hot Topics in Cloud Computing}, 2019.

\bibitem[Banaie and Djemame(2022)]{banaie2022serverless}
Fatemeh Banaie and Karim Djemame.
\newblock A serverless computing platform for software defined networks.
\newblock In \emph{International Conference on the Economics of Grids, Clouds, Systems, and Services}, pages 113--123. Springer, 2022.

\bibitem[Manvi and Shyam(2014)]{manvi2014resource}
Sunilkumar~S Manvi and Gopal~Krishna Shyam.
\newblock Resource management for infrastructure as a service (iaas) in cloud computing: A survey.
\newblock \emph{Journal of network and computer applications}, 41:\penalty0 424--440, 2014.

\bibitem[Beimborn et~al.(2011)Beimborn, Miletzki, and Wenzel]{beimborn2011platform}
Daniel Beimborn, Thomas Miletzki, and Stefan Wenzel.
\newblock Platform as a service (paas).
\newblock \emph{Wirtschaftsinformatik}, 53:\penalty0 371--375, 2011.

\bibitem[Cusumano(2010)]{cusumano2010cloud}
Michael Cusumano.
\newblock Cloud computing and saas as new computing platforms.
\newblock \emph{Communications of the ACM}, 53\penalty0 (4):\penalty0 27--29, 2010.

\bibitem[Shahrad et~al.(2019)Shahrad, Balkind, and Wentzlaff]{shahrad2019architectural}
Mohammad Shahrad, Jonathan Balkind, and David Wentzlaff.
\newblock Architectural implications of function-as-a-service computing.
\newblock In \emph{Proceedings of the 52nd annual IEEE/ACM international symposium on microarchitecture}, pages 1063--1075, 2019.

\bibitem[Balla et~al.(2020)Balla, Maliosz, and Simon]{balla2020open}
David Balla, Markosz Maliosz, and Csaba Simon.
\newblock Open source faas performance aspects.
\newblock In \emph{2020 43rd International Conference on Telecommunications and Signal Processing (TSP)}, pages 358--364. IEEE, 2020.

\bibitem[Le et~al.(2022)Le, Pal, and Pattnaik]{le2022openfaas}
Dac-Nhuong Le, Souvik Pal, and Prasant~Kumar Pattnaik.
\newblock Openfaas.
\newblock \emph{Cloud Computing Solutions: Architecture, Data Storage, Implementation and Security}, pages 287--303, 2022.

\bibitem[Quevedo et~al.(2019)Quevedo, Merch{\'a}n, Rivadeneira, and Dominguez]{quevedo2019evaluating}
Sebasti{\'a}n Quevedo, Freddy Merch{\'a}n, Rafael Rivadeneira, and Federico~X Dominguez.
\newblock Evaluating apache openwhisk-faas.
\newblock In \emph{2019 IEEE fourth ecuador technical chapters meeting (ETCM)}, pages 1--5. IEEE, 2019.

\bibitem[Malawski et~al.(2020)Malawski, Gajek, Zima, Balis, and Figiela]{malawski2020serverless}
Maciej Malawski, Adam Gajek, Adam Zima, Bartosz Balis, and Kamil Figiela.
\newblock Serverless execution of scientific workflows: Experiments with hyperflow, aws lambda and google cloud functions.
\newblock \emph{Future Generation Computer Systems}, 110:\penalty0 502--514, 2020.

\bibitem[Chappell et~al.(2008)]{chappell2008introducing}
David Chappell et~al.
\newblock Introducing the azure services platform.
\newblock \emph{White paper, Oct}, 1364\penalty0 (11), 2008.

\bibitem[Schwarzl et~al.(2022)Schwarzl, Borrello, Kogler, Varda, Schuster, Schwarz, and Gruss]{schwarzl2022robust}
Martin Schwarzl, Pietro Borrello, Andreas Kogler, Kenton Varda, Thomas Schuster, Michael Schwarz, and Daniel Gruss.
\newblock Robust and scalable process isolation against spectre in the cloud.
\newblock In \emph{European Symposium on Research in Computer Security}, pages 167--186. Springer, 2022.

\bibitem[Hafeez et~al.(2018)Hafeez, Wajahat, and Gandhi]{hafeez2018elmem}
Ubaid~Ullah Hafeez, Muhammad Wajahat, and Anshul Gandhi.
\newblock Elmem: Towards an elastic memcached system.
\newblock In \emph{2018 IEEE 38th International Conference on Distributed Computing Systems (ICDCS)}, pages 278--289. IEEE, 2018.

\bibitem[Tulving and Psotka(1971)]{tulving1971retroactive}
Endel Tulving and Joseph Psotka.
\newblock Retroactive inhibition in free recall: Inaccessibility of information available in the memory store.
\newblock \emph{Journal of experimental Psychology}, 87\penalty0 (1):\penalty0 1, 1971.

\bibitem[Karthikeyan(2017)]{karthikeyan2017azure}
Shijimol~Ambi Karthikeyan.
\newblock \emph{Azure automation using the ARM model: an in-depth guide to automation with Azure resource manager}.
\newblock Apress, 2017.

\bibitem[Kiener et~al.(2021)Kiener, Chadha, and Gerndt]{kiener2021towards}
Michael Kiener, Mohak Chadha, and Michael Gerndt.
\newblock Towards demystifying intra-function parallelism in serverless computing.
\newblock In \emph{Proceedings of the Seventh International Workshop on Serverless Computing (WoSC7) 2021}, pages 42--49, 2021.

\bibitem[Martens and Teuteberg(2012)]{martens2012decision}
Benedikt Martens and Frank Teuteberg.
\newblock Decision-making in cloud computing environments: A cost and risk based approach.
\newblock \emph{Information Systems Frontiers}, 14:\penalty0 871--893, 2012.

\bibitem[Figiela et~al.(2018)Figiela, Gajek, Zima, Obrok, and Malawski]{figiela2018performance}
Kamil Figiela, Adam Gajek, Adam Zima, Beata Obrok, and Maciej Malawski.
\newblock Performance evaluation of heterogeneous cloud functions.
\newblock \emph{Concurrency and Computation: Practice and Experience}, 30\penalty0 (23):\penalty0 e4792, 2018.

\bibitem[Cao et~al.(2015)Cao, Simonin, Cooperman, and Morin]{cao2015checkpointing}
Jiajun Cao, Matthieu Simonin, Gene Cooperman, and Christine Morin.
\newblock Checkpointing as a service in heterogeneous cloud environments.
\newblock In \emph{2015 15th IEEE/ACM International Symposium on Cluster, Cloud and Grid Computing}, pages 61--70. IEEE, 2015.

\bibitem[Chowhan(2018)]{chowhan2018hands}
Kuldeep Chowhan.
\newblock \emph{Hands-on Serverless Computing: Build, Run and Orchestrate Serverless Applications Using AWS Lambda, Microsoft Azure Functions, and Google Cloud Functions}.
\newblock Packt Publishing Ltd, 2018.

\bibitem[M{\"a}enp{\"a}{\"a}(2009)]{maenpaa2009cloud}
Jouni M{\"a}enp{\"a}{\"a}.
\newblock Cloud computing with the azure platform.
\newblock In \emph{TKK T-110.5190 Seminar on Internet Working}, 2009.

\bibitem[Ferraris et~al.(2012)Ferraris, Franceschelli, Gioiosa, Lucia, Ardagna, Di~Nitto, and Sharif]{ferraris2012evaluating}
Filippo~Lorenzo Ferraris, Davide Franceschelli, Mario~Pio Gioiosa, Donato Lucia, Danilo Ardagna, Elisabetta Di~Nitto, and Tabassum Sharif.
\newblock Evaluating the auto scaling performance of flexiscale and amazon ec2 clouds.
\newblock In \emph{2012 14th International Symposium on Symbolic and Numeric Algorithms for Scientific Computing}, pages 423--429. IEEE, 2012.

\end{thebibliography}

\end{document}